\documentclass{aastex701}

\newcommand{\ie}{i.e.\/}

\usepackage{amsmath}
\usepackage{amssymb}
\usepackage{graphicx}

\shorttitle{Dust Continuum Emission}
\shortauthors{Shirley et al.}
\graphicspath{{./}{figures/}}

\begin{document}

\title{How To Use Thermal Dust Continuum Emission To Measure The Physical Properties Of Dusty Astrophysical Objects}

\correspondingauthor{Yancy L. Shirley}
\email{yshirley@arizona.edu}

\author{Yancy L. Shirley}
\email{yshirley@arizona.edu}
\affiliation{Department of Astronomy \& Steward Observatory \\
The University of Arizona \\
933 N Cherry Ave. \\
Tucson, AZ 85721, USA}
\affiliation{Herzberg Astronomy and Astrophysics Research Centre \\ National Research Council of Canada \\ 5071 West Saanich Road \\ Victoria, BC, V9E 2E7, Canada}


\author{Jeffrey G. Mangum}
\email{jeff.mangum@nrao.edu}
\affiliation{National Radio Astronomy Observatory \\ 520 Edgemont Rd \\
Charlottesville, VA 22903, USA}

\author{Desika Narayanan}
\email{desika.narayanan@ufl.edu}
\affiliation{Department of Astronomy \\ University of Florida \\ 211 Bryant Space Sciences Center \\ Gainesville, FL 32611, USA}

\author{James Di Francesco}
\email{James.DiFrancesco@nrc.ca}
\affiliation{Herzberg Astronomy and Astrophysics Research Centre \\ National Research Council of Canada \\ 5071 West Saanich Road \\ Victoria, BC, V9E 2E7, Canada}



\begin{abstract}

Dust grains in the interstellar medium interact with photons across the electromagnetic spectrum. 
They are generally photon energy converters, absorbing short wavelength radiation and emitting long wavelength radiation.
Sixty years ago in 1965, thermal emission from dust grains in the interstellar medium was discovered.
This tutorial is a summary of the physics of thermal dust continuum emission and how to use observations of the intensity and flux density of dusty objects to calculate physical properties such as mass, column density, luminosity, dust temperature, and dust opacity spectral index.
Equations are derived, when feasible, from first principles with all limits and assumptions explicitly stated.
Properties of dust opacities appropriate for different astrophysical environments (e.g. diffuse ISM, dense cores, protoplanetary disks) are discussed and tabulated for the wavelengths of past, current, and future bolometer cameras.
Corrections for observations at high redshift as well as the effects of telescope measurement limitations are derived.
We also update the calculation of the mean molecular weight in different ISM environments and find that it is $1.404$ per H atom, $2.809$ per H$_2$ molecule, and $2.351$ per gas particle assuming protosolar metallicity and the latest values of the ISM gas phase abundances of metals.

\end{abstract}



\section{Introduction} \label{sec:intro}

Dust grains are solid particles that float in interplanetary \citep{2019SSRv..215...34K} and interstellar space \citep{2003ARA&A..41..241D}. 
They are composed of the refractory elements of the periodic table with $>99$\% of the interstellar dust grain mass consisting of the elements H, C, O, Mg, Al, Si, S, Ca, Fe, and Ni \citep{2021ApJ...906...73H}.  
Amorphous silicates and carbonaceous materials dominate the composition of interstellar dust grains.
They come in various shapes and sizes, from just a few atoms arranged in regular, highly symmetric patterns (\ie\ polycyclic aromatic hydrocarbons or  PAHs, \citealt{1989ApJS...71..733A,2008ARA&A..46..289T}) to amorphous collections of atoms that range in size up to pebbles in protoplanetary disks \citep{ 2014prpl.conf..339T}.

\begin{figure}[tbh]
\centering 
\includegraphics[scale=0.66, trim= 0mm 71.8mm 0mm 71.8mm, clip]{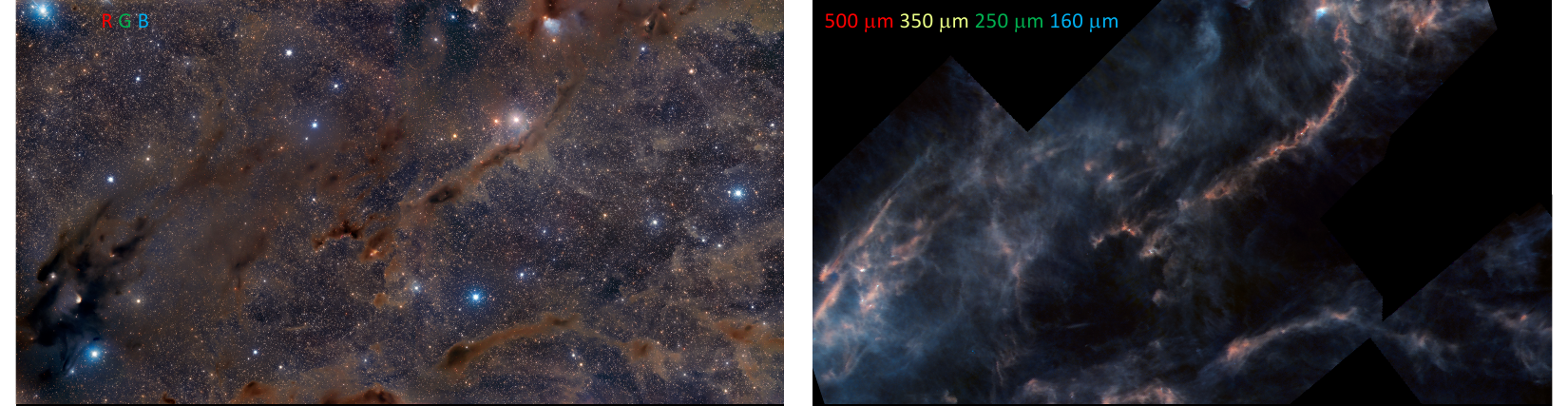}
\caption{LEFT: Optical color image of dust absorption and scattering in the Taurus Molecular Cloud. The optical image is a 21 hour exposure taken by Adam Block (Steward Observatory/University of Arizona) 
using the Pomenis Astrograph (Takahashi E-180 - Epsilon f2.8 180mm ED) with an APOGEE ALTA F9000 
camera and equal integration time with R, G, B filters.  RIGHT: \textit{Herschel Space Observatory} image of thermal dust emission in the Taurus Molecular Cloud.  The colors correspond to 160 $\mu$m (blue), 250 $\mu$m (green), 350 $\mu$m (split between green and red), and 500 $\mu$m (red).  Regions with clear dust absorption in the optical image glow with thermal dust emission at far-infrared and submillimeter wavelengths. Credit: ESA/Herschel/NASA/JPL-Caltech, CC BY-SA 3.0 IGO; Acknowledgement: R. Hurt (JPL-Caltech).   
Note that some portions covered by the optical image were not imaged by the \textit{Herschel Space Observatory}.
}
\label{fig:OptSubmmComp}
\end{figure}

\begin{figure}[h]
\includegraphics[scale=0.42, trim= 30mm 3mm 30mm 10mm, clip]{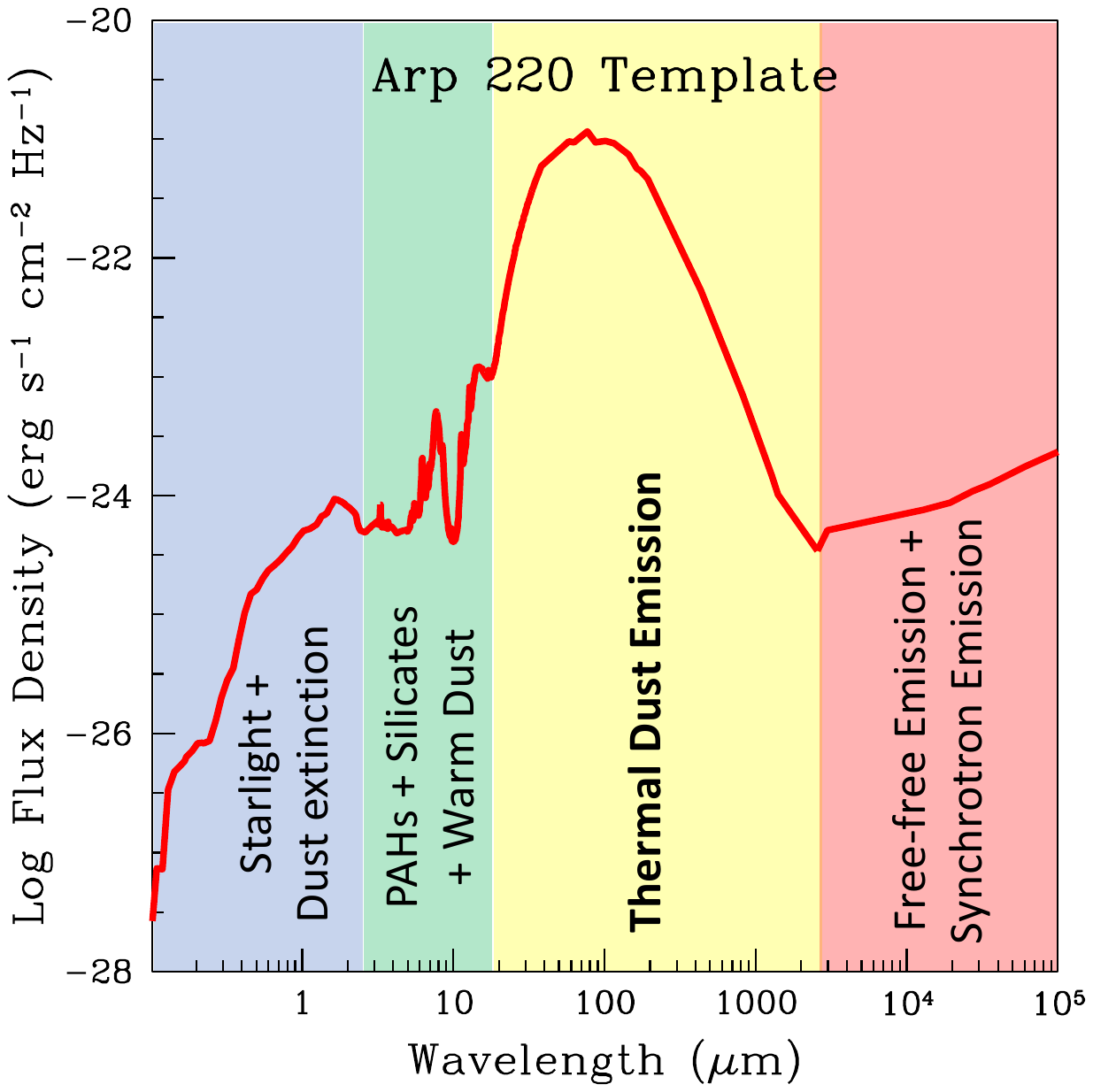}
\includegraphics[scale=0.23, trim= 40mm 40mm 40mm 40mm, clip]{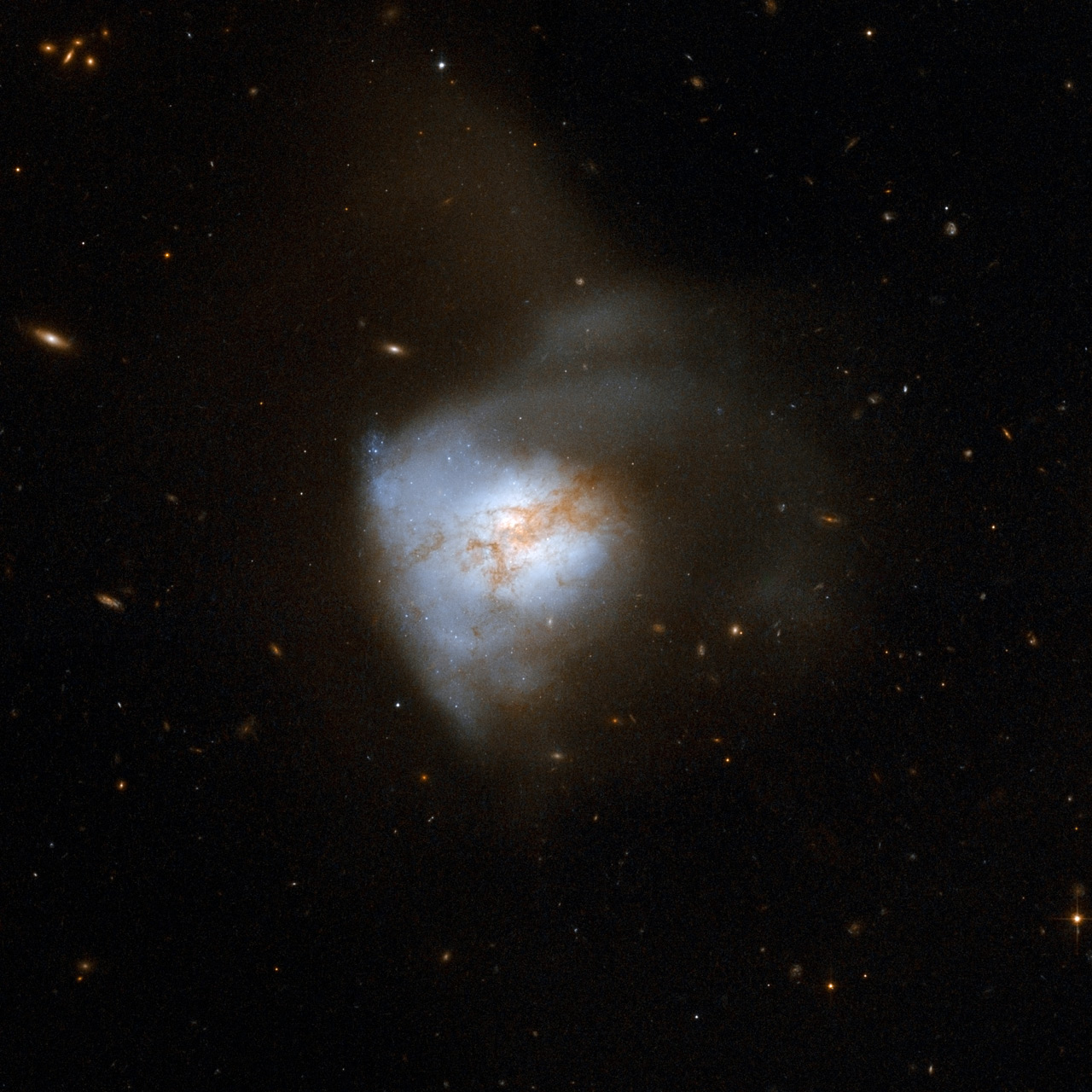}
\caption{\textbf{LEFT:} The Spectral Energy Distribution (SED) of a template Ultra Luminous InfraRed Galaxy (ULIRG) is dominated by thermal dust emission at far-infrared and submillimeter wavelengths (shaded yellow region from $\sim 20$ $\mu$m to $3$ mm). 
The Arp 220 template (red curve) is from \citealt{2007ApJ...660..167D}. \textbf{RIGHT}: \textit{Hubble Space Telescope} image of Arp 220. The image is a combination of filters F435W (B) and F814W (I). Credit: NASA, ESA, the Hubble Heritage (STScI/AURA)-ESA/Hubble Collaboration, and A. Evans (University of Virginia, Charlottesville/NRAO/Stony Brook University)\footnote{This image may be downloaded from \url{https://science.nasa.gov/asset/hubble/hubble-interacting-galaxy-arp-220/}.}} 
\label{fig:Arp220SED}
\end{figure}

Photons across the entire electromagnetic spectrum, from high energy gamma rays to low energy radio waves, interact with interstellar dust grains by being absorbed, scattered, or emitted.
The efficiency with which a dust grain absorbs or scatters light varies with the size and composition of the dust grain and the wavelength of light.  
A dust grain's absorption and scattering efficiencies are typically highest when the wavelength of light is comparable, within factors of a few, of the dust grain size \citep{Draine2011}.
Energy that is absorbed by dust grains, generally from short wavelength radiation, is re-emitted at longer wavelengths depending on the dust grain temperature and opacity\footnote{Also known as  emissivity, although we shall use the term opacity throughout this article.}. 
One environment where the processes of dust absorption, scattering, and thermal emission are easily observable is in dusty molecular clouds, the sites of star formation in galaxies (see Figure \ref{fig:OptSubmmComp}).
Even though dust grains only comprise about $0.1$\% of the total baryonic matter in the Milky Way Galaxy, approximately half of the energy radiated by stars in the Milky Way is absorbed by dust grains in the interstellar medium and re-radiated at infrared wavelengths  \citep{2003ARA&A..41..241D}.
In extreme dusty galaxies undergoing a burst of star formation (\ie\ ULIRGS or Ultra Luminous InfraRed Galaxies), their spectral energy distributions are dominated by the thermal emission from dust grains (see Figure \ref{fig:Arp220SED}).
Given the ubiquity of dust in different astrophysical environments within galaxies (\ie, diffuse cirrus, molecular clouds, dense cores, protoplanetary disks, debris disks, and the late stages of stellar evolution), and the importance of dust in reprocessing radiation, a detailed understanding of dust radiative processes is necessary to interpret many astronomical observations.

Photographic plates taken toward nearby molecular clouds at the beginning of the 20th century helped convince E. E. Barnard that ``vacancies" (of stars) in the Milky Way were actually obscuring clouds of material  (\citealt{Barnard1907ApJ}, see Figure \ref{fig:OptSubmmComp} for a modern image of the region that E. E. Barnard observed in 1907.).  
Robert Trumpler is formally credited with the discovery of interstellar dust in 1930 from his observations of the variation in the size and color excess of open clusters at different distances \citep{1930LicOB..14..154T}. 
He deduced that the Galactic  interstellar medium (ISM) was imbued with a ``selective absorber" (meaning varying with wavelength) that could be due to ``fine cosmic dust"  \citep{1930PASP...42..214T}.  
His conclusion was supported by his synthesis of prior observations of dark clouds   (e.g.,  \citealt{Barnard1907ApJ, 1913ApJ....38..496B, 1919ApJ....49....1B,  1927pasr.book.....B}), of star counts \citep{1919MNRAS..80..162H, 1929AN....236..249S}, of stellar reddening  \citep{1909ApJ....30..284K, 1909ApJ....30..398K, 1914MNRAS..75....4S, 1916ApJ....43...36V}, and of color excess in stellar clusters\footnote{See Table 1 of \citealt{1930PASP...42..214T} for a reference list.}.
Early work on dust focused on its absorption and scattering properties at optical wavelengths  
(see \citealt{2005JPhCS...6..229L} for a history of early dust studies). 
It was not until the 1960s that infrared  observations revealed thermal emission from dust grains heated by  nearby stars \citep{1965ApJ...142..399N, 1965ApJ...142..808J}.
Most interstellar dust grains in the ISM are not in the immediate vicinity of a star and are therefore heated by the ambient interstellar radiation field.
As a result, their equilibrium temperature is cold ($T_d \sim 10 - 20$ K), resulting in the bulk of their emission being at far-infrared through millimeter wavelengths (see Figures \ref{fig:OptSubmmComp} and \ref{fig:Arp220SED}).
This cold dust emission was surveyed at far-infrared through millimeter wavelengths in the ISM of the Milky Way \citep{1984ApJ...278L..19L, 2013ApJS..208...14G,2016A&A...594A..28P,2018MNRAS.473.1059U,2021MNRAS.504.2742E}.
Thermal dust emission is even detected in the ISM of nascent galaxies emitted within the first 650 million years after the Big Bang (\ie, $z > 8$,  \citealt{2017ApJ...837L..21L,2022MNRAS.515.3126I,2023ApJ...952....9T}).

In this article we focus on the radiative processes of thermal dust emission.
This article is intended to be a tutorial for students and a review for researchers.
It began as notes and homework problems for ASTR 300B, a third year (sixth semester) undergraduate class studying radiative processes in astrophysics at The University of Arizona.
There is extensive literature on the properties of dust absorption, scattering, and emission (\ie,  \citealt{1983QJRAS..24..267H,1993duun.book.....E,1998ppim.book.....S,2003ARA&A..41..241D,Draine2011,2022dge..book.....W}).
The purposes of this tutorial are to collect the equations found in various places in the literature into one document (see Table \ref{tab:eqsummary} in Appendix \ref{AppendixEquationSummary}) and to fill in gaps in the derivations of those equations.
Wherever pedagogically practical, we derive equations from first principles or start from the most general case with minimal assumptions and then show the effect of those assumptions on the equations.
The limits that apply to each equation are given explicitly.
It is assumed that the reader has a basic understanding of radiative transfer (\ie, Chapter 1 of  \citealt{1986rpa..book.....R}).
We stick to 1D radiative transfer for the majority of the article with a brief introduction to 3D radiative transfer with numerical codes in the final section.
We do not consider polarized emission and scattering since this would require 3D radiative transfer from the outset  (see \citealt{2021ApJ...906...73H} for a summary of polarized dust emission).

This tutorial begins by presenting the geometry of the 1D radiative transfer problem for thermal dust emission and defining fundamental variables (Section \ref{sec:RadTrans}).
We update the calculation of the mean molecular weight appropriate for different astrophysical environments (Section \ref{sec:RadTrans} and Appendix \ref{sec:AppendixMuZ}).
The solution to the 1D radiative transfer equation for thermal dust emission is derived (Section \ref{sec:1DSolution}) with the conditions established for when the optically thin approximation is appropriate (Section \ref{sec:OpticallyThinSection}).
Formulae are derived for the monochromatic specific intensity (Section \ref{sec:1DSolution} along with an example solution for spherical geometry in Section \ref{SphGeoApp} and Appendix \ref{AbelApp}), interferometric visibilities (Appendix \ref{AppendixVisibilities}), flux density (Section \ref{sec:FluxDensity} and Appendix \ref{AppendixAvgTau}), mass surface density, column density, mass (Section \ref{sec:MassDerive}), and luminosity (Section \ref{sec:Luminosity}).
The modifications of the equations for sources at high redshift (Section \ref{Redshift}) are shown, as well as the effects of observing dust emission with telescope beam patterns and filters (Section \ref{Caveats}).
A summary of dust mass opacities appropriate for galactic ISMs, dense molecular cores, and protoplanetary disks is presented (Section \ref{OpacitySection}) 
along with a table of dust mass opacities in commonly observed filters (Appendix \ref{AppendixOpacity}, Table \ref{tab:dustopacity}).
Modeling of spectral energy distributions (SEDs) is described (Section \ref{sec:SEDModels}), including finding the bolometric temperature (Appendix \ref{AppendixTbol}), calculating the spectral index (Appendix \ref{AppendixSpectralIndex}), and finding the peak of the SED with a generalized Wien's Law (Appendix \ref{SEDPeakAppendix}).
Formulae are derived for the equilibrium dust temperature of dust either heated by a nearby, central luminous source (Section \ref{sec:TdCentralSource}) or by the ambient interstellar radiation field (Appendix \ref{AppendixISRF}).
We conclude with a brief summary of 3D numerical radiative transfer modeling with Monte Carlo techniques (Section \ref{sec:CodesforRadTrans}).
Appendix \ref{AppendixEquationSummary} includes a summary of the most important equations derived in this tutorial in Table \ref{tab:eqsummary}.

\section{Optically Thin Emission} \label{sec:emission}

\subsection{Radiative Transfer}\label{sec:RadTrans}

\begin{figure}[h!]
\includegraphics[scale=0.6]{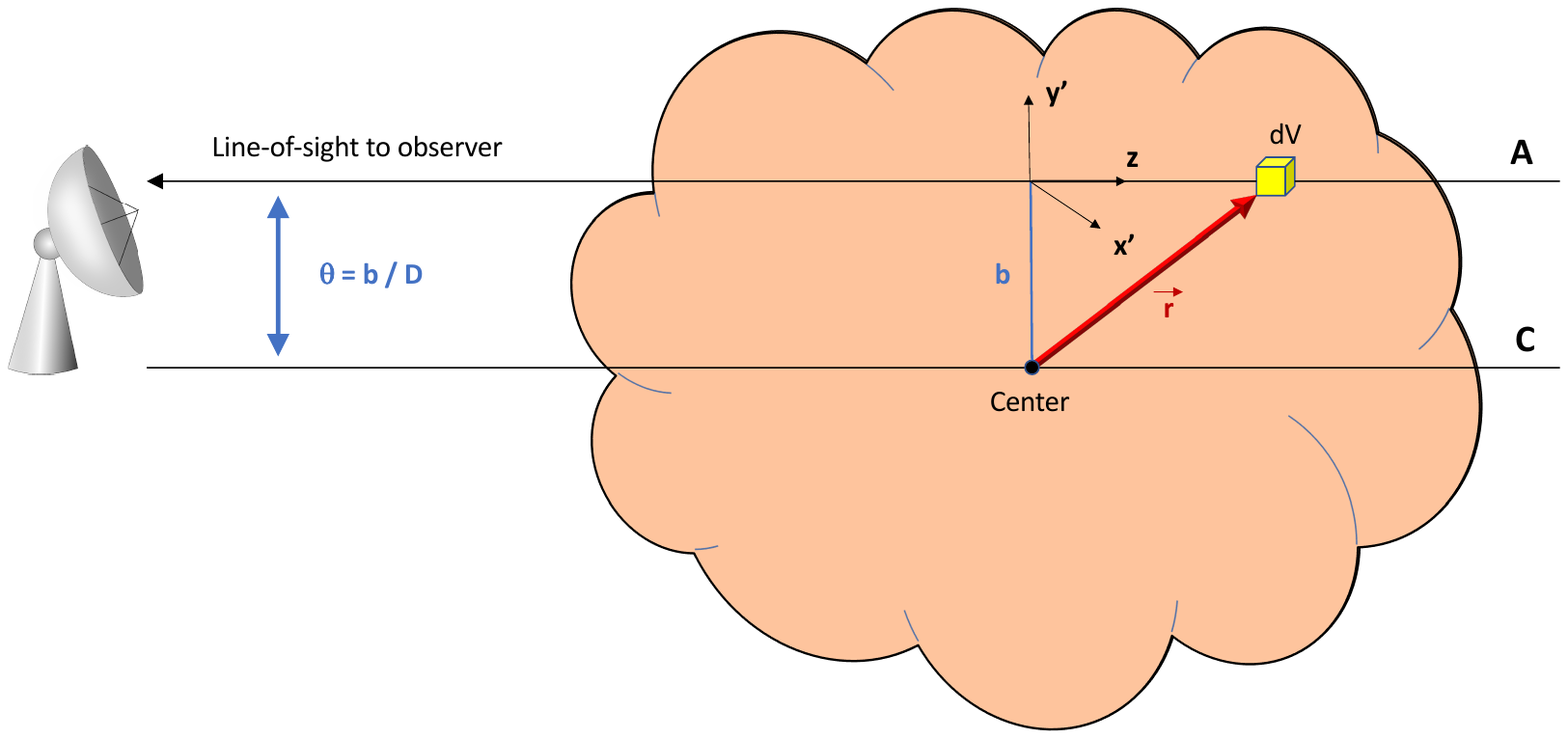}
\vspace{-2cm}
\caption{Geometry of an emitting dusty object with two lines-of-sight labeled A (observed line-of-sight) and C (line-of-sight through object center). The observer is located to the left and is looking toward the dusty object. A differential volume, $dV$, is located a position $\vec{r}$ from the center of the object.  The $z$-direction is measured along the observed line-of-sight through the cloud that passes through $dV$ (labeled A).  The coordinates $(x^{\prime},y^{\prime})$ are perpendicular coordinates in the plane of the sky of the observer ($x^{\prime}$ points out of the page).  The impact parameter, $b$, is the shortest distance between the observed line-of-sight (A) and the cloud center (C).  For an object located at a distance, $D$, that is very far away from the observer compared to the size of the object, the angle between lines-of-sight A and C, the angular impact parameter, is $\theta = b/D$ (and lines-of-sight A and C are very close to parallel).  This Figure is drawn in the limit $D \gg \rm{max}\{| \vec{r}| \}$.}
\label{fig:Geo1}
\end{figure}

The three-dimensional geometry of a general dust emission radiative transfer problem is shown in Figure \ref{fig:Geo1}.
While Figure \ref{fig:Geo1} is generically drawn as a single cloud, this object can be any dusty astrophysical object such as a protoplanetary disk, a dense core or clump, a molecular cloud, or an entire galaxy with a collection of molecular clouds.
One position within the object is designated the center.
This could be a geometric center of the dusty object (most appropriate for geometries with symmetry), or it could be the center-of-mass, or it could be located along the line-of-sight of the projected peak emission on the sky.
From the center position, a Cartesian coordinate system $(x,y,z)$ is defined where we choose the $z$ axis to be perpendicular to the tangent plane of the sky in the direction C.
Every position within the object can be measured by a radius vector from the center position given by
\begin{equation}
    \vec{r} = x \hat{x} + y \hat{y} + z\hat{z} \;\;\rm{cm},
\end{equation}
where $\hat{x}$, $\hat{y}$, and $\hat{z}$ are the unit vectors in the x, y, and z directions.
For line-of-sight A that does not pass through the center of the object, we can translate the coordinate system to $(x^{\prime},y^{\prime},z)$ and define an impact parameter $b$ as
\begin{equation}
    b = \sqrt{ x^{2} + y^{2}} \;\; \rm{cm} \;,
\end{equation}
where the impact parameter $b$ is the shortest distance between the line-of-sight A and the center of the object, connecting $(x=0,y=0,z=0)$ to $(x^{\prime}=0,y^{\prime}=0,z=0)$.
In the limit that the distance to the object, $D$ (cm), is much farther away than the largest physical size of the object, then lines-of-sight A and C are essentially parallel and $b$ is essentially perpendicular to both A and C.

We use the monochromatic specific intensity, $I_{\nu}$ (erg s$^{-1}$ cm$^{-2}$ ster$^{-1}$ Hz$^{-1}$), as the fundamental photometric quantity in radiative transfer calculations since, in a vacuum, it is a distance-independent quantity \citep{1978stat.book.....M}\footnote{This statement assumes that there is no cosmological expansion (\ie\ redshift due to the expansion of spacetime) relevant on the scales of the problem.  See Section \ref{Redshift} for a discussion of the effects of spacetime expansion.}.
Changes in $I_{\nu}$ are the result of physical processes such as emission, absorption, or scattering.
At each physical position in the object, $\vec{r}$, the monochromatic specific intensity is a directional quantity, $I_{\nu}(\vec{r},\hat{r}^{\prime})$, that can vary depending on the direction you are looking through the object, $\hat{r}^{\prime}$, if you were located at position $\vec{r}$. 
Thus, the unit vector $\hat{r}^{\prime}$ is located at the position of $\vec{r}$, and can point in any direction in space from that position.
We shall consider the radiative transfer for position vectors, $\vec{r}$, that point to positions along the line-of-sight A.
If we ignore scattering into or out of the line of sight, then in one dimension, the radiative transfer equation is a differential equation that only depends on the intensity in the $\hat{r}^{\prime} = \hat{z}$ direction (see Figure \ref{fig:Geo1}).
In steady state (no variation with time), the monochromatic specific intensity varies along the line of sight A as,
\begin{equation}
    \frac{dI_{\nu}(\vec{r},\hat{z})}{dz} = j_{\nu}(\vec{r}) - \alpha_{\nu}(\vec{r}) I_{\nu}(\vec{r},\hat{z}) \;\;\;\; \rm{erg}\, \rm{s}^{-1}\, \rm{cm}^{-3}\, \rm{ster}^{-1}\, \rm{Hz}^{-1} \;\; ,
\label{eq:dIdx}
\end{equation}
where $j_{\nu}$ is the dust emissivity coefficient (erg s$^{-1}$ cm$^{-3}$ ster$^{-1}$ Hz$^{-1}$) and $\alpha_{\nu}$ is the dust absorption coefficient (cm$^{-1}$).
The absorption coefficient can generally be written as a number density times a cross section or as mass density times cross section per unit mass (which is called mass opacity)
\begin{eqnarray}
    \alpha_{\nu}(\vec{r}) & = & n_d(\vec{r}) \sigma_{\nu}(\vec{r}) \; \rm{cm}^{-1} \; , \nonumber \\ 
    & = & \rho_d(\vec{r}) \kappa_{\nu}(\vec{r}) \; \rm{cm}^{-1} \; ,
\label{eq:alphanu}
\end{eqnarray}
where $n_d$ is the number density of dust grains (cm$^{-3}$), $\sigma_{\nu}$ is the cross-section for absorption (cm$^2$), $\rho_d$ is the mass density of dust grains (g cm$^{-3}$), and $\kappa_{\nu}$ is the dust mass opacity (cm$^2$ g$^{-1}$ of dust).
Historically, for analysis of thermal dust emission the use of $\kappa_{\nu}$ is generally preferred over $\sigma_{\nu}$ (see Section \ref{OpacitySection} for a summary of the properties of $\kappa_{\nu}$).
Since dust emission is typically a thermal emission process, we can use Kirchoff's law to relate the absorption and emissivity coefficeints with the Planck function at the equilibrium dust temperature
\begin{equation}
    j_{\nu}(\vec{r}) = \alpha_{\nu}(\vec{r}) B_{\nu}[T_d(\vec{r})] \;\;\;\; \rm{erg}\, \rm{s}^{-1}\, \rm{cm}^{-3}\, \rm{ster}^{-1}\, \rm{Hz}^{-1} \;\; ,
\label{eq:jnu}
\end{equation}
where 
\begin{equation}
    B_{\nu}[T_d(\vec{r})] = \frac{2 h \nu^3}{c^2} \frac{1}{e^{h\nu/kT_d(\vec{r})} - 1}  \;\;\;\; \rm{erg}\, \rm{s}^{-1}\, \rm{cm}^{-2}\, \rm{ster}^{-1}\, \rm{Hz}^{-1} \;\;
\label{eq:planck}
\end{equation}
is the Planck function with units of per unit frequency calculated at the dust temperature $T_d$ (K) at a position, $\vec{r}$, in the object\footnote{The Rayleigh-Jeans limit occurs when $h\nu/k \ll T_d(\vec{r})$. In this limit, the exponential term may be expanded in a Maclaurin series (a Taylor series centered on $h\nu/kT_d(\vec{r}) = 0$) as $e^{h\nu/kT_d(\vec{r})} \approx 1 + h\nu/kT_d(\vec{r})$.  The Planck function then becomes $\lim_{h\nu/k \ll T_d(\vec{r})} \, B_{\nu}[T_d(\vec{r})] = \frac{2k\nu^2}{c^2}T_d(\vec{r})$ erg s$^{-1}$ cm$^{-2}$ ster$^{-1}$ Hz$^{-1}$.  We caution that the Rayleigh-Jeans limit should only be used when $h\nu/k \ll T_d(\vec{r})$, otherwise the full expression for the Planck function in Equation \ref{eq:planck} should be used.}.

The total optical depth, $\tau_{\nu}$, is the integral of the absorption coefficent along the line of sight
\begin{eqnarray}
    \tau_{\nu} & = & \int_{\tau_{\nu}^{\prime} = 0}^{\tau_{\nu}} d \tau_{\nu}^{\prime} \label{eq:taunu1} \\ 
               & = & \int_{\rm{A}} \alpha_{\nu}(\vec{r}) dz \;\;\;\; \rm{unitless} \; ,
\label{eq:taunu}
\end{eqnarray}
where the notation $\int_{\rm{A}}$ means integration of all $z$ along the line-of-sight A ($z \in \{\rm{A} \bigcap \rm{Object}\}$).
The optical depth is a unitless quantity.
If we assume the dust along the line of sight has the same cross-section or dust mass opacity, then
\begin{eqnarray}
 \lim_{\sigma_{\nu}(z) = \sigma_{\nu}}   \tau_{\nu} & = & N_d \sigma_{\nu} \;\; , \nonumber \\ 
  \lim_{\kappa_{\nu}(z) = \kappa_{\nu}}   \tau_{\nu} & = & \Sigma_d \kappa_{\nu} \;\; ,
\label{eq:taunu_limit}
\end{eqnarray}
where 
\begin{equation}
    N_d = \int_A n_d(\vec{r}) dz \;\;\;\; \rm{cm}^{-2}  
\end{equation}
is the column density of dust\footnote{In general, the column density of objects (dust grains, H atoms, H$_2$ molecules, etc.) has units of number of objects per square cm, but this is usually shortened to just cm$^{-2}$ and the ``number of objects" is implied.  In this tutorial, we will indicate what type of object is being counted in column densities and mass surface densities by the subscript on $N$ or $\Sigma$.} and
\begin{equation}
   \Sigma_d = \int_A \rho_d(\vec{r}) dz \;\;\;\; \rm{g} \; \rm{of} \; \rm{dust} \; \rm{cm^{-2}} \; 
\label{eq:DefnSigmad}
\end{equation}
is the mass surface density of dust.

The dust density is related to the gas density, $\rho_g(\vec{r})$ (g cm$^{-3}$ of gas) through the empirically-determined gas to dust mass ratio, $R_{gd}(\vec{r})$, by
\begin{equation}
    \rho_d(\vec{r}) = \frac{\rho_g(\vec{r}) }{R_{gd}(\vec{r})} \;\;\;\;\;\;\;\;\;\;\;\;\;\;\; \rm{g} \; \rm{of} \; \rm{dust} \; \rm{cm^{-3}} \; .
\label{eq:gastodust}
\end{equation}
The gas density is related to the gas number density, $n_g(\vec{r})$ (cm$^{-3}$), through
\begin{equation}
    \rho_g(\vec{r}) = \mu_g(\vec{r}) m_{\rm{H}} n_g(\vec{r})  \;\;\;\;\;\;\; \rm{g} \; \rm{of} \; \rm{gas} \; \rm{cm^{-3}} \; ,
\label{eq:gastonumber}
\end{equation}
where $m_{\rm{H}} = 1.6735 \times 10^{-24}$ g = $1.0078$ amu is the mass of a hydrogen atom and $\mu_{g}$ is the mean molecular weight of the gas\footnote{$\mu_g$ is defined as the mean mass of objects in the gas in grams divided by the mass of a hydrogen atom in grams where the subscript of $\mu$ indicates the type of objects: $\mu_{\rm{objects}}m_{\rm{H}} = M_g/\mathcal{N}_{\rm{objects}}$.
It is referenced to $m_{\rm{H}}$, meaning that it should be used in the combination $\mu_g m_{\rm{H}}$.}.
The dust mass surface density is found by integrating along the line-of-sight
\begin{equation}
    \Sigma_d = \int_A \frac{\mu_g(\vec{r}) m_{\rm{H}} n_g(\vec{r})}{R_{gd}(\vec{r})} dz  \;\;\;\; \rm{g} \; \rm{of} \; \rm{dust} \; \rm{cm^{-2}} \; .
\label{eq:gastodustint}
\end{equation}
If the gas to dust mass ratio and the mean molecular weight of the gas does not vary along the line-of-sight, then the dust mass surface density is related to the gas column density, $N_{g} = \int_A n_g(\vec{r}) dz$ (cm$^{-2}$) through
\begin{equation}
    \lim_{R_{gd}(z) = R_{gd}} \, \lim_{\mu_{g}(z) = \mu_{g}} \,
    \Sigma_d = \frac{\mu_{g} m_{\rm{H}} N_{g}}{R_{gd}} \;\;\;\; \rm{g} \; \rm{of} \; \rm{dust} \; \rm{cm^{-2}} \;\; .
\label{eq:sigmad}
\end{equation}
This gas to dust mass ratio may also be used to convert between the dust mass surface density and the gas mass surface density, $\Sigma_g = \int_A \rho_g(\vec{r}) dz$ (g cm$^{-2}$),
\begin{equation}
    \lim_{R_{gd}(z) = R_{gd}} \, \lim_{\mu_{g}(z) = \mu_{g}} \, \Sigma_{g} = R_{gd} \Sigma_d \;\;\;\; \rm{g} \; \rm{of} \; \rm{gas} \; \rm{cm^{-2}} \;\; .
\label{eq:sigmag}
\end{equation}

Since dust clouds can be found in different phases of the interstellar medium (\ie\ ionized, atomic, and molecular), then one has to carefully pair the correct $\mu_{g}$ with the gas component for which the column density is calculated.
We update and modify the calculation of the mean molecular weight of the gas from Appendix A.1 of \cite{Kauffmann2008A&A} using the proto-solar mass fractions found in Table 4 of \cite{2021A&A...653A.141A} with the mass fraction of hydrogen of $M(\rm{H})/M_g = 0.7121$, the mass fraction of helium of $M(\rm{He})/M_g = 0.2725$, and the mass fraction of metals heavier than helium (denoted by Z) of $M(\rm{Z})/M_g = 0.0154$. 
The total mass of gas is $M_g = M(\rm{H}) + M(\rm{He}) + M(Z)$.
If $N_{g} = N_{\rm{H}}$, the column density of H atoms (cm$^{-2}$), then $\mu_g = \mu_{\rm{H}}$ and
\begin{equation}
    \mu_{\rm{H}} m_{\rm{H}} = \frac{M_g}{\mathcal{N}(\rm{H})}  \;,
    \label{eq:muh}
\end{equation}
where $\mathcal{N(\rm{H})}$ is the number of hydrogen atoms.
Since $m_{\rm{H}} \mathcal{N(\rm{H})} = M(\rm{H})$ then we find that
\begin{equation}
    \mu_{\rm{H}} = \frac{M_g}{M(\rm{H})}   \;,
    \label{eq:muhfinal}
\end{equation}
and we have $\mu_{\rm{H}} = 1/0.7121 = 1.404$. 
If $N_{g} = N_{\rm{H}_2}$, the column density of H$_2$ molecules (cm$^{-2}$), then $\mu_g = \mu_{\rm{H}_2}$ and
\begin{eqnarray}
    \mu_{\rm{H}_2} m_{\rm{H}} & = & \frac{M_g}{\mathcal{N}(\rm{H}_2)} \\ 
    & = & \frac{M_g}{\frac{1}{2}\mathcal{N}(\rm{H})} \\
    \mu_{\rm{H}_2} & = & \frac{2M_g}{M(\rm{H})}   =  2 \mu_{\rm{H}} \;\; ,
    \label{eq:muh2final}
\end{eqnarray}
since the number of H atoms is twice the number of H$_2$ molecules ($2\mathcal{N(\rm{H}_2)} = \mathcal{N(\rm{H})}$).
As a result, $\mu_{\rm{H}_2} = 2.809$ (see also \citealt{2022ApJ...929L..18E}).
If $N_{g}$ is the total column density (cm$^{-2}$) of atoms and molecules in a molecular environment, then $\mu_{g} = \mu_{p}$, the mean molecular weight per gas particle, and is calculated as
\begin{eqnarray}
    \mu_p m_{\rm{H}}  & = &  \frac{M_g}{\mathcal{N}(\rm{H}_2)  + \mathcal{N}(\rm{He}) + \sum_i \mathcal{N}(Z_i)} \\ 
    \mu_p & = & \frac{\frac{M_g}{m_{\rm{H}}\mathcal{N}(H)}}{\frac{\mathcal{N}(\rm{H}_2)}{\mathcal{N}(\rm{H})}  + \frac{\mathcal{N}(\rm{He})}{\mathcal{N}(\rm{H})} + \sum_i \frac{\mathcal{N}(Z_i)}{\mathcal{N}(\rm{H})} } \;\; ,
    \label{eq:mupfinal}
\end{eqnarray}
where $\mathcal{N}(\rm{Z}_i)$ is the number of each gas phase metal atom heavier than helium and we have divided the numerator and denominator by $\mathcal{N}(\rm{H})$ to arrive at Equation \ref{eq:mupfinal}.
The fraction in the numerator of Equation \ref{eq:mupfinal} is equal to Equation \ref{eq:muh}, $M_g/m_{\rm{H}}\mathcal{N}(\rm{H}) = \mu_{\rm{H}} = 1.404$.
We show in Appendix \ref{sec:AppendixMuZ} that only the most abundant gas phase metals of O, C, Ne, and N make a significant contribution to the calculation of $\mu_p$ to 3 decimal places.
If we use the gas phase ISM abundances of oxygen and carbon from  \cite{2021ApJ...906...73H} plus the protosolar abundances of nitrogen and neon from \cite{2021A&A...653A.141A},
then we find that $\mu_p = 2.351$ for the mean molecular weight per gas particle (see Appendix \ref{sec:AppendixMuZ} for more details).

$R_{gd} = 100$ is often assumed for convenience because this value is similar to measured values in a wide variety of environments in the ISM.
In the diffuse ISM, measurements point to a slightly higher value of $R_{gd} \sim 150$ \citep{Draine2011}.
In molecular clouds, there are a wide range of physical conditions from relatively ice-free dust grains at gas densities of $\sim 10 - 100$ cm$^{-3}$ to very dense cores ($> 10^6$ cm$^{-3}$) with dust grains possessing substantial ice mantles.
At $A_{\rm{V}} > 3$ mag, ices, such as H$_2$O, form on the surfaces of dust grains \citep{2001ApJ...547..872W, 2011ApJ...731....9C} and the grain ice mass should be included in the gas to solid (dust + ice) ratio.
$R_{gd}$ may vary from the low density molecular cloud value of $132$ to dense core environments with values as low as $\sim 60$ for solar metallicity 
\citep{2025ApJ...983..133P}.
Interestingly, if the mass ratio of ice to dust of $1.3$ from \cite{2014prpl.conf..363P} is included in the gas to solids (dust + ice) ratio, then $R_{gd} = 102$ \citep{2025ApJ...983..133P}.
In protoplanetary disks, the dust to gas ratio can vary by orders of magnitude, especially in the mid-plane of disks with cavities, gaps, rings, or asymmetries (e.g., see   \citealt{2023ASPC..534..501M,2023A&A...670A..12S, 2023A&A...677A..76P,2024A&A...682A.149S,2025ApJ...984L..18T}).
On galactic scales, the dust to gas ratio varies with metallicity with the general trend that the dust to gas ratio decreases as metallicity decreases (e.g., see Figure 12 in both \citealt{2022ApJ...928...90R} and  \citealt{2024ApJ...966...80H}). 

\subsubsection{Solution to Equation of Radiative Transfer in 1D}\label{sec:1DSolution} 

\begin{figure}[h!]
\includegraphics[scale=0.6]{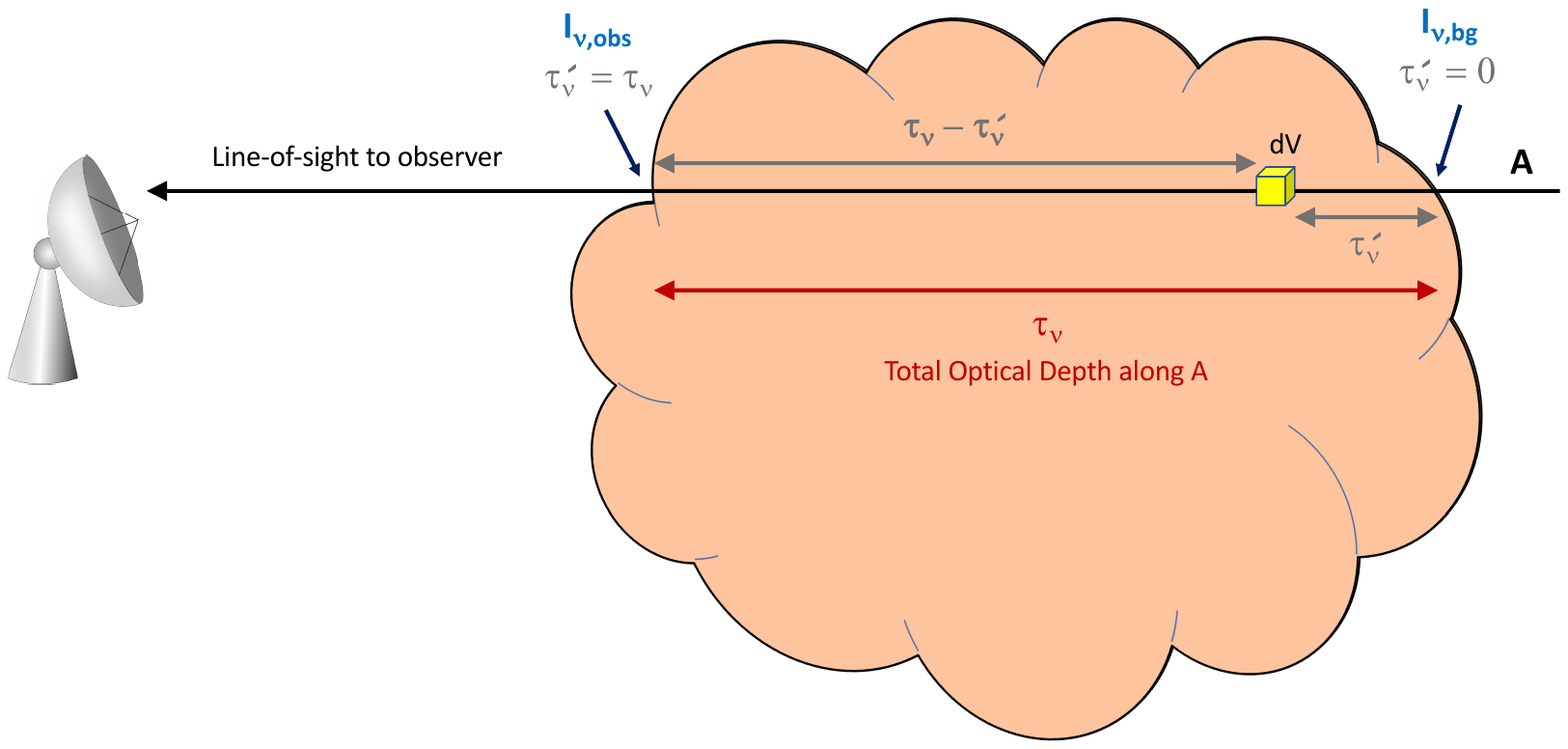}
\vspace{-2cm}
\caption{Geometry of the optical depth coordinates in the radiative transfer problem.  The arrow towards the observer along line-of-sight A indicates the direction of integration in the radiative transfer problem.  $\tau_{\nu}$ is the total optical depth along A through the object. If $dV$ emits radiation into the line-of-sight A, then this radiation will see an optical depth of $\tau_{\nu} - \tau_{\nu}^{\prime}$ as it passes out of the object.}
\label{fig:TauGeo}
\end{figure}

If we divide both sides of Equation~\ref{eq:dIdx} by $\alpha_{\nu}$ and use Kirchoff's law to substitute for $j_{\nu}/\alpha_{\nu} = B_{\nu}(T_d)$, then
we transform each term of the differential equation into units of monochromatic specific intensity
\begin{equation}
       \frac{dI_{\nu}(\tau_{\nu}^{\prime},\hat{z})}{d\tau_{\nu}^{\prime}} + I_{\nu}(\tau_{\nu}^{\prime},\hat{z}) =  B_{\nu}[T_d(\tau_{\nu}^{\prime})] \;\;\;\; \rm{erg}\, \rm{s}^{-1}\, \rm{cm}^{-2}\, \rm{ster}^{-1}\, \rm{Hz}^{-1} \;\; ,
\label{eq:dIdtauprime}
\end{equation}
where we have introduced the variable $\tau_{\nu}^{\prime}$ that is the optical depth coordinate ($d\tau_{\nu}^{\prime} = \alpha_{\nu}(\vec{r}) dz$, see Equation~\ref{eq:taunu1}).
This unitless coordinate has the limits of
$\tau_{\nu}^{\prime} = 0$ when the intensity is equal to the (initial) background intensity, $I_{\nu}(\tau_{\nu}^{\prime} = 0,\hat{z}) = I_{\nu,\rm{bg}}$,  and goes to $\tau_{\nu}^{\prime} = \tau_{\nu}$ when the intensity is equal to the (final) observed intensity, $I_{\nu}(\tau_{\nu}^{\prime} = \tau_{\nu},\hat{z}) = I_{\nu,\rm{obs}}$, along the line-of-sight A (see Figure \ref{fig:TauGeo}).  
Note that $\tau_{\nu}$ without the prime superscript is the total optical depth (Equation~\ref{eq:taunu1}) and is a constant.
The integrating factor for this linear first order differential equation is $e^{\tau_{\nu}^{\prime}}$ which is multiplied into Equation~\ref{eq:dIdtauprime} to yield
\begin{equation}
    \frac{d[I_{\nu}(\tau_{\nu}^{\prime},\hat{z}) e^{\tau_{\nu}^{\prime}}]}{d \tau_{\nu}^{\prime}} = B_{\nu}[T_d(\tau_{\nu}^{\prime})] e^{\tau_{\nu}^{\prime}}   \;\; \rm{erg}\, \rm{s}^{-1}\, \rm{cm}^{-2}\, \rm{ster}^{-1}\, \rm{Hz}^{-1} \;\; ,
\label{eq:dIetauprime}
\end{equation}
where we have used the chain rule to convert the left-hand side of Equation~\ref{eq:dIetauprime} into a single derivative\footnote{$\frac{d[I_{\nu}(\tau_{\nu}^{\prime},\hat{z}) e^{\tau_{\nu}^{\prime}}]}{d \tau_{\nu}^{\prime}} = \frac{dI_{\nu}(\tau_{\nu}^{\prime},\hat{z})}{d\tau_{\nu}^{\prime}} e^{\tau_{\nu}^{\prime}} + I_{\nu}(\tau_{\nu}^{\prime},\hat{z}) e^{\tau_{\nu}^{\prime}}$}.
If we multiply Equation \ref{eq:dIetauprime} by $d \tau_{\nu}^{\prime}$, then integrating the left-hand side of Equation~\ref{eq:dIetauprime} and applying boundary conditions yields
\begin{equation}
    \int_{I_{\nu,bg}e^{0}}^{I_{\nu,obs}e^{\tau_{\nu}}} d[I_{\nu}(\tau_{\nu}^{\prime},\hat{z}) e^{\tau_{\nu}^{\prime}} ] = (I_{\nu,\rm{obs}}e^{\tau_{\nu}} - I_{\nu,\rm{bg}}) \;\; \rm{erg}\, \rm{s}^{-1}\, \rm{cm}^{-2}\, \rm{ster}^{-1}\, \rm{Hz}^{-1} \;\; .
\label{eq:intdIetauprime}
\end{equation}
Since $I_{\nu}$ is a distance-independent quantity in a vacuum, then $I_{\nu,\rm{obs}}$ does not change between the near object surface  (where $\tau_{\nu}^{\prime} = \tau_{\nu}$) and the observer as long as there is no absorption, emission, or scattering at the frequency of the observations outside of the object\footnote{We again note that this statement is true if redshift is not important.  See Section \ref{Redshift} for a discussion of the breakdown of this assumption.}.
Multiplying both sides of the integrated Equation \ref{eq:dIetauprime} by $e^{-\tau_{\nu}}$ and inserting Equation \ref{eq:intdIetauprime}, we have
\begin{eqnarray}
    (I_{\nu,\rm{obs}}e^{\tau_{\nu}} - I_{\nu,\rm{bg}})e^{-\tau_{\nu}} & = &  \int_{\tau_{\nu}^{\prime} = 0}^{\tau_{\nu}} B_{\nu}[T_d(\tau_{\nu}^{\prime})] e^{-(\tau_{\nu} - \tau_{\nu}^{\prime})}  d \tau_{\nu}^{\prime} \nonumber \\
    I_{\nu,\rm{obs}} - I_{\nu,\rm{bg}}e^{-\tau_{\nu}} & = &\int_{\tau_{\nu}^{\prime} = 0}^{\tau_{\nu}} B_{\nu}[T_d(\tau_{\nu}^{\prime})] e^{-(\tau_{\nu} - \tau_{\nu}^{\prime})}  d \tau_{\nu}^{\prime} \nonumber \\
   I_{\nu,\rm{obs}} & = & I_{\nu,\rm{bg}}e^{-\tau_{\nu}} + \int_{\tau_{\nu}^{\prime} = 0}^{\tau_{\nu}} B_{\nu}[T_d(\tau_{\nu}^{\prime})] e^{-(\tau_{\nu} - \tau_{\nu}^{\prime})}  d \tau_{\nu}^{\prime} \;\;\;\; \rm{erg}\, \rm{s}^{-1}\, \rm{cm}^{-2}\, \rm{ster}^{-1}\, \rm{Hz}^{-1} \;\; .
\label{eq:Inu}
\end{eqnarray}
The first term on the right side of Equation \ref{eq:Inu} indicates that the background intensity, $I_{\nu,\rm{bg}}$, is reduced by a factor of $e^{-\tau_{\nu}}$ as it passes through the object.
The second term on the right side of Equation \ref{eq:Inu} is emission from the source.
If there is emission from within the object at a depth corresponding to  $\tau_{\nu}^{\prime}$ (see Figure \ref{fig:TauGeo}), then the new intensity that it has added along the line of sight will be reduced by the remaining amount of optical depth from its position within the object to the edge of the object ($e^{-(\tau_{\nu} - \tau_{\nu}^{\prime})}$).
Thus, the integral on the right-hand side of Equation~\ref{eq:Inu} is integrating the new emission through the object corrected for how much optical depth that emission experiences as it exits the object.
We define this integral to be equal to total the monochromatic integrated intensity coming solely from the source
\begin{equation}
    I_{\nu,\rm{src}} = \int_{\tau_{\nu}^{\prime} = 0}^{\tau_{\nu}} B_{\nu}[T_d(\tau_{\nu}^{\prime})] e^{-(\tau_{\nu} - \tau_{\nu}^{\prime})}  d \tau_{\nu}^{\prime} \;\;\;\; \rm{erg}\, \rm{s}^{-1}\, \rm{cm}^{-2}\, \rm{ster}^{-1}\, \rm{Hz}^{-1} \;\; ,
\label{eq:Inusrcdefinition}
\end{equation}
such that the formal solution to the 1D raditive transfer equation is
\begin{equation}
I_{\nu,\rm{obs}}  =  I_{\nu,\rm{bg}}e^{-\tau_{\nu}} + I_{\nu,\rm{src}} \;\;\;\; \rm{erg}\, \rm{s}^{-1}\, \rm{cm}^{-2}\, \rm{ster}^{-1}\, \rm{Hz}^{-1} \;\; .
\label{eq:InuFormalSolution}
\end{equation}

In the limit that the object is isothermal at dust temperature $T_d$ (K) along the line of sight, the source integral can be solved explicitly\footnote{Let $u = -(\tau_{\nu} - \tau_{\nu}^{\prime})$.  Then $d\tau_{\nu}^{\prime} = du$ and $I_{\nu,\rm{src}} = B_{\nu}(T_d) \int_{-\tau_{\nu}}^0 e^u du$ = $B_{\nu}(T_d) (1 - e^{-\tau_{\nu}})$.} to find that
\begin{equation}
 \lim_{T_d(z) = T_d}  I_{\nu,\rm{obs}} = I_{\nu,\rm{bg}}e^{-\tau_{\nu}} + B_{\nu}(T_d) (1 - e^{-\tau_{\nu}}) \;\;\;\; \rm{erg}\, \rm{s}^{-1}\, \rm{cm}^{-2}\, \rm{ster}^{-1}\, \rm{Hz}^{-1} \;\; .
\label{eq:Inuiso}
\end{equation}
It is instructive to explore the limits of optical depth in this equation.
If the source is very optically thick ($\tau_{\nu} \gg 1$), then the exponential terms become $e^{-\tau_{\nu}} \rightarrow 0$ and the observed intensity is
\begin{equation}
   \lim_{\tau_{\nu} \gg 1} \lim_{T_d(z) = T_d} I_{\nu,\rm{obs}} = B_{\nu}(T_d) \;\;\;\; \rm{erg}\, \rm{s}^{-1}\, \rm{cm}^{-2}\, \rm{ster}^{-1}\, \rm{Hz}^{-1} \;\; .
\label{eq:Inuisothicklimit}
\end{equation}
In the strict optically thick limit, the observed intensity is just equal to the source function\footnote{An exception to this equation is that observed intensity in the optically thick limit can be less than the Planck function when scattering is important \citep{1986rpa..book.....R}.  A good example is dust self-scattering in protoplanetary disks \citep{2019ApJ...877L..18Z}.} and the background cannot be seen through the source.
As a result, an optically thick intensity may be used to determine the dust temperature of the source
\begin{equation}
    \lim_{\tau_{\nu} \gg 1} \; T_d = \frac{h\nu}{k} \left[\ln \left(1 + \frac{2h\nu^2}{c^2 I_{\nu,\rm{obs}}}  \right) \right]^{-1} \;\;\;\; K \;\; .
\label{eq:TdthickInu}    
\end{equation}
If the source is very optically thin ($\tau_{\nu} \ll 1$), then we can simplify the exponential terms with the Maclaurin expansion $e^{-\tau_{\nu}} \approx 1 - \tau_{\nu}$ and Equation \ref{eq:Inuiso} becomes
\begin{equation}
   \lim_{\tau_{\nu} \ll 1} \lim_{T_d(z) = T_d} I_{\nu,\rm{obs}} = I_{\nu,\rm{bg}}(1 -\tau_{\nu}) + B_{\nu}(T_d)\tau_{\nu} \;\;\;\; \rm{erg}\, \rm{s}^{-1}\, \rm{cm}^{-2}\, \rm{ster}^{-1}\, \rm{Hz}^{-1} \;\; .
\label{eq:Inuisothinlimit}
\end{equation}
In the strict optically thin limit, the observed intensity includes contributions from nearly all ($1 - \tau_{\nu})$ of the background intensity plus the source function scaled by the optical depth of the source.
There is a degeneracy between $T_d$ and $\tau_{\nu}$ for the source intensity.
If an estimate of $T_d$ can be made from additional observations or assumptions, then due to the dependence on $\tau_{\nu}$ in Equation \ref{eq:Inuisothinlimit}, which itself depends on the number density of dust grains ($n_d(\vec{r})$ cm$^{-3}$; see Equations \ref{eq:alphanu} and \ref{eq:taunu}), an optically thin intensity may be used to determine the column density and mass of the source (see Section \ref{sec:MassDerive}).
We shall explore the optically thin limit in detail, including derivations with a variable source function along the line-of-sight, in Section \ref{sec:OpticallyThinSection}.

\begin{figure}[h!]
\includegraphics[scale=0.58]{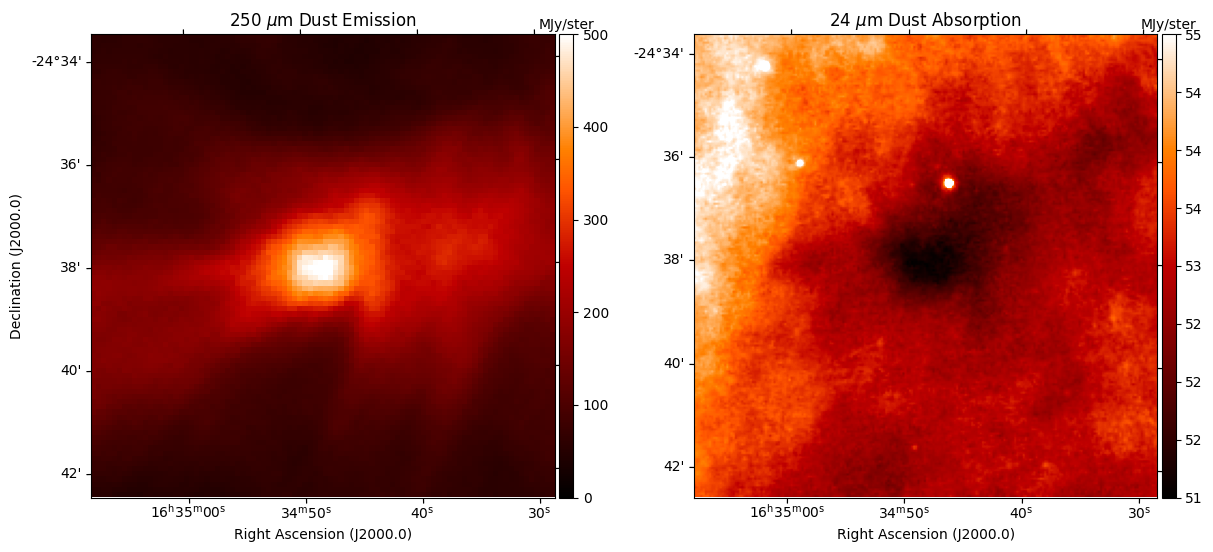}
\vspace{0.2cm}
\caption{Examples of thermal dust emission at $250$ $\mu$m (left) and dust absorption against a $24$ $\mu$m continuum (right) toward the cold ($T_d \sim 10$ K) starless core, L1689B, in the Ophiuchus Molecular Cloud.  
The $250$ $\mu$m \textit{Herschel Space Observatory} and the $24$ $\mu$m \textit{Spitzer Space Telescope} images were downloaded from the NASA/IPAC Infrared Science Archive.
}
\label{fig:AbsEmit}
\end{figure}

If we define the background subtracted intensity as $\Delta I_{\nu} = I_{\nu,\rm{obs}} - I_{\nu,\rm{bg}}$, then we find that
\begin{equation}
 \lim_{T_d(z) = T_d}  \Delta I_{\nu} = [B_{\nu}(T_d) - I_{\nu,\rm{bg}}]  (1 - e^{-\tau_{\nu}}) \;\;\;\; \rm{erg}\, \rm{s}^{-1}\, \rm{cm}^{-2}\, \rm{ster}^{-1}\, \rm{Hz}^{-1} \;\; .
\label{eq:deltaInu}
\end{equation}
An object is observed in emission above the background intensity ($\Delta I_{\nu} > 0$) if $B_{\nu}(T_d) > I_{\nu,\rm{bg}}$, while an object is observed in absorption against the background intensity ($\Delta I_{\nu} < 0$) if $B_{\nu}(T_d) < I_{\nu,\rm{bg}}$, at a particular frequency.
See Figure \ref{fig:AbsEmit} for examples of dust absorption and emission.
If we assign a temperature to the background (e.g. the Cosmic Microwave Background), then we have the well known result that a hotter source that is in front of a colder background will emit above that background and that a colder source that is in front of a hotter background will be seen in absorption against that background.

Equations \ref{eq:Inuiso} and \ref{eq:deltaInu} were derived assuming that the dust temperature is constant along the line-of-sight (\ie\ that the source function is constant along the line-of-sight).
We can still define an equivalent line-of-sight isothermal temperature, $\overline{T_d}$, even if the dust temperature varies along the line-of-sight by finding the $\overline{T_d}$ that satisfies the equation 
\begin{equation}
    I_{\nu,\rm{src}} = B_{\nu}(\overline{T_d}) (1 - e^{-\tau_{\nu}}) =  \int_{\tau_{\nu}^{\prime} = 0}^{\tau_{\nu}} B_{\nu}[T_d(\tau_{\nu}^{\prime})] e^{-(\tau_{\nu} - \tau_{\nu}^{\prime})}  d \tau_{\nu}^{\prime} \;\;\;\; \rm{erg}\, \rm{s}^{-1}\, \rm{cm}^{-2}\, \rm{ster}^{-1}\, \rm{Hz}^{-1} \; .
\label{eq:IsoDustTemp}
\end{equation}
The equivalent line-of-sight isothermal dust temperature is the single dust temperature that goes into a Planck function such that the expression $B_{\nu}(\overline{T_d})(1 - e^{-\tau_{\nu}})$ is equal to the integral of the source function term in the radiative transfer equation for that line-of-sight.

\subsubsection{Optically Thin Limit}\label{sec:OpticallyThinSection}

Equations \ref{eq:Inu}, \ref{eq:Inuiso}, \ref{eq:deltaInu}, and \ref{eq:IsoDustTemp} are applicable for any optical depth.
In this section, we will explore the optically thin limit in more detail by investigating the definitions of optically thick versus optically thin radiative transfer.
We derive the solution to the radiative transfer equation in the optically thin limit for the more general case of an object where the source function may vary along the line-of-sight.

Generally, emission is said to be optically thick when $\tau_{\nu} \geq 1$ but optical depth effects can be apparent even at $\tau_{\nu} < 1$.
For optical depth effects to be less than 10\% of the intensity in typical radiative transfer problems, then $\tau_{\nu} < 0.2$ is required at any given frequency\footnote{Estimated from $\tau / (1 - \rm{exp}(-\tau)) \sim 1.1$ for $\tau \sim 0.2$.}.
To put this limit in practical terms, we can solve for the dust mass surface density in Equation \ref{eq:taunu_limit},
\begin{equation}
    \Sigma_d = \frac{\tau_{\nu}}{\kappa_{\nu}}  \;\; \rm{g} \; \rm{of} \; \rm{dust}\; \rm{cm}^{-2} , 
\end{equation}
finding the dust mass surface density that corresponds to any optical depth for a given dust mass opacity.
We can also use Equations \ref{eq:sigmad} and \ref{eq:sigmag} to convert this expression into the corresponding gas surface density and gas column density.
Plugging in an optical depth of $\tau_{\nu, \rm{thick}} = 0.2$, then we find the criteria for when the dust surface density, gas surface density, and gas column density are optically thin to be
\begin{eqnarray}
    \Sigma_d^{\rm{thin}} & < & \frac{\tau_{\nu,\rm{thick}}}{ \kappa_{\nu}} = 0.02 \;\; \left( \frac{\tau_{\nu,\rm{thick}}}{0.2} \right) \left( \frac{10 \, \rm{cm}^2 \, \rm{g}^{-1} }{\kappa_{\nu}} \right) \;\; \rm{g} \; \rm{of} \; \rm{dust}\; \rm{cm}^{-2} , \\
    \Sigma_g^{\rm{thin}} & < & \frac{\tau_{\nu,\rm{thick}} R_{gd}}{ \kappa_{\nu}} = 2.0 \;\; \left( \frac{\tau_{\nu,\rm{thick}}}{0.2} \right) \left( \frac{R_{gd}}{100} \right) \left( \frac{10 \, \rm{cm}^2 \, \rm{g}^{-1} }{\kappa_{\nu}} \right) \;\; \rm{g} \; \rm{of} \; \rm{gas}\; \rm{cm}^{-2} , \\
    N_{\rm{H}_2}^{\rm{thin}} & < & \frac{\tau_{\nu,\rm{thick}} R_{gd}}{\mu_{\rm{H}_2} m_{\rm{H}} \kappa_{\nu}} = 4.25 \times 10^{23}
     \;\; \left( \frac{\tau_{\nu,\rm{thick}}}{0.2} \right) \left( \frac{R_{gd}}{100} \right)  \left( \frac{10 \, \rm{cm}^2 \, \rm{g}^{-1} }{\kappa_{\nu}} \right) \;\; \;\; \rm{H}_2 \; \rm{molecules} \; \rm{cm}^{-2} ,
\label{eq:NH2thin}
\end{eqnarray}
where $\mu_{\rm{H}_2} = 2.809$ is the mean molecular weight per H$_2$ molecule (Equation \ref{eq:muh2final}) and $N_{\rm{H}_2}^{\rm{thin}}$ is the column density of H$_2$  in cm$^{-2}$ required to be in the optically thin limit.
At a frequency of 1\,THz ($300$\,$\mu$m), dust opacities have a wide range of values depending on the grain composition and grain size distribution. 
The dust mass opacity will be discussed in more detail in Section \ref{OpacitySection}. 
If we pick an intermediate value for $\kappa \sim 10$ cm$^2$ g$^{-1}$ of dust\footnote{The  ``AstroDust" dust model for Milky Way ISM $R_V = 3.1$ has $\kappa_{1 \, \rm{THz}} = 4.5$ cm$^2$ g$^{-1}$ of dust \citep{2023ApJ...948...55H}, the icy dense core dust model ``KP5" has a $\kappa_{1\, \rm{THz}} = 7.1$ cm$^2$ g$^{-1}$ of dust \citep{2024RNAAS...8...68P}, and the icy dense core coagulated dust model ``OH5" with ``thin ice" has a $\kappa_{1\, \rm{THz}} = 13.2 $ cm$^2$ g$^{-1}$ of dust \citep{1994A&A...291..943O}.  These opacity models are discussed in detail in Section \ref{OpacitySection} and Appendix \ref{AppendixOpacity}.}, then assuming $R_{gd} = 100$ and that $\tau_{\nu,\rm{thick}} = 0.2$, Equation~\ref{eq:NH2thin} becomes $N_{\rm{H}_2}^{\rm{thin}} < 4.25 \times 10^{23}$ cm$^{-2}$.
This column density of H$_2$ molecules is very large. 
For the conversion between column density of H$_2$ molecules and dust extinction of $N_{\rm{H_2}}/A_{\rm{V}} = 1.07 \times 10^{21}$ cm$^{-2}$ mag$^{-1}$ \citep{Rachford2009ApJS}, this corresponds to 
\begin{equation}
    A_{\rm{V}}^{\rm{thin}} < \frac{N_{\rm{H}_2}^{\rm{thin}}}{1.07 \times 10^{21}} = 398
     \;\; \left( \frac{\tau_{\nu,\rm{thick}}}{0.2} \right) \left( \frac{R_{gd}}{100} \right)  \left( \frac{10 \, \rm{cm}^2 \, \rm{g}^{-1} }{\kappa_{\nu}} \right) \;\; \;\; \rm{mag} \, .
\label{eq:avthinlimit}
\end{equation}
Since the dust opacity drops at lower frequencies (longer wavelengths), then these limits will become larger at longer wavelengths (see Section \ref{OpacitySection}).
At submillimeter and longer wavelengths ($\lambda > 300$ $\mu$m, $\nu < 1$ THz), thermal emission from dust clouds is typically optically thin except for environments with more than several hundred magnitudes of extinction (e.g., viewing the midplane of a protostellar disk or the central regions near an AGN).
Care must be taken, however, at far-infrared wavelengths as the opacity is larger at higher frequencies (shorter wavelengths) and optical depth effects may begin to affect the thermal dust emission.
If the reader prefers a different criteria than $\tau_{\nu,\rm{thick}} = 0.2$ for the radiative transfer problem to be affected by optical depth effects, then since $\Sigma_d^{thin} \propto N_{H_2}^{thin} \propto \tau_{\nu}$, the relationships derived above may be linearly scaled.
For example, it would take $A_{\rm{V}} \sim 2000 $ mag to reach $\tau_{1\,\rm{THz}} = 1.0$ (i.e., from multiplying Equation \ref{eq:avthinlimit} by a factor of $5$).

If we assume that we are in the optically thin ($\tau_{\nu} \ll 1$) limit, then we can use a Maclaurin expansion of the exponential terms in Equation~\ref{eq:Inu} up to linear order, $e^{-\tau_{\nu}} \approx 1 - \tau_{\nu}$ and $ e^{-(\tau_{\nu} - \tau_{\nu}^{\prime})}\approx 1 - (\tau_{\nu} - \tau_{\nu}^{\prime}) $.
Since the second exponential is inside an integral in Equation~\ref{eq:Inu}, we only need to keep the leading term of its expansion which is $1$.
An easy way to see why we can ignore the linear term in parenthesis in the Maclaurin expansion within the integral is to temporarily assume that the Planck function is constant along the line of sight and can be brought outside the integral, then the integral of the $-(\tau_{\nu} - \tau_{\nu}^{\prime})$ term will result in terms proportional to $\tau_{\nu}^2$ which will be negligible in the optically thin limit\footnote{$\int_o^{\tau_{\nu}} (\tau_{\nu} - \tau_{\nu}^{\prime}) d\tau_{\nu}^{\prime} =  \tau_{\nu}^2 - \tau_{\nu}^2/2$.}.
Substituting the Maclaurin expansions into Equation \ref{eq:Inuiso}, we find 
\begin{eqnarray}
       \lim_{\tau_{\nu} \ll 1}  I_{\nu,\rm{obs}} & = &  I_{\nu,\rm{bg}}(1 -\tau_{\nu}) + \int_{\tau_{\nu}^{\prime} = 0}^{\tau_{\nu}} B_{\nu}[T_d(\tau_{\nu}^{\prime})] d \tau_{\nu}^{\prime} \label{eq:Inu_bg_thin_aA}  \nonumber \\ 
       & = & I_{\nu,\rm{bg}}\left( 1 - \int_{\rm{A}} \alpha_{\nu}(\vec{r}) dz \right) + \int_{\rm{A}}  B_{\nu}[T_d(\vec{r})] \alpha_{\nu}(\vec{r}) dz \label{eq:Inu_bg_thin_aB} \nonumber \\ 
       & = & I_{\nu,\rm{bg}}\left( 1 - \int_{\rm{A}} \kappa_{\nu}(\vec{r}) \rho_d(\vec{r}) dz \right) + \int_{\rm{A}} B_{\nu}[T_d(\vec{r})] \kappa_{\nu}(\vec{r}) \rho_d(\vec{r}) dz  \;\;\;\; \rm{erg}\, \rm{s}^{-1}\, \rm{cm}^{-2}\, \rm{ster}^{-1}\, \rm{Hz}^{-1} \;.
\label{eq:Inu_bg_thin}
\end{eqnarray}
If the background intensity has been subtracted then 
\begin{equation}
    \lim_{\tau_{\nu} \ll 1} \Delta I_{\nu} = I_{\nu,obs} - I_{\nu,bg} = \int_{\rm{A}} B_{\nu}[T_d(\vec{r})] \kappa_{\nu}(\vec{r}) \rho_d(\vec{r}) dz - I_{\nu,bg}\int_{\rm{A}} \kappa_{\nu}(\vec{r}) \rho_d(\vec{r}) dz \;\;\;\; \rm{erg}\, \rm{s}^{-1}\, \rm{cm}^{-2}\, \rm{ster}^{-1}\, \rm{Hz}^{-1} \; .
\label{eq:Inu_bgsub_thin}
\end{equation}
If the background intensity is negligible, then Equations \ref{eq:Inu_bg_thin} and \ref{eq:Inu_bgsub_thin} reduce to
\begin{equation}
    \lim_{I_{\nu,\rm{bg}} \rightarrow 0} \; \lim_{\tau_{\nu} \ll 1}  I_{\nu,\rm{obs}} = I_{\nu,\rm{src}} = \int_{\rm{A}} B_{\nu}[T_d(\vec{r})] \kappa_{\nu}(\vec{r}) \rho_d(\vec{r}) dz  \;\;\;\; \rm{erg}\, \rm{s}^{-1}\, \rm{cm}^{-2}\, \rm{ster}^{-1}\, \rm{Hz}^{-1} \;,
\label{eq:Inu_nobg_thin}
\end{equation}
where the observed intensity is equal to the source intensity since this is the emission coming solely from the object with no background.
Equation \ref{eq:Inu_nobg_thin} is the standard expression for the monochromatic specific intensity of optically thin thermal dust continuum emission.
It depends on the line-of-sight profiles of the dust temperature, dust mass opacity, and dust mass density.
Since $j_{\nu}(\vec{r}) = B_{\nu}[T_d(\vec{r})] \kappa_{\nu}(\vec{r}) \rho_d(\vec{r}) $ (see Equation \ref{eq:jnu}), Equation \ref{eq:Inu_nobg_thin} is also  the integral of the emissivity coefficient along the line-of-sight A, $I_{\nu,\rm{src}} = \int_A j_{\nu}(\vec{r}) dz$.\footnote{We note that if you ignore the absorption term in Equation \ref{eq:dIdx} and integrate both sides of the equation you will find that $I_{\nu,src} = I_{\nu,bg} + \int_A j_{\nu}(\vec{r}) dz$ which is different from Equation \ref{eq:Inu_bg_thin} for the background intensity term.  While this shortcut in the derivation arrives at the correct integral over the emissivity coefficient for optically thin emission, its treatment of the background term is incorrect.  Equation \ref{eq:Inu_bg_thin} is the formally correct solution in the optically thin limit.}

If we assume that the object is isothermal and that the dust opacity is constant along the line of sight, then the monochromatic specific intensity integral for optically thin emission is linearly proportional to the mass surface density of dust (g of dust cm$^{-2}$) along line-of-sight A,
\begin{eqnarray}
    \lim_{\tau_{\nu} \ll 1} \; \lim_{T_d(z) = T_d} \;  \lim_{\kappa_{\nu}(z) = \kappa_{\nu}} I_{\nu,\rm{src}} & = & B_{\nu}(T_d) \kappa_{\nu} \int_{\rm{A}}  \rho_d(\vec{r}) dz  \nonumber \\ 
    & = & B_{\nu}(T_d) \kappa_{\nu} \Sigma_d  \;\;\;\; \rm{erg}\, \rm{s}^{-1}\, \rm{cm}^{-2}\, \rm{ster}^{-1}\, \rm{Hz}^{-1} \;.
\label{eq:Inu_nobg_thin_iso_constk}
\end{eqnarray}

\subsection{On-sky Intensity Distributions}\label{sec:OnSkyCoord}

\begin{figure}[h]
\includegraphics[scale=0.6]{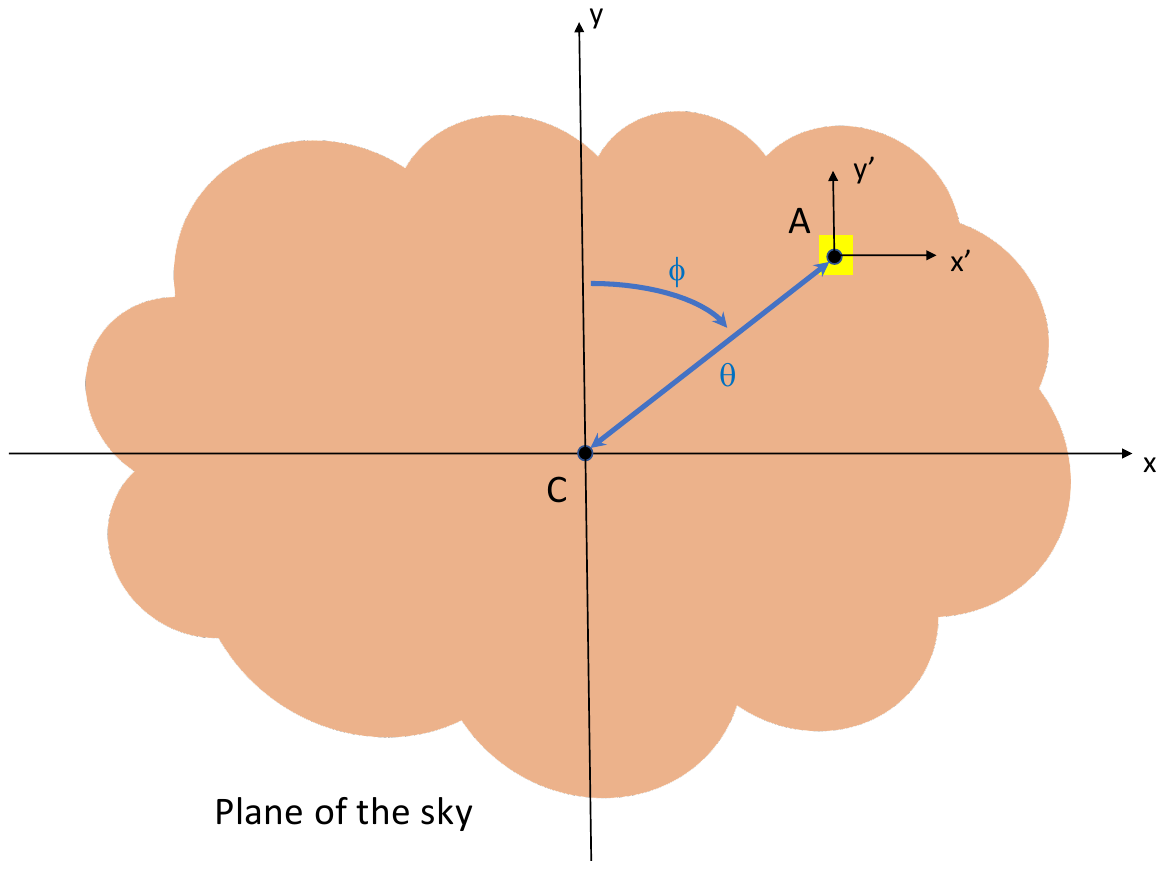}
\vspace{-2cm}
\caption{Observer's view of the tangent plane to the sky in the direction of the line-of-sight, C, through the center of the cloud.  The observer is above the page, looking down on to it.  C defines the pole of a spherical polar coordinate system where $\theta = 0$.  We have aligned the $y$-axis of the cloud with a constant right ascension pointing north such that any line-of-sight, A, with coordinates $(x=0,y > 0)$ has $\phi = 0$ and with coordinates $(x=0,y < 0)$ has $\phi = \pi$.  The $z$-axis is perpendicular, pointing into the page.  Since the maximum $\theta$ is often a very small angle for astrophysical sources, the spherical polar coordinate system can often be approximated as a planar polar coordinate system tangent to the celestial sphere at the direction defined by line-of-sight C.}
\label{fig:Geo2}
\end{figure}

We define a spherical polar coordinate system $(\theta,\phi)$ from the point of view of an observer (\ie\ centered on the observer) that is located a distance, D (cm), away from the object. 
The angle $\theta = 0$ (rad) is a pole of the coordinate system and is defined as the direction toward the line-of-sight C through the center of the cloud (Figure \ref{fig:Geo2}).
The angle $\theta$ (rad) is the angular impact parameter between lines-of-sight A and C (see Figure \ref{fig:Geo1}).
The angle $\phi$ is an azimuthal angle measured such that $\phi = 0$ is defined along a great circle of constant right ascension pointing north on the sky (Figure \ref{fig:Geo2}).
At fixed $\theta$, $\phi$ running from $0$ to $2\pi$ will trace a circle on the sphere that is centered on the direction in the sky toward the line-of-sight C.
Each unique line-of-sight, A, through the object will correspond to a unique $(\theta,\phi)$ position on the sky from the point-of-view of an observer.
The impact parameter, b (cm), of line-of-sight A is related to $\theta$ by $\tan(\theta) = b/D$ .
If the distance to the object is much larger than the size of the object, then using the small angle approximation for $\tan(\theta) \approx \theta$, gives $b = D\theta$ (cm).
Figure \ref{fig:Geo1} is drawn in this limit such that lines A and C are essentially parallel.
The intensity we have derived in Equations \ref{eq:Inu}, \ref{eq:Inu_bg_thin},  \ref{eq:Inu_nobg_thin}, and \ref{eq:Inu_nobg_thin_iso_constk} is the observed intensity, $I_{\nu,\rm{obs}}$, along the line-of-sight, A, which corresponds to a unique $I_{\nu}(\theta,\phi)$ from the observer's point of view (see Figure \ref{fig:Geo2}).

\subsubsection{Example Intensity Calculation: Spherical Dust Source} 
\label{SphGeoApp}

In general, to calculate how the monochromatic specific intensity varies with the coordinates $\theta$ and $\phi$ on the sky, we must connect the source geometry to the observer's spherical (or plane-polar in the small angle approximation) coordinate system. 
This problem depends on the geometry of the emitting source.
As an example, we show the solution for a spherical dust cloud.

\begin{figure}[h!]
\includegraphics[scale=0.6]{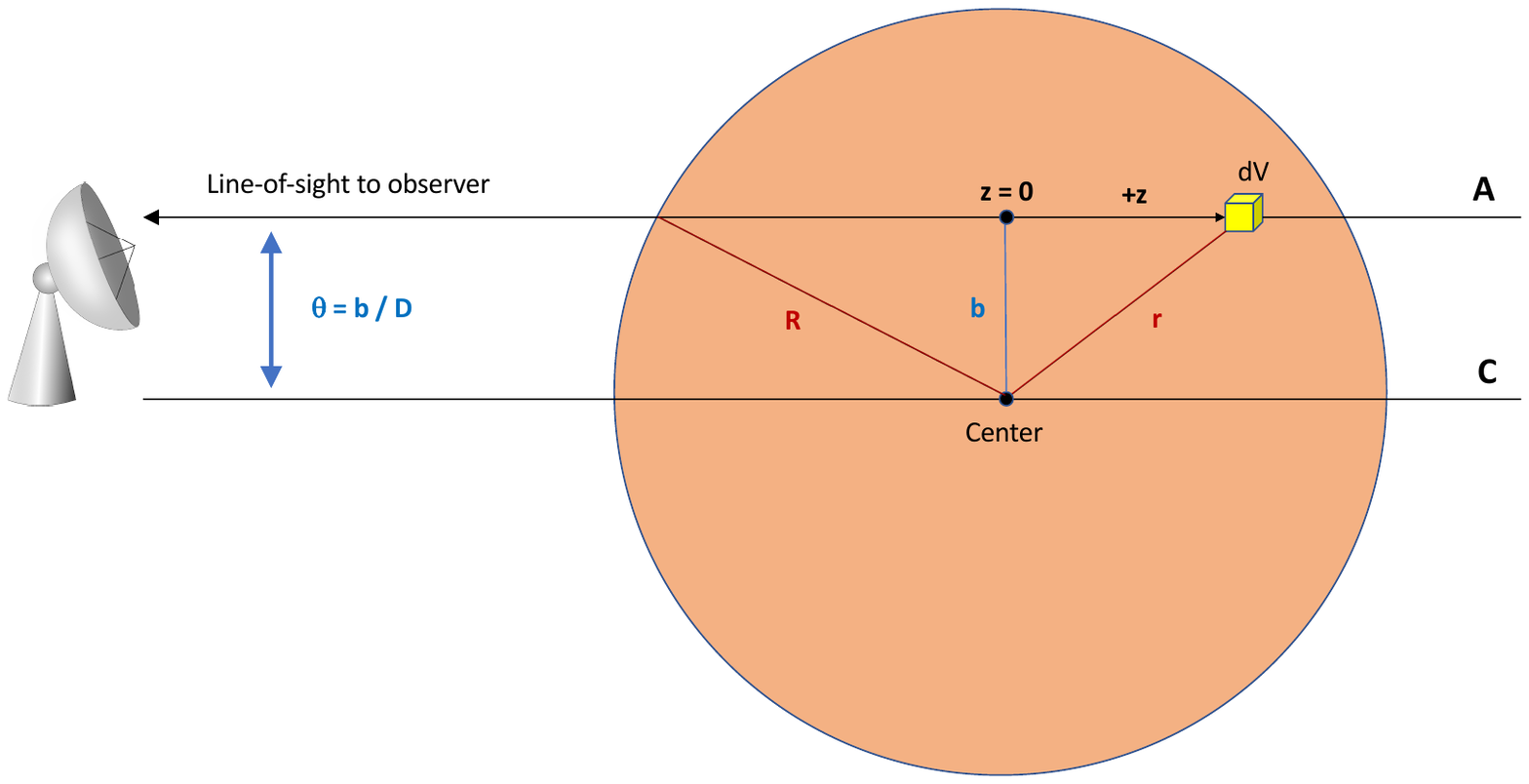}
\vspace{-2cm}
\caption{Geometry of a spherical dust cloud.}
\label{fig:GeoSph}
\end{figure}

A spherical dust cloud with outer radius, $R$, has azimuthal symmetry in the $\phi$ coordinate, therefore the specific intensity is only a function of $\theta$.
The line-of-sight, A, through the sphere may be offset from the center by the impact parameter, b (cm), the distance measured from the center of the sphere perpendicular to the line-of-sight A (Figure \ref{fig:GeoSph}).
At each impact parameter, b (cm), the geometric variables are related through 
\begin{equation}
    z = \sqrt{r^2 - b^2} = \sqrt{r^2 - D^2 \theta^2} \;\; \rm{cm} \;,
\label{eq:zsph}
\end{equation}
where we define $z = 0$ as the point where the impact parameter intersects A. 

The monochromatic specific intensity integral becomes (Equation~\ref{eq:Inu_nobg_thin})
\begin{equation}
         \lim_{\tau_{\nu} \ll 1} \, I_{\nu,\rm{src}}(\theta) = 2 \int_{z = 0}^{\sqrt{R^2 - D^2\theta^2}} B_{\nu}[T_d(\vec{r})] \kappa_{\nu}(\vec{r}) \rho_d(\vec{r}) dz \;\;\;\; \rm{erg}\, \rm{s}^{-1}\, \rm{cm}^{-2}\, \rm{ster}^{-1}\, \rm{Hz}^{-1} \;.
\label{eq:Inu_nobg_thin_full2}
\end{equation}
There is a factor of $2$ in Equation \ref{eq:Inu_nobg_thin_full2} for integration through the entire line-of-sight (both $\pm z$ directions).
The integration along the line-of-sight variable, z (cm), can be transformed into an integral along the radius, r (cm)
if we differentiate Equation \ref{eq:zsph} at a fixed impact parameter, b,
\begin{equation}
    dz = \frac{r}{\sqrt{r^2 - b^2}} dr =  \frac{r}{\sqrt{r^2 - D^2 \theta^2}} dr \;\; \rm{cm} \;.
\end{equation}
Substituting into Equation \ref{eq:Inu_nobg_thin_full2} and expanding the Planck formula (Equation \ref{eq:planck}), we find
\begin{equation}
       \lim_{\tau_{\nu} \ll 1}  I_{\nu,\rm{src}}(\theta) = \frac{4 h \nu^3}{c^2} \int_{r = D\theta}^R  \frac{\kappa_{\nu}(r) \rho_d(r)}{e^{T_{\nu}/T_d(r)}  - 1} \frac{r}{\sqrt{r^2 - D^2 \theta^2}} dr  \;\;\;\; \rm{erg}\, \rm{s}^{-1}\, \rm{cm}^{-2}\, \rm{ster}^{-1}\, \rm{Hz}^{-1} \;,
\label{eq:Inu_Abel}
\end{equation}
where the limits of the integral run from $r = b = D \theta$ to $r = R$ and we define $T_{\nu} = h\nu/k$.
In general, Equation \ref{eq:Inu_Abel} must be solved numerically  and analytical solutions require special limits.
For example, if the spherical dust cloud is isothermal ($T_d(r) = T_d$) with a constant dust opacity ($\kappa_{\nu}(r) = \kappa_{\nu}$),
then the specific intensity reduces to (see Equation \ref{eq:Inu_nobg_thin_iso_constk})
\begin{equation}
       \lim_{T_d(\vec{r}) = T_d}  \, \lim_{\kappa_{\nu}(\vec{r})} \, \lim_{\tau_{\nu} \ll 1}\, I_{\nu,src}(\theta)  =  B_{\nu}(T_d) \kappa_{\nu} \Sigma_d(\theta) \;\;\;\; \rm{erg}\, \rm{s}^{-1}\, \rm{cm}^{-2}\, \rm{ster}^{-1}\, \rm{Hz}^{-1} \;,
\label{eq:Inu_isotherm_Abel}       
\end{equation}
where the dust mass surface density depends on $\theta$ as (see Equation \ref{eq:Inu_Abel})
\begin{equation}
    \Sigma_d(\theta) = 2 \int_{r = D\theta}^R \frac{\rho_d(r)\, r}{\sqrt{r^2 - D^2 \theta^2}} dr  \;\;\;\; \rm{g} \; \rm{of} \; \rm{dust}\; \rm{cm}^{-2} \;\;.
\label{eq:MassSurfDensitySph}
\end{equation}

We point out that the integrals in Equations \ref{eq:Inu_Abel} and  \ref{eq:MassSurfDensitySph} for spherical geometry are Abel integral transforms if the outer radius of the sphere goes to infinity ($\lim R \rightarrow \infty$). 
Their inverse Abel transforms are discussed in more detail in Appendix \ref{AbelApp}.
In Appendix \ref{AppendixVisibilities}, we also discuss how to calculate interferometric visibilities of the observed intensity distribution by taking the Fourier transform.

\subsection{Flux Density}\label{sec:FluxDensity}

The flux density, $F_{\nu}$ (erg s$^{-1}$ cm$^{-2}$ Hz$^{-1}$)\footnote{The notation $S_{\nu}$ is sometimes used in the literature for flux density.  We prefer $F_{\nu}$ for flux density.}, is the monochromatic specific intensity integrated over angular factors.
The spectral energy distribution (SED) of a source is the collection of observed flux densities at different frequencies. 
The popular unit of flux density at infrared and radio wavelengths is the Jansky where 1 Jy = $10^{-23}$ erg s$^{-1}$ cm$^{-2}$ Hz$^{-1}$. 
Using the spherical polar coordinate system defined in Figure \ref{fig:Geo2}, the differential solid angle is $d \Omega = \sin{\theta} d\theta d\phi$ (ster).
The flux density (also known as the monochromatic flux) is given by
\begin{eqnarray}
    F_{\nu,\rm{obs}} & = & \int_{\Omega} I_{\nu,\rm{obs}}(\theta,\phi) \cos{\theta} d \Omega \nonumber \\ 
    & = & \int_{\phi = 0}^{2\pi} \int_{\theta=0}^{\theta_{ap}} I_{\nu,\rm{obs}}(\theta,\phi) \cos{\theta} \sin{\theta} d\theta d\phi  \;\;\;\; \rm{erg}\, \rm{s}^{-1}\, \rm{cm}^{-2}\, \rm{Hz}^{-1} \;\; ,
\label{eq:Fmono}
\end{eqnarray}
where the solid angle is calculated over an aperture (a region on the sky) with maximal angular extent, $\theta_{ap}$ from the center of the object.
The solid angle of the aperture could be the telescope beam size (\ie, the FWHM or full width half-maximum beam width), a fixed aperture size, or encompass the entire source on the sky (\ie, in the most general case for a source with no symmetry, $\theta_{ap}(\phi)$ could be a function of $\phi$).
Note that for integration over the, typically, small solid angles of astronomical sources,  $\cos{\theta} \approx 1$ and $\sin{\theta} \approx \theta$ such that $d \Omega \approx \theta d\theta d\phi$ (ster) and Equation \ref{eq:Fmono} becomes
\begin{equation}
    F_{\nu,\rm{obs}} = \int_{\phi = 0}^{2\pi} \int_{\theta=0}^{\theta_{ap}} I_{\nu,\rm{obs}}(\theta,\phi) \theta d\theta d\phi  \;\;\;\; \rm{erg}\, \rm{s}^{-1}\, \rm{cm}^{-2}\, \rm{Hz}^{-1} \;\; .
\label{eq:Fmonosa}
\end{equation}
In the small angle approximation, the integration on the celestial sphere (spherical angular coordinates) is approximated as integration in a tangent plane to the celestial sphere at C (plane polar angular coordinates).
We shall assume we are in the small angle approximation for the remainder of the tutorial unless otherwise specified.

Plugging in the solution to the radiative transfer equation (Equation \ref{eq:InuFormalSolution}) in its most general form gives an observed flux of  
\begin{equation}
 F_{\nu,obs} = \int_{\Omega} I_{\nu,obs}(\theta,\phi) d\Omega =  \int_{\Omega} I_{\nu,bg}(\theta, \phi) e^{-\tau_{\nu}(\theta,\phi)} d\Omega + \int_{\Omega} I_{\nu,src}(\theta,\phi)  d\Omega \;\;\;\; \rm{erg}\, \rm{s}^{-1}\, \rm{cm}^{-2}\, \rm{Hz}^{-1} \; .
\label{eq:MostGeneral_flux_iso}
\end{equation}
Expanding the $I_{\nu,src}$ term, we can also write this equation as 
\begin{equation}
 F_{\nu,obs} = \int_{\Omega} I_{\nu,bg}(\theta, \phi) e^{-\tau_{\nu}(\theta,\phi)} d\Omega + \int_{\Omega} B_{\nu}[\overline{T_d}(\theta,\phi)] (1 - e^{-\tau_{\nu}(\theta,\phi)}) d\Omega \;\;\;\; \rm{erg}\, \rm{s}^{-1}\, \rm{cm}^{-2}\, \rm{Hz}^{-1} \; ,
\label{eq:General_flux_iso}
\end{equation}
where $\overline{T_d}(\theta,\phi)$ is the equivalent isothermal line-of-sight dust temperature\footnote{Equation \ref{eq:IsoDustTemp} is defined in terms of the optical depth coordinates from the radiative transfer problem.  As an example of a more practical form, in the optically thin limit, we can use Equation \ref{eq:Inu_nobg_thin} in Equation \ref{eq:IsoDustTemp} to find that $\overline{T_d} = \frac{h\nu}{k} \left[ 1 + \frac{\tau_{\nu}}{\int_{A} \frac{\kappa_{\nu}(\vec{r}) \rho_d(\vec{r})}{e^{h\nu/kT_d(\vec{r})} - 1}  dz} \right]^{-1}$.  This expression can only be formally calculated if the distributions of dust temperature, dust density, and dust mass opacity are known along the line-of-sight. See \citealt{2011AJ....141...39S} for an example of this method also accounting for the telescope beam.} toward coordinates $(\theta,\phi)$ defined by Equation \ref{eq:IsoDustTemp}.

Formal solutions to Equations \ref{eq:MostGeneral_flux_iso} or \ref{eq:General_flux_iso} require knowing the on-sky intensity distribution, $I_{\nu}(\theta,\phi)$ (e.g., see Section \ref{SphGeoApp}) and, in general, the flux density integrals are calculated numerically.
It is possible, however, to derive analytic expressions in simplified cases. 
For instance, if the intensity distribution is a constant, $I_{\nu,obs}(\theta,\phi) = I_0$, then the observed flux density is just the intensity times the solid angle, $F_{\nu,obs} = I_0 \Omega_{ap}$.
For more complicated intensity distributions, it is convenient to define the solid angle-averaged specific intensity as 
\begin{equation}
        \overline{I_{\nu,obs}} = \frac{ \int_{\phi = 0}^{2\pi} \int_{\theta=0}^{\theta_{ap}} I_{\nu,obs}(\theta,\phi) \theta d\theta d\phi}{\Omega_{ap}}  \;\;\;\; \rm{erg}\, \rm{s}^{-1}\, \rm{cm}^{-2}\, \rm{ster}^{-1}\, \rm{Hz}^{-1} \;,
\label{eq:Inu_angleavg}
\end{equation}
where the solid angle in the denominator is defined as
\begin{equation}
    \Omega_{ap} = \int_{\phi = 0}^{2\pi} \int_{\theta=0}^{\theta_{ap}} \theta d\theta d\phi
\label{eq:ApertureSolidAngle}
\end{equation}
The observed flux density is then related to the solid angle-averaged specific intensity by
\begin{equation}
F_{\nu,obs} = \overline{I_{\nu,obs}} \Omega_{ap} \;\;\;\; \rm{erg}\, \rm{s}^{-1}\, \rm{cm}^{-2}\, \rm{Hz}^{-1} \;\; .
\label{eq:FnuInu}
\end{equation}
This equation also permits calculation of the solid angle-averaged specific intensity from the observed flux density over the solid angle of the aperture, $\overline{I_{\nu,obs}} = F_{\nu,obs} / \Omega_{ap}$.
Image pixels that are in units of Jy/beam, MJy/ster, or Jy/pixel are reporting $\overline{I_{\nu,obs}}$ within the solid angle of the aperture that corresponds to the telescope beam, steradians, or a pixel (see Section \ref{Caveats} for more details).
Aperture photometry in the image is performed using Equation \ref{eq:FnuInu}.
For example, if the $i^{\rm th}$ pixel in an image has average intensity  $\overline{I_{\nu,obs}}_i$ and a solid angle of $\Omega_{pix,i}$ that lies within the aperture\footnote{Assuming all pixels in the image have the same solid angle, this notation accounts for the possibility that the aperture boundary intersects pixels resulting in different fractions of the pixel solid angle.}, then the flux density within the aperture is found by summing up the flux densities within each pixel
\begin{equation}
    F_{\nu,obs} = \sum_{i \;\rm{within} \;\Omega_{ap}} \overline{I_{\nu,obs}}_i \Omega_{pix,i} \;\;\;\; \rm{erg}\, \rm{s}^{-1}\, \rm{cm}^{-2}\, \rm{Hz}^{-1} \;\; .
\label{eq:Fnupixels}
\end{equation}

We can derive compact analytical expressions for the observed flux density when the source is isothermal and the background intensity is constant ($I_{\nu,bg}(\theta,\phi) = I_{\nu,bg}$).
The solid angle-averaged source optical depth at each frequency, $\overline{\tau_{\nu}}$, is defined as
\begin{equation}
  1 - e^{-\overline{\tau_{\nu}}} = \frac{ \int_{\phi = 0}^{2\pi} \int_{\theta=0}^{\theta_{ap}} 1 - e^{-\tau_{\nu}(\theta,\phi)}  \theta d\theta d\phi}{\Omega_{ap}} \;\;.
\label{eq:angleavgtau}
\end{equation}
In the limit that the source is isothermal and that the background intensity is constant, then the angle-averaged specific intensity is given by the 1D solution to the radiative transfer (Equation \ref{eq:Inuiso})
\begin{equation}
    \lim_{I_{\nu,\rm{bg}}(\theta,\phi) = I_{\nu,bg}} \; \lim_{T_d(\vec{r}) = T_d} \overline{I_{\nu,obs}} =  I_{\nu,bg}e^{-\overline{\tau_{\nu}}} + B_{\nu}(T_d) (1 - e^{-\overline{\tau_{\nu}}}) \;\;\;\; \rm{erg}\, \rm{s}^{-1}\, \rm{cm}^{-2}\, \rm{ster}^{-1}\, \rm{Hz}^{-1} \;,
\end{equation}
and the observed flux density is
\begin{equation}
    \lim_{I_{\nu,\rm{bg}}(\theta,\phi) = I_{\nu,bg}} \; \lim_{T_d(\vec{r}) = T_d}  \; F_{\nu,obs} = [I_{\nu,bg}e^{-\overline{\tau_{\nu}}} + B_{\nu}(T_d)(1 - e^{-\overline{\tau_{\nu}}}) ]\Omega_{ap} \;\;\;\; \rm{erg}\, \rm{s}^{-1}\, \rm{cm}^{-2}\, \rm{Hz}^{-1} \;\; .
\label{eq:FluxSEDwithbg}
\end{equation}
In the limit that the background intensity is negligible, then this equation becomes
\begin{equation}
   \lim_{I_{\nu,\rm{bg}} \rightarrow 0} \; \lim_{T_d(\vec{r}) = T_d}  \; F_{\nu,obs} =  B_{\nu}(T_d)(1 - e^{-\overline{\tau_{\nu}}}) \Omega_{ap} \;\;\;\; \rm{erg}\, \rm{s}^{-1}\, \rm{cm}^{-2}\, \rm{Hz}^{-1} \;\; .
\label{eq:FluxSED}
\end{equation}
We can further simplify this equation if we are in the optically thick limit($\overline{\tau_{\nu}} \gg 1$),
\begin{equation}
    \lim_{\tau_{\nu} \gg 1}  \; F_{\nu,obs} =  B_{\nu}(T_d) \Omega_{ap} \;\;\;\; \rm{erg}\, \rm{s}^{-1}\, \rm{cm}^{-2}\, \rm{Hz}^{-1} \;\;, 
\label{eq:FluxSEDThick}
\end{equation}
or if we are in the optically thin limit ($\overline{\tau_{\nu}} \ll 1$),
\begin{equation}
    \lim_{\tau_{\nu} \ll 1}  \; F_{\nu,obs} =  B_{\nu}(T_d) \overline{\tau_{\nu}} \Omega_{ap} \;\;\;\; \rm{erg}\, \rm{s}^{-1}\, \rm{cm}^{-2}\, \rm{Hz}^{-1} \;\;.  
\end{equation}

Equation \ref{eq:FluxSED} is often the starting point when analyzing an object's SED. 
We must stress, however, that Equation \ref{eq:FluxSED} has very simplifying assumptions such as isothermality and negligible background emission.
SED fitting is discussed in detail in Section \ref{ModelingSection} and the calculation of $\overline{\tau_{\nu}}$ for various geometries is discussed in Appendix \ref{AppendixAvgTau}.

\subsection{Mass Surface Density, Column Density, and Mass}\label{sec:MassDerive}

\begin{figure}[h!]
\includegraphics[scale=0.5]{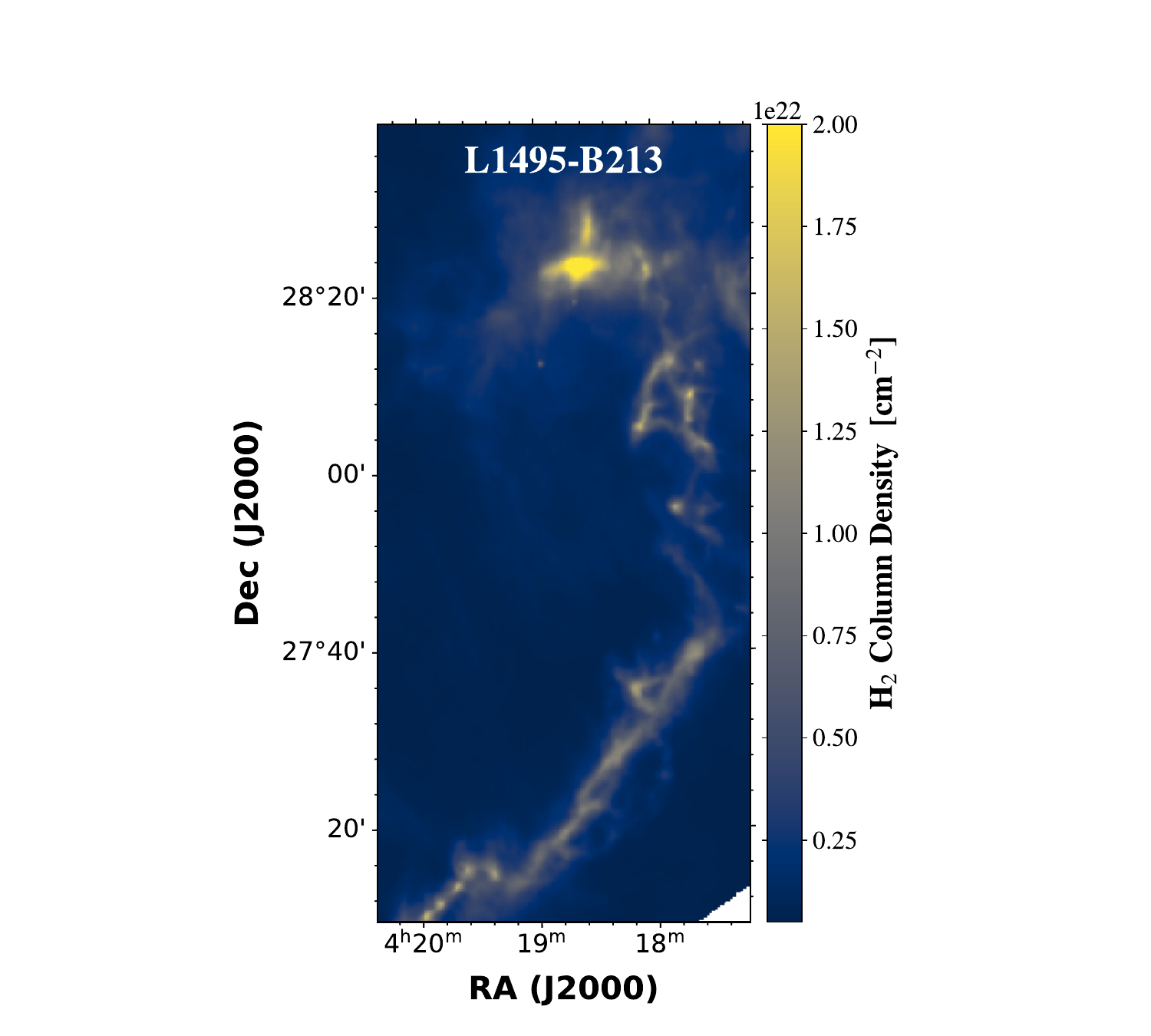}
\caption{The H$_2$ column density of the L1495-B213 Regions of the Taurus Molecular Cloud determined from \textit{Herschel} and \textit{Planck} observations  \citep{2022ApJ...941..135S}.  The optical depth at $1$ THz and dust temperature were determined from fitting the spectral energy distribution.  Equation \ref{eq:tauNg} was then used to convert to $N_{\rm{H}_2}$ assuming $\kappa_{1\,\rm{THz}} = 10$ cm$^2$ per gram of dust, $\mu_{\rm{H}_2} = 2.8$, and $R_{gd} = 100$.}
\label{fig:H2Column}
\end{figure}

In this section, we show that the observed flux density is linearly proportional to the dust mass surface density, the gas column density, and the mass in the limits:
\begin{itemize}
    \item the background intensity does not vary across the solid angle of the aperture,
    \item the emission is optically thin,
    \item the source is isothermal, and 
    \item the source has constant opacity.
\end{itemize}
If we assume that the emission is optically thin, then in Equation \ref{eq:General_flux_iso} we can use the Maclaurin expansion of $e^{-\tau_{\nu}(\theta,\phi)} \approx 1 - \tau_{\nu}(\theta,\phi)$ to simplify this expression to
\begin{equation}
\lim_{\tau_{\nu} \ll 1} \;     F_{\nu,obs} = \int_{\Omega} I_{\nu,bg}(\theta, \phi) [1 - \tau_{\nu}(\theta,\phi)] d\Omega + \int_{\Omega}  B_{\nu}[\overline{T_d}(\theta,\phi)] \tau_{\nu}(\theta,\phi)) d\Omega \;\;\;\; \rm{erg}\, \rm{s}^{-1}\, \rm{cm}^{-2}\, \rm{Hz}^{-1} \; .
\label{eq:General_thin_flux}
\end{equation}
Applying the additional limits to Equation \ref{eq:General_thin_flux} and using Equation \ref{eq:taunu_limit}, we find that
\begin{eqnarray}
\lim_{I_{\nu,\rm{bg}}(\theta,\phi) = I_{\nu,bg}} \; \lim_{\tau_{\nu} \ll 1} \; \lim_{T_d(\vec{r}) = T_d}  \; \lim_{\kappa_{\nu}(\vec{r}) = \kappa_{\nu}}    F_{\nu,obs}  =   I_{\nu,bg}\Omega_{ap} \,  & - &  \, I_{\nu,bg} \kappa_{\nu} \int_{\Omega} \Sigma_d(\theta,\phi) d\Omega \nonumber \\ & + & B_{\nu}(T_d)\kappa_{\nu} \int_{\Omega} \Sigma_d(\theta,\phi) d\Omega \;\;\;\; \rm{erg}\, \rm{s}^{-1}\, \rm{cm}^{-2}\, \rm{Hz}^{-1} \;,
\label{eq:Flux_iso_constbg}
\end{eqnarray}
where $\Omega_{ap}$ is the solid angle of the aperture (or beam) that the observed flux density is determined within.
We define the background subtracted flux density as 
\begin{equation}
\Delta F_{\nu,obs} = F_{\nu,obs} - I_{\nu,bg}\Omega_{ap} \;\;\;\; \rm{erg}\, \rm{s}^{-1}\, \rm{cm}^{-2}\, \rm{Hz}^{-1} \;.
\label{eq:DeltaF_Ibg}
\end{equation}
If the background flux density is determined from a different solid angle (\ie, $\Omega_{bg}$) than the solid angle of the observed flux density ($\Omega_{ap}$), then this background flux density must be corrected by the ratio of solid angles ($\Omega_{ap}/\Omega_{bg}$) before being used in Equation \ref{eq:DeltaF_Ibg}.
Using the definition of $\Delta F_{\nu,obs}$, we can rewrite Equation \ref{eq:Flux_iso_constbg} as
\begin{equation}
\lim_{I_{\nu,\rm{bg}}(\theta,\phi) = I_{\nu,bg}} \; \lim_{\tau_{\nu} \ll 1} \; \lim_{T_d(\vec{r}) = T_d}  \; \lim_{\kappa_{\nu}(\vec{r}) = \kappa_{\nu}}    \Delta F_{\nu,obs}  =  [B_{\nu}(T_d) - I_{\nu,bg}]\kappa_{\nu} \int_{\Omega} \Sigma_d(\theta,\phi) d\Omega \;\;\;\; \rm{erg}\, \rm{s}^{-1}\, \rm{cm}^{-2}\, \rm{ster}^{-1}\, \rm{Hz}^{-1} \;.
\label{eq:DeltaFlux_iso_constbg}
\end{equation}
The integral of $\Sigma_d$ over solid angle can be solved explicitly if the density distribution, $\rho_d(\vec{r})$ is known but this distribution is often an unknown quantity.
Therefore, we define the solid angle-averaged mass surface density over an aperture as 
\begin{equation}
 \overline{\Sigma_d} =   \frac{ \int_{\phi = 0}^{2\pi} \int_{\theta=0}^{\theta_{ap}} \Sigma_d(\theta,\phi) \theta d\theta d\phi}{\Omega_{ap}} \;\;\;\; \rm{g} \; \rm{of} \; \rm{dust}\; \rm{cm}^{-2}  \;\; .
\label{eq:MassSurfDenAvg}  
\end{equation}
Solving for the solid angle-averaged mass surface density in Equation~\ref{eq:DeltaFlux_iso_constbg} and inserting the result into Equation~\ref{eq:MassSurfDenAvg} we find that
\begin{equation}
\lim_{I_{\nu,\rm{bg}}(\theta,\phi) = I_{\nu,bg}} \; \lim_{\tau_{\nu} \ll 1} \; \lim_{T_d(\vec{r}) = T_d}  \; \lim_{\kappa_{\nu}(\vec{r}) = \kappa_{\nu}}    \overline{\Sigma_d} = \frac{\Delta F_{\nu,obs}}{[B_{\nu}(T_d) - I_{\nu,bg}] \kappa_{\nu} \Omega_{ap}}  \;\;\;\; \rm{g} \; \rm{of} \; \rm{dust}\; \rm{cm}^{-2} \;\; .
\label{eq:MassSurfDenAvg_final}
\end{equation}
Therefore, the flux density received from an isothermal, constant opacity dusty object is linearly proportional to the solid angle-averaged dust mass surface density.
Equation \ref{eq:sigmad} may also be used to relate $\overline{\Sigma_d}$ to the gas column density, $N_g$ (cm$^{-2}$), in the limit that $\mu_{g}$ and $R_{gd}$ is constant within the solid angle of the aperture
\begin{equation}
\lim_{I_{\nu,\rm{bg}}(\theta,\phi) = I_{\nu,bg}} \; \lim_{\tau_{\nu} \ll 1} \; \lim_{T_d(\vec{r}) = T_d}  \; \lim_{\kappa_{\nu}(\vec{r}) = \kappa_{\nu}} \; \lim_{R_{gd}(\vec{r})=R_{gd}} \; \lim_{\mu_g(\vec{r})=\mu_g} \;    \overline{N_{g}} = \frac{R_{gd} \Delta F_{\nu,obs}}{\mu_{g} m_{\rm{H}} [B_{\nu}(T_d) - I_{\nu,bg}] \kappa_{\nu} \Omega_{ap}} \;\;\;\; \rm{cm}^{-2} \;\; .
\label{eq:GasColumnDen_final}
\end{equation}
The solid angle-averaged mass surface density is also related to the solid angle-averaged optical depth in the limits that the emission is optically thin and that the source has a constant dust mass opacity,
\begin{equation}
\lim_{\tau_{\nu} \ll 1} \; \lim_{\kappa_{\nu}(\vec{r}) = \kappa_{\nu}} \;  \overline{\tau_{\nu}} = \kappa_{\nu} \overline{\Sigma_d}  \;\; , 
\label{eq:taukappasigma}
\end{equation}
(see Equation \ref{eq:angleavgtau} for the definition of $\overline{\tau_{\nu}}$).
Using Equation \ref{eq:sigmad}, we can also relate $\overline{\tau_{\nu}}$ to $\overline{N_g}$ with
\begin{equation}
    \lim_{\tau_{\nu} \ll 1} \; \lim_{\kappa_{\nu}(\vec{r}) = \kappa_{\nu}} \; \lim_{R_{gd}(\vec{r})=R_{gd}} \; \lim_{\mu_g(\vec{r})=\mu_g} \;  \overline{\tau_{\nu}} = \kappa_{\nu} \frac{\mu_g m_{\rm{H}}}{R_{gd}}  \overline{N_g} \;\;\;\; .
\label{eq:tauNg}
\end{equation}
Equation \ref{eq:tauNg} is a useful conversion since $\overline{\tau_{\nu}}$ can be determined from fitting the spectral energy distribution of observed flux densities (see Section \ref{sec:SEDModels} for a discussion of SED fitting).
For example, the H$_2$ column density map shown in Figure \ref{fig:H2Column} is generated from measurements of $\overline{\tau_{\nu}}$ at 1 THz by solving Equation \ref{eq:tauNg} for $\overline{N_g} = \overline{N_{\rm{H}_2}}$ \citep{2022ApJ...941..135S}.

The solid angle of the aperture is proportional to the projected area of the aperture, $A_{ap}$ (cm$^2$), at the distance, $D$ (cm), of the object by the definition of solid angle
\begin{equation}
    \Omega_{ap} = \frac{A_{ap}}{D^2} \;\;\;\; \rm{ster} \;.
\label{eq:solidang}
\end{equation}
We can then calculate the dust mass within the solid angle of the aperture by multiplying the solid angle-averaged mass surface density by the projected area of the aperture 
\begin{equation}
   M_d = \overline{\Sigma_d} A_{ap} = \overline{\Sigma_d} \Omega_{ap} D^2 \;\;\;\; \rm{g}\; \rm{of}\; \rm{dust} \;\; .
\label{eq:Md_thin}
\end{equation}
Substituting for the dust mass surface density (Equation~\ref{eq:MassSurfDenAvg_final}), we find that
\begin{equation}
\lim_{I_{\nu,\rm{bg}}(\theta,\phi) = I_{\nu,bg}} \; \lim_{\tau_{\nu} \ll 1} \; \lim_{T_d(\vec{r}) = T_d}  \; \lim_{\kappa_{\nu}(\vec{r}) = \kappa_{\nu}}    M_d = \frac{\Delta F_{\nu,obs} D^2}{[B_{\nu}(T_d) - I_{\nu,bg}] \kappa_{\nu}} \;\;\;\; \rm{g} \; \rm{of} \; \rm{dust} \;\;.
\label{eq:Md_thin_final}
\end{equation}
This mass formula is valid for an optically thin, isothermal, constant opacity dust source with a constant intensity background.
If the aperture encompasses the entire object, then this equation gives the total dust mass of the object.
If, however, the flux is calculated from an aperture smaller than the source, then this equation gives the dust mass only within the solid angle of the aperture.

We can convert the dust mass into the gas mass using the gas-to-dust mass ratio (see Equation \ref{eq:sigmag})
\begin{equation}
    M_g = R_{gd} M_d \;\;\;\; \rm{g} \; \rm{of} \; \rm{gas} \;\;.
\label{eq:Md_thin_final_gas}
\end{equation}
The total mass of the source is then the sum of the gas mass and the dust mass 
\begin{equation}
M_{\rm{tot}} = M_d + M_g = (R_{gd} + 1)M_d \;\;\;\; g \;\; .
\label{eq:Md_thin_final_tot}
\end{equation}
If $R_{gd}$ is large, then the gas mass alone is a good approximation to the total mass of the source.

We note that our derivation is a more general approach to deriving the dust mass than the classic \citealt{1977ApJ...216..698H} and \citealt{1983QJRAS..24..267H} articles which include no background.
In the limit that the background intensity is negligible, then our equations for dust mass surface density, gas column density, and mass, are modified by replacing the terms  $[B_{\nu}(T_d) - I_{\nu,bg}] \rightarrow B_{\nu}(T_d)$ and $\Delta F_{\nu,obs} \rightarrow F_{\nu,src}$ where
\begin{equation}
   F_{\nu,src} =  \int_{\Omega_{src}} I_{\nu,src}(\theta,\phi) d\Omega \;\;\;\; \rm{erg}\, \rm{s}^{-1}\, \rm{cm}^{-2}\, \rm{Hz}^{-1} \; ,
\label{eq:Fsrc_definition}
\end{equation}
which is just the flux density intrinsic to only the source.
The equations then become
\begin{equation}
   \lim_{I_{\nu,\rm{bg}} \rightarrow 0} \; \lim_{\tau_{\nu} \ll 1} \; \lim_{T_d(\vec{r}) = T_d}  \; \lim_{\kappa_{\nu}(\vec{r}) = \kappa_{\nu}} \overline{\Sigma_d} = \frac{F_{\nu,src}}{B_{\nu}(T_d) \kappa_{\nu} \Omega_{ap}}  \;\;\;\; \rm{g} \; \rm{of} \; \rm{dust}\; \rm{cm}^{-2} \;\; ,
\label{eq:MassSurfDenAvg_nobg_final}
\end{equation}
\begin{equation}
    \lim_{I_{\nu,\rm{bg}} \rightarrow 0} \; \lim_{\tau_{\nu} \ll 1} \; \lim_{T_d(\vec{r}) = T_d}  \; \lim_{\kappa_{\nu}(\vec{r}) = \kappa_{\nu}} \; \lim_{R_{gd}(\vec{r})=R_{gd}} \; \lim_{\mu_g(\vec{r})=\mu_g} \; \overline{N_{g}} = \frac{R_{gd} F_{\nu,src}}{\mu_{g} m_{\rm{H}} B_{\nu}(T_d) \kappa_{\nu} \Omega_{ap}} \;\;\;\; \rm{cm}^{-2} \;\; ,
\label{eq:GasColumnDen_nobg_final}
\end{equation}
and
\begin{equation}
    \lim_{I_{\nu,\rm{bg}} \rightarrow 0} \; \lim_{\tau_{\nu} \ll 1} \; \lim_{T_d(\vec{r}) = T_d}  \; \lim_{\kappa_{\nu}(\vec{r}) = \kappa_{\nu}} M_d = \frac{F_{\nu,src} D^2}{B_{\nu}(T_d) \kappa_{\nu}} \;\;\;\; \rm{g} \; \rm{of} \; \rm{dust} \;\;.
\label{eq:Md_thin_nobg_final}
\end{equation}
Equation \ref{eq:Md_thin_nobg_final} is the formula most commonly found in the literature for the dust mass.
We note that if the background flux density ($F_{\nu,bg} = I_{\nu,bg}\Omega_{ap}$) is just subtracted from $F_{\nu,obs}$ to calculate $F_{\nu,src}$ and then is used in Equation \ref{eq:Md_thin_nobg_final},  then the dust mass will be underestimated.  This is because only a fraction of the background intensity, $1 - \tau_{\nu} < 1$, passes through the object in the optically thin limit. 
The correct forms of the dust mass equations are Equation \ref{eq:Md_thin_final} in the limit of a constant intensity background and Equation \ref{eq:Md_thin_nobg_final} in the limit of a negligible background.
We also note that the derivations in this article have assumed there is no foreground emission.
In situations where there is foreground emission, the equivalent of Equation \ref{eq:General_thin_flux} must first be derived.

\subsection{Luminosity}\label{sec:Luminosity}

The monochromatic luminosity, $L_{\nu}$ (erg s$^{-1}$ Hz$^{-1}$), at each frequency is proportional to the total observed flux density of an isotropic emitting source (integrated over the full solid angle of the source and not including any background emission),
\begin{equation}
 L_{\nu} = 4 \pi D^2 F_{\nu,src} \;\;\;\; \rm{erg}\, \rm{s}^{-1}\,  \rm{Hz}^{-1} \;\; .
\label{eq:Lnuflux}
\end{equation}
The bolometric luminosity of the source is found by integrating the monochromatic luminosity over all frequencies
\begin{eqnarray}
    L_{\rm{bol}} & = & \int_{\nu = 0}^{\infty} L_{\nu} d\nu \label{eq:Lbol1} \nonumber \\ 
    & = & 4 \pi D^2 \int_{\nu = 0}^{\infty} F_{\nu,src} \, d\nu \;\; \rm{erg} \; \rm{s}^{-1}\;\; .
\label{eq:Lbol}
\end{eqnarray}
If the SED is well sampled, then the bolometric luminosity is calculated from the integral of the observed total flux densities.
In practice, this integration is often accomplished through a coarse numerical integration technique, such as the trapezoid rule, that does not require regularly-spaced flux densities in frequency (as would be required by Simpson's Rule).

We can derive a simple expression for $L_{\nu}$ in the limit of optically thin emission.
If we assume that emission from dust grains is isotropic, then each differential volume element, $dV$, in a source is emitting with emissivity coefficient, $j_{\nu}$, into $4 \pi$ steradians (see Figure \ref{fig:Geo1}).
Thus, for an optically thin emitting source, the monochromatic luminosity is given by the volume integral (see Homework problem 1.8 of \citealt{1986rpa..book.....R}), using Equations~\ref{eq:alphanu} and \ref{eq:jnu},
\begin{eqnarray}
 \lim_{\tau_{\nu} \ll 1}   L_{\nu} & = & 4 \pi \int_V  j_{\nu}(\vec{r}) dV \label{eq:Lnu_mono1}\\ L_{\nu} & = & 4 \pi \int_{V}  B_{\nu}[T_d(\vec{r})]\kappa_{\nu}(\vec{r}) \rho_d(\vec{r}) dV \;\;\;\; \rm{erg}\, \rm{s}^{-1}\, \rm{Hz}^{-1} \;\;.
\label{eq:Lnu_mono2}
\end{eqnarray}
This equation is valid for an arbitrary geometry of the source. 
If we assume that the emitting object is isothermal and has constant opacity, then the monochromatic luminosity is proportional to the total dust mass of the object,
\begin{eqnarray}
    \lim_{\tau_{\nu} \ll 1} \lim_{T_d(\vec{r}) = T_d}   \lim_{\kappa_{\nu}(\vec{r}) = \kappa_{\nu}} L_{\nu} & = & 4 \pi   B_{\nu}(T_d) \kappa_{\nu} \int_{V} \rho_d(\vec{r}) dV  \nonumber \\ 
    & = & 4 \pi  B_{\nu}(T_d) \kappa_{\nu} M_d \;\;\;\; \rm{erg}\, \rm{s}^{-1}\, \rm{Hz}^{-1} \;\;.
\label{eq:Lnu_mono_thin_isotherm}
\end{eqnarray}
The dust mass equation we derived above (Equation~\ref{eq:Md_thin_final}) can also be derived by substituting for the monochromatic luminosity from Equation \ref{eq:Lnu_mono_thin_isotherm} into Equation \ref{eq:Lnuflux}, then we find the same equation for the dust mass (Equation \ref{eq:Md_thin_final}; $M_{d,\rm{tot}} = F_{\nu,src}D^2/B_{\nu}(T_d)\kappa_{\nu}$), but now for the total object dust mass.

\subsection{Modification of Equations for Redshift, Gravitational Lensing and the Cosmic Microwave Background} \label{Redshift}

Observations of dust emission from distant galaxies have the additional complication that the observed frequency, $\nu_o$, is not the same as the emitted frequency, $\nu_e$, due to the expansion of spacetime.
These two frequencies are related by the redshift of the object, $z_{\rm{red}}$, through the equation
\begin{equation}
    \frac{\nu_e}{\nu_o} = 1 + z_{\rm{red}} \;\; .
\label{eq:redshift}
\end{equation}
When $\lim z_{\rm{red}} \rightarrow 0$, the observed frequency is equal to the emitted frequency and all of the equations derived in previous sections apply.
When $z_{\rm{red}}$ becomes non-negligible though, then we have to consider the effect of cosmological redshift on the observed monochromatic specific intensity, solid angle, and flux density. 
We started our derivation of the equation of radiative transfer by stating that the monochromatic specific intensity is constant along a line of sight so long as there is no absorption, emission, or scattering. When the frequency of the light is reduced by the cosmological expansion, then $I_{\nu}$ is no longer an invariant quantity.
Instead, it can be shown that $I_{\nu}/\nu^3$ is the proper Lorentz invariant quantity (see Section 1.6 of \citealt{goodman2013topics})\footnote{The total number of photons, with two possible spin states, in 6-dimensional position-momentum phase space is $\mathcal{N}_{\gamma} = 2 \int n_{\gamma} \frac{d^3x d^3p}{h^3}$ where $n_{\gamma}$ is the Bose-Einstein distribution (with chemical potential zero) describing the photon occupation number, $n_{\gamma} = \frac{1}{e^{h\nu/kT}-1}$. Therefore, $n_{\gamma} \propto d\mathcal{N}_{\gamma}/d^3x d^3p$.  
Since each phase space volume element, $d^3x d^3p$, is Lorentz invariant (see Section 1.5 of \citealt{goodman2013topics}), then $n_{\gamma}$ is Lorentz invariant. $I_{\nu} = \frac{2h\nu^3}{c^2} n_{\gamma}$, therefore $n_{\gamma} \propto I_{\nu}/\nu^3$. Thus, $I_{\nu}/\nu^3$ is also Lorentz invariant.}.
This invariance means that
\begin{eqnarray}
    \frac{I_{\nu,obs}}{\nu_o^3} & = & \frac{I_{\nu_e}}{\nu_e^3} \;\; \\
    I_{\nu,obs} & = & \frac{I_{\nu_e}}{(1 + z_{\rm{red}})^3} \;\;\;\; \rm{erg}\, \rm{s}^{-1}\, \rm{cm}^{-2}\, \rm{ster}^{-1}\, \rm{Hz}^{-1} \; .
\label{eq:Inucubed}
\end{eqnarray}
Again, in the $\lim z_{\rm{red}} \rightarrow 0$, then $I_{\nu,obs} = I_{\nu_e}$ and our previous equations are valid.
Equation \ref{eq:Inucubed} gives the general result for how the emitted specific intensity is modified by redshift. 
Depending on the distribution of mass along the line-of-sight, however, gravitational lensing may also increase the observed specific intensity by a magnification factor of $\mu_{\rm{lens}} \geq 1$, 
\begin{equation}
    I_{\nu,obs}  =  \mu_{\rm{lens}} \frac{I_{\nu_e}}{(1 + z_{\rm{red}})^3} \;\;\;\; \rm{erg}\, \rm{s}^{-1}\, \rm{cm}^{-2}\, \rm{ster}^{-1}\, \rm{Hz}^{-1} \; ,
\end{equation}
(see \citealt{2017grle.book.....D} and \citealt{2018pgl..book.....C} for a summary of gravitational lensing).

The Cosmic Microwave Background (CMB) temperature increases with redshift as 
\begin{equation}
    T_{\rm{CMB}_e} = T_{\rm{CMB}_o} (1 + z_{\rm{red}}) \;\; \rm{K} \;,
\end{equation}
where $T_{\rm{CMB}_o} = 2.7255$ K is the observed CMB temperature at $z_{\rm{red}} = 0$ \citep{2009ApJ...707..916F}.
For instance, at $z_{\rm{red}} = 2.7$, $T_{\rm{CMB}_e} \sim 10$ K which is comparable to the typical dust temperature in dense cores in the Milky Way \citep{2014prpl.conf...27A}. 
The CMB background, as well as the radiative transfer effects on the equilibrium dust temperature due to the CMB, must be taken into account at high redshift (see \citealt{2013ApJ...766...13D}). 
Over the small angular scale of distant galaxies, the CMB is an approximately constant background intensity (\ie\ it is in the limit that $I_{\nu,bg}(\theta,\phi) = I_{\nu,bg}$).

One additional quantity that varies with redshift is the observed solid angle of the source, $\Omega(z_{\rm{red}})$. 
The observed solid angle is related to the angular-diameter distance\footnote{We use angular-diameter distance here because it is the distance measure that preserves the relationship between linear distance within the source and the observed angle that distance subtends.}, $D_A$ (cm), and the true cross-sectional area of the emitting source, $A_{src}$ (cm$^2$), by
\begin{equation}
    \Omega(z_{\rm{red}}) = \frac{A_{src}}{D_A^2} \;\;\;\; \rm{ster} \; .
\end{equation}
The angular-diameter distance depends on the exact cosmological model used for the expansion history of the Universe (see the \citealt{1999astro.ph..5116H} review for a summary of different cosmological distance measures).
The most appropriate distance for flux density observations though is the luminosity distance\footnote{This is because the luminosity distance preserves the relationship between luminosity and flux, $F_{obs} = L_e/4 \pi D_L^2$, where the $e$ subscript refers to emitted luminosity.}, $D_L$ (cm), which is related to $D_A$ by $D_L = D_A (1 + z_{\rm{red}})^2$ \citep{1999astro.ph..5116H}.
We find that the observed solid angle depends on redshift and luminosity distance by
\begin{equation}
    \Omega(z_{\rm{red}}) = \frac{A_{src} (1 + z_{\rm{red}} )^4}{D_L^2} \;\;\;\; \rm{ster} \; .
\label{eq:SolidAng_redshift}
\end{equation}

We can then modify Equation \ref{eq:FluxSEDwithbg} for isothermal emission by including a CMB background and correcting for redshift and gravitational lensing to find that
\begin{equation}
   \lim_{T_d(\vec{r}) = T_d}  \;    F_{\nu, obs} = \mu_{\rm{lens}} \frac{B_{\nu_e}(T_{\rm{CMB}_e}) e^{-\overline{\tau_{\nu_e}}} +  B_{\nu_e}(T_{d_e}) (1 - e^{-\overline{\tau_{\nu_e}}})}{(1 + z_{\rm{red}})^3}  \Omega(z_{\rm{red}}) \;\;\;\; \rm{erg}\, \rm{s}^{-1}\, \rm{cm}^{-2}\, \rm{Hz}^{-1} \; .
\label{eq:fluxred4}
\end{equation}
where $T_{d_e}$ is the equilibrium dust temperature emitted by the source at redshift $z_{\rm{red}}$ (see \citealt{2013ApJ...766...13D} for a derivation of the CMB effects on the equilibrium dust temperature, $T_{d_e}$, at different redshifts). 
If we define the observed CMB-subtracted flux density as 
\begin{eqnarray}
    \Delta F_{\nu,obs} & = &  F_{\nu,obs} - B_{\nu_o}(T_{\rm{CMB}_o})\Omega(z_{\rm{red}}) \nonumber \\
     & = & F_{\nu,obs} - \mu_{\rm{lens}} \frac{B_{\nu_e}(T_{\rm{CMB}_e})\Omega(z_{\rm{red}})}{(1 + z_{\rm{red}})^3} \;\;\;\; \rm{erg}\, \rm{s}^{-1}\, \rm{cm}^{-2}\, \rm{Hz}^{-1} \; ,
\label{eq:DelFnuz}
\end{eqnarray}
where we have subtracted off the observed flux density of the CMB at the observed frequency in the same solid angle subtended by the observed source, then we find that (substituting for the solid angle with Equation \ref{eq:SolidAng_redshift}) that
\begin{equation}
   \lim_{T_d(\vec{r}) = T_d}  \;   \Delta F_{\nu,obs} =  \mu_{\rm{lens}} (1 + z_{\rm{red}}) \left[B_{\nu_e}(T_{d_e}) - B_{\nu_e}(T_{\rm{CMB}_e}) \right] (1 - e^{-\overline{\tau_{\nu_e}}} ) \frac{A_{src}}{D_L^2}  \;\;\;\; \rm{erg}\, \rm{s}^{-1}\, \rm{cm}^{-2}\, \rm{Hz}^{-1} \; .
\label{eq:fluxred7}
\end{equation}
If we are in the optically thin limit, $\overline{\tau_{\nu_e}} \ll 1$, with a constant opacity then we can approximate $e^{-\overline{\tau_{\nu_e}}} \approx 1 - \overline{\tau_{\nu_e}}$ and relate the average optical depth to the average mass surface density of the dust by using Equation \ref{eq:taukappasigma}. 
We find that Equation \ref{eq:fluxred7} becomes 
\begin{equation}
   \lim_{T_d(\vec{r}) = T_d}  \;  \lim_{\tau_{\nu_e} \ll 1}  \; \lim_{\kappa_{\nu}(\vec{r}) = \kappa_{\nu}} \;    \Delta F_{\nu,obs} =  \mu_{\rm{lens}} (1 + z_{\rm{red}}) \left[B_{\nu_e}(T_{d_e}) - B_{\nu_e}(T_{\rm{CMB}_e}) \right] \kappa_{\nu_e} \overline{\Sigma_d}  \frac{A_{src}}{D_L^2}  \;\;\;\; \rm{erg}\, \rm{s}^{-1}\, \rm{cm}^{-2}\, \rm{Hz}^{-1} \; .
\label{eq:fluxred6}
\end{equation}
Converting $\overline{\Sigma_d} A_{src}$ to dust mass and solving for the dust mass we find that
\begin{equation}
   \lim_{T_d(\vec{r}) = T_d}  \;  \lim_{\tau_{\nu_e} \ll 1}  \; \lim_{\kappa_{\nu}(\vec{r}) = \kappa_{\nu}} \;   M_d = \frac{1}{\mu_{\rm{lens}} (1 + z_{\rm{red}})} \frac{\Delta F_{\nu,obs} D_L^2}{[B_{\nu_e}(T_{d_e}) -  B_{\nu_e}(T_{\rm{CMB}_e})] \kappa_{\nu_e}} \;\;\;\; \rm{g} \; \rm{of} \; \rm{dust} \;\;.
\label{eq:Md_thin_red_cmb}
\end{equation}
This can also be written as 
\begin{equation}
   \lim_{T_d(\vec{r}) = T_d}  \;  \lim_{\tau_{\nu_e} \ll 1}  \; \lim_{\kappa_{\nu}(\vec{r}) = \kappa_{\nu}} \;  M_d = \frac{1}{\mu_{\rm{lens}} (1 + z_{\rm{red}}) f_{\rm{CMB}}} \frac{\Delta F_{\nu,obs} D_L^2}{B_{\nu_e}(T_{d_e}) \kappa_{\nu_e}} \;\;\;\; \rm{g} \; \rm{of} \; \rm{dust} \;\;.
\label{eq:Md_thin_red_cmb2}
\end{equation}
where 
\begin{equation}
   f_{\rm{CMB}} = 1 - \frac{B_{\nu_e}(T_{\rm{CMB}_e})}{B_{\nu_e}(T_{d_e})}  \;\; .
\end{equation}
Equation \ref{eq:Md_thin_red_cmb2} is only valid for emission if $f_{\rm{CMB}} > 0$ which simply means that the equilibrium dust temperature must be larger than the temperature of the CMB at that redshift for the source to be detected in emission.
Comparison with Equation \ref{eq:Md_thin_final} finds that the distance in that equation has been replaced with the luminosity distance and that the expression is divided by the gravitational lensing magnification and one factor of $(1 + z_{\rm{red}})$. 
If the CMB background is negligible, then $f_{\rm{CMB}} \rightarrow 1$ and $\Delta F_{\nu,obs} \rightarrow F_{\nu,obs}$.
In this limit, the mass equation simplifies to 
\begin{equation}
    \lim_{I_{\nu,\rm{bg}} \rightarrow 0} \; \lim_{T_d(\vec{r}) = T_d}  \;  \lim_{\tau_{\nu_e} \ll 1}  \; \lim_{\kappa_{\nu}(\vec{r}) = \kappa_{\nu}} \;  M_d = \frac{1}{\mu_{\rm{lens}} (1 + z_{\rm{red}})} \frac{F_{\nu,obs} D_L^2}{B_{\nu_e}(T_{d_e}) \kappa_{\nu_e}} \;\;\;\; \rm{g} \; \rm{of} \; \rm{dust} \;\;.
\label{eq:Md_thin_red}
\end{equation}

\subsection{Caveats About Real Observations} \label{Caveats}

\subsubsection{Telescope Power Patterns}
All real telescopes have a diffraction pattern that represents the angular sensitivity of the telescope to light coming from different angles away from the direction that it is pointing.
The resulting power pattern, $P_n(\theta,\phi)$, is normalized such that $P_n(\theta = 0) = 1$ in the direction that the telescope is pointing.

The power pattern depends on the optical illumination of the telescope.
For an unblocked circular telescope aperture with uniform illumination, the power pattern is given by the Airy pattern (in the small angle approximation) as 
\begin{equation}
 P_n(\theta) = 4 \frac{ J_1^2\left( \frac{\pi D_t \theta}{\lambda} \right) }{ \left( \frac{\pi D_t \theta }{\lambda} \right)^2}  \;\;\;\;, 
\label{eq:AiryPn}
\end{equation}
where $J_1()$ is a first-order Bessel function of the first kind, $\lambda = c/\nu$ (cm) is the wavelength of the observation, and $D_t$ (cm) is the physical diameter of the telescope aperture \citep{toolsradioastro, HechtOptics}.
If we let $u = \pi D_t \theta / \lambda$, then the Half Width Half Maximum (HWHM) of the Airy Pattern is found when $P_n(u_{1/2}) = 1/2$ which occurs at $u_{1/2} \approx 1.616$.  The Full Width Half-Maxiumum (FWHM) of the main beam of an Airy pattern, $\theta_{mb}$ (rad), is given by 
\begin{equation}
  \theta_{mb} = \frac{2 u_{1/2} \lambda}{ \pi D_t} \approx 1.029 \frac{\lambda}{D_t}  \;\;\;\; \rm{rad} \;. 
\label{eq:AiryMB}
\end{equation}
Radio telescopes are not uniformly illuminated, but instead are illuminated with an edge taper, $T_e$ (dB).
For an unblocked telescope aperture, the edge taper is defined from the ratio of the power at the edge of the telescope aperture, $P_{edge}$, to the power at the center of the telescope, $P_{center}$, as
\begin{equation}
    T_e = 10 \log_{10} \left( \frac{P_{edge}}{P_{center}} \right) \;\;\;\; \rm{dB} \; ,
\end{equation}
(see \cite{2016teas.book.....W}).
The resulting radio telescope power patterns are often well described by Gaussian functions in the angle $\theta$ and the Full Width Half-Maxiumum (FWHM) $\theta_{mb}$
\begin{equation}
    P_n(\theta) = \exp \left[ -4\ln(2) \frac{\theta^2}{\theta_{mb}^2}
 \right] \;\;\;\; . 
\label{eq:GaussPn}
\end{equation}
The main beam FWHM is then given by the approximate equation
\begin{equation}
    \theta_{mb} \approx (1.02 + 0.0135 T_e) \frac{\lambda}{D_t} \;\;\;\; \rm{rad} \; ,
\end{equation}
\citep{goldsmith1998quasi}.
In the limit of uniform illumination ($\lim T_e \rightarrow 0$ dB), this equation approximately reduces (to within $1$\%) to the FWHM of an Airy pattern (Equation \ref{eq:AiryMB}).

In the case of a single pointing observation toward the center of the source (e.g., the telescope pointing toward line-of-sight C in Figures \ref{fig:Geo1} and \ref{fig:Geo2}), the flux density equation (Equation~\ref{eq:Fmono}) is modified to calculate the observed flux density intercepted by the telescope power pattern as
\begin{equation}
    F_{\nu, obs} = \int_{\Omega} I_{\nu}(\theta,\phi) P_n(\theta,\phi) \cos{\theta} d \Omega \;\;\;\; \rm{erg}\, \rm{s}^{-1}\, \rm{cm}^{-2}\, \rm{Hz}^{-1} \;\; .
\end{equation}
If both the source intensity distribution on the sky and the telescope power pattern are small enough that we can use the small angle approximation for $\theta$ (transforming the spherical polar coordinate system to a planar polar coordinate system), then we can re-write this equation as 
\begin{equation}
    F_{\nu, obs} = \int_{\phi = 0}^{2\pi} \int_{\theta = 0}^{\theta_{ap}} I_{\nu}(\theta,\phi) P_n(\theta,\phi) \theta d \theta d \phi \;\;\;\; \rm{erg}\, \rm{s}^{-1}\, \rm{cm}^{-2}\, \rm{Hz}^{-1} \;\; ,
\label{eq:Fmonosapn}
\end{equation}
where $\theta_{ap}$ is the maximal angular extent of the aperture or the source.
During mapping observations, the telescope will scan across different coordinates on the sky and the observed flux density at each position, $F_{\nu, obs}(\theta_0,\phi_0)$, is given by the convolution of the source intensity distribution and the telescope beam pattern (see Chapter 3 of  \citealt{Kraus1966} for a detailed discussion).\footnote{The spherical coordinate system we have defined, with a pole toward the center of the object, presents mathematical challenges to calculating the convolution.
\citealt{Kraus1966} introduces a new coordinate system with a $\pi/2$ phase shift in the $\theta$ coordinate to avoid this issue.
In some problems, it may be easier to perform this convolution in a Cartesian coordinate system centered at $\theta = 0$ (see Figures \ref{fig:Geo1} and \ref{fig:Geo2}) given by $x_{\alpha} = x/D$ (rad) and $y_{\delta} = y/D$ (rad) where $D$ here is the distance to the object. This new planar coordinate system is connected to our original spherical coordinate system by the angular Cartesian variables $x_{\alpha} = \theta \sin(\phi)$, $y_{\delta} = \theta \cos(\phi)$.
The differential solid angle is $d\Omega = dx dy / D^2 = dx_{\alpha} dy_{\alpha}$ (see Figure \ref{fig:Geo1}) and the flux density at position $(x_{\alpha}^{\prime},y_{\delta}^{\prime})$ is given by $F_{\nu,obs}(x_{\alpha}^{\prime},y_{\delta_0}^{\prime}) = \int_{x_{\alpha}} \int_{y_{\delta}} I_{\nu}(x_{\alpha},y_{\delta}) P_n(x_{\alpha} - x_{\alpha}^{\prime}, y_{\delta} - y_{\delta}^{\prime}) dx_{\alpha} dy_{\delta}$.}

Dust continuum images are typically gridded into pixels with the units of intensity such as MJy/ster, Jy/arcsec$^2$, Jy/beam, or Jy/pixel.  
This is the average intensity within the pixel calculated from $\overline{I_{\nu}} = F_{\nu,obs} / \Omega_{ap}$.
The choice of solid angle defines which intensity units are used.
Standard cgs units are erg s$^{-1}$ cm$^{-2}$ ster$^{-1}$ Hz$^{-1}$.
These are often converted into MJy/ster using the conversion factor that $1$ MJy/ster = $10^{-17}$ erg s$^{-1}$ cm$^{-2}$ ster$^{-1}$ Hz$^{-1}$.
If $\overline{I_{\nu}}$ is in units of MJy/ster, then it can also be converted to Jy/beam or Jy/pixel by multiplying by $10^6 \Omega_{beam}$ or $10^6 \Omega_{pixel}$ where $\Omega_{beam}$ is the solid angle of the telescope beam (ster) and $\Omega_{pixel}$ is the solid angle of a pixel (ster).
The average intensity can also be converted between Jy/beam and Jy/pixel by multiplying by the ratio of solid angles $\Omega_{beam}/\Omega_{pixel}$.
The solid angle of the telescope beam is determined from integrating the telescope power pattern over solid angle (assuming the small angle approximation, planar polar coordinates)
\begin{equation}
    \Omega_{beam} = \int_{\phi = 0}^{2\pi} \int_{\theta = 0}^{\theta_{ap}} P_n(\theta,\phi) \theta d \theta d \phi \;\;\;\; \rm{ster} \; .
    \label{eq:omegaairy}
\end{equation}
For mathematical convenience, $\theta_{ap}$ is often set to infinity.
While this would appear to violate the small angle approximation, $P_n$ typically decreases toward zero quickly enough with $\theta$ such that this approximation is cromulent.
The solid angle of an Airy Pattern (Equation \ref{eq:AiryPn}) is\footnote{If we let $u = \pi D_t \theta / \lambda$, then $\theta = (\lambda/\pi D_t) u$ and $d\theta = (\lambda/\pi D_t) du$. The Airy pattern solid angle is $\Omega_{beam} = 2 \pi (\lambda/\pi D_t)^2 \int_0^{\infty} [4 J_1^2(u)/u^2] \, u \,  du$ (ster). The solution to the integral is $\int [4 J_1^2(u)/u] du  = -2[J_0^2(u) + J_1^2(u)] + C$ where C is an integration constant and $J_0()$ is a zeroth-order Bessel function of the first kind. 
At the limits of integration, $J_0(0) = 1$, $J_1(0) = 0$, $\lim_{u \rightarrow \infty} J_0(u) = 0$, and $\lim_{u \rightarrow \infty} J_1(u) = 0$ resulting in $\int_0^{\infty} [4J_1^2(u)/u] \,  du = 2$.
The factor of $(\lambda/\pi D_t)^2$ is eliminated using Equation \ref{eq:AiryMB}.
} 
\begin{equation}
   \Omega_{beam} = \frac{\pi \theta_{mb}^2}{u_{1/2}^2} \;\;\;\; \rm{ster} \;.
   \label{eq:omegagaussian}
\end{equation}
If the beam pattern is a single Gaussian power pattern (Equation \ref{eq:GaussPn}),
then the integrals reduce to\footnote{If we let $u = -4 \ln(2) \theta^2/\theta_{mb}^2$, then $\theta d\theta = [-\theta_{mb}^2/8 \ln(2)] du$.  The Gaussian beam solid angle is $\Omega_{beam} = 2 \pi [\theta_{mb}^2/8 \ln(2)] \int_{-\infty}^0 e^u du$ (ster) where $\int_{-\infty}^0 e^u du = 1$.}
\begin{equation}
    \Omega_{beam} = \frac{\pi \theta_{mb}^2}{4 \ln{2}} \;\;\;\; \rm{ster} \;.
\end{equation}
The solid angle of a square pixel with angular side, $\theta_{pixel}$ (rad), is given by 
\begin{equation}
    \Omega_{pixel} = \theta_{pixel}^2 \;\;\;\; \rm{ster} \;\; .
\end{equation}
Useful conversions are that $206264.8$ arcsec corresponds to $1$ rad, therefore $1$ ster $= 4.254517 \times 10^{10}$ arcsec$^2$ and $1$ arcsec$^2$ = $2.350443 \times 10^{-11} $ ster.

\subsubsection{Filters}

\begin{figure}[h]
\includegraphics[scale=0.8, trim= 0mm 55mm 0mm 90mm, clip]{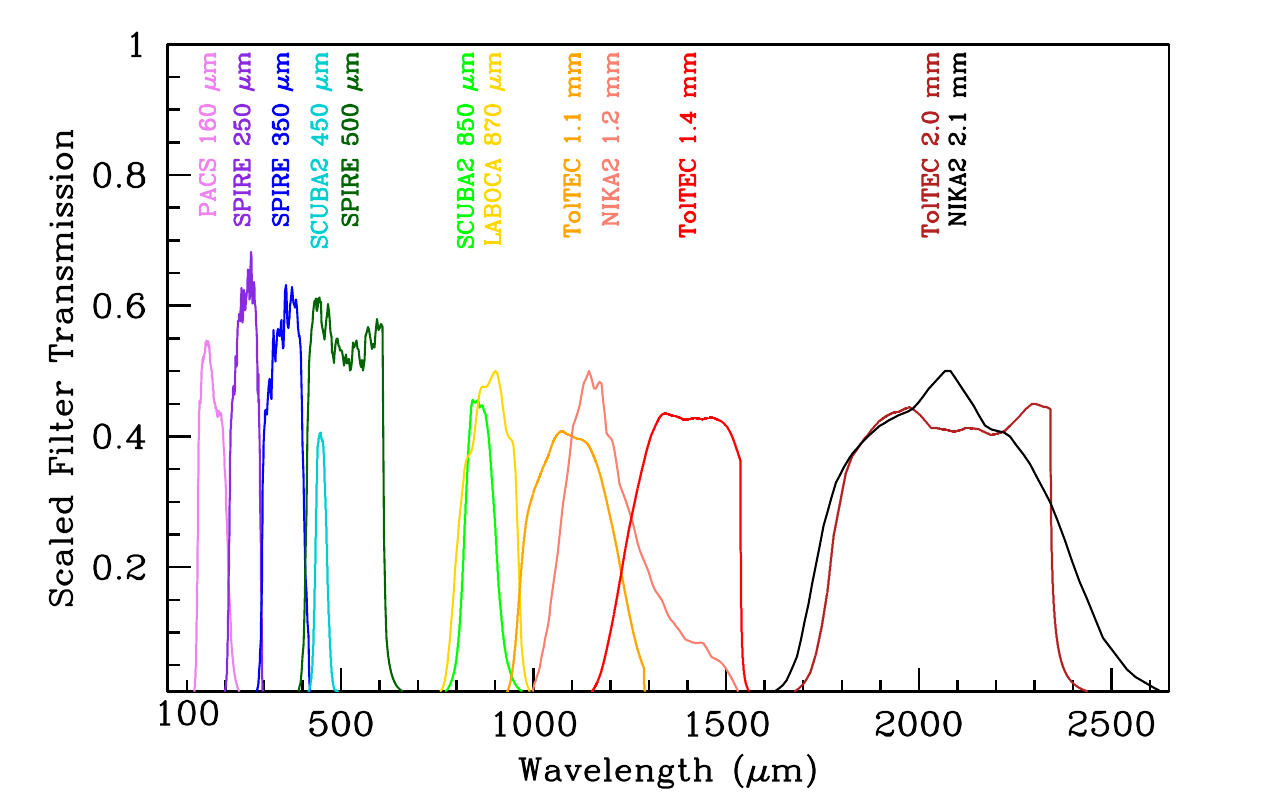}
\caption{Filter transmission curves, $\mathcal{T}(\lambda)$, from bolometer cameras at submillimeter and millimeter wavelengths.  The instruments plotted include PACS and SPIRE from the \textit{Herschel Space Observatory}, SCUBA2 at the JCMT, LABOCA at the APEX telescope, TolTEC at the LMT, and NIKA2 at the IRAM 30m telescope. The plotted curves have been scaled to fit in the plot window and are available from the Spanish Virtual Observatory Filter Profile Service (\citealt{2012ivoa.rept.1015R, 2020sea..confE.182R}; \url{http://svo2.cab.inta-csic.es/theory/fps/}).}
\label{fig:BolometerFilters}
\end{figure}

Real observations are also not monochromatic but are obtained over a specified range of frequencies.
For broadband instruments, such as bolometers, filters are used to constrain the range of frequencies detected.
Each filter has a transmission curve, $\mathcal{T}(\nu)$, that is proportional to the fraction of incident light transmitted at each frequency.
The observed flux densities are the continuous weighted-average of the monochromatic flux densities with the filter transmission curve acting as a weighting function
\begin{equation}
    F_{\overline{\nu},obs} =  \frac{\int_{0}^{\infty} F_{\nu,obs} \mathcal{T}(\nu) d\nu }{\int_{0}^{\infty} \mathcal{T}(\nu) d\nu} \;\;\;\; \rm{erg}\, \rm{s}^{-1}\, \rm{cm}^{-2}\, \rm{Hz}^{-1} \; , 
\label{eq:FilterFlux}
\end{equation}
where $\overline{\nu}$ is the weighted average frequency of the filter
\begin{equation}
   \overline{\nu} =  \frac{\int_{0}^{\infty} F_{\nu,obs} \mathcal{T}(\nu) \nu d\nu }{\int_{0}^{\infty} F_{\nu,obs} \mathcal{T}(\nu) d\nu} \;\;\;\; \rm{Hz} \; . 
\label{eq:avgnu}
\end{equation}
See Figure \ref{fig:BolometerFilters} for examples of submillimeter and millimeter bolometer filter transmission curves.

The weighted average frequency of the filter depends on the shape of the SED of the observed object.  
For example, at 3.3 mm, the MUSTANG2 bolometer on the 100m Green Bank Telescope has an weighted average frequency (wavelength) that ranges from $85.454$ GHz ($3508$ $\mu$m) to $91.879$ GHz ($3263$ $\mu$m) for SEDs, $F_{\nu} \propto \nu^{\alpha_s}$, whose spectral indices range from $\alpha_s = -4.0$ to $+4.0$ respectively\footnote{see   \url{https://gbtdocs.readthedocs.io/en/latest/references/receivers/mustang2/mustang2\_bandpass.html} }.
If the SED is unknown or if a generic value is needed for different SEDs, then $F_{\nu,obs}$ is often set to $1$ in Equation \ref{eq:avgnu}.
This is the assumption made in Table \ref{tab:dustopacity}.

Throughout this tutorial, we have assumed that flux densities are measured in units that are per unit frequency, $F_{\nu}$ (erg s$^{-1}$ cm$^{-2}$ Hz$^{-1}$).
Flux densities may also be measured in units that are per unit wavelength, $F_{\lambda}$ (erg s$^{-1}$ cm$^{-2}$ cm$^{-1}$).
If the flux densities are measured per unit wavelength, then Equations \ref{eq:FilterFlux} and \ref{eq:avgnu} can be re-written as 
\begin{equation}
    F_{\overline{\lambda},obs} =  \frac{\int_{0}^{\infty} F_{\lambda,obs} \mathcal{T}(\lambda) d\lambda }{\int_{0}^{\infty} \mathcal{T}(\lambda) d\lambda} \;\;\;\; \rm{erg}\, \rm{s}^{-1}\, \rm{cm}^{-2}\, \rm{cm}^{-1} \;\;\;\; \rm{erg}\, \rm{s}^{-1}\, \rm{cm}^{-2}\, \rm{Hz}^{-1} \; ,
\label{eq:FilterFluxLam}
\end{equation}
where $\overline{\lambda}$ is the weighted average wavelength of the filter
\begin{equation}
   \overline{\lambda} =  \frac{\int_{0}^{\infty} F_{\lambda,obs} \mathcal{T}(\lambda) \lambda d\lambda }{\int_{0}^{\infty} F_{\lambda,obs} \mathcal{T}(\lambda) d\lambda} \;\;\;\; \rm{cm} \; . 
\label{eq:avglambda}
\end{equation}
The conversion from $\mathcal{T}(\lambda)$ to $\mathcal{T}(\nu)$ is simply to convert each $\lambda$ to $\nu = c/\lambda$ and vice versa.
The conversion between $F_{\overline{\lambda},obs}$ and $F_{\overline{\nu},obs}$ is more complicated, however, because of the non-linear differential relationship\footnote{For our definitions of the mean frequency (Equation \ref{eq:avgnu}) and mean wavelength (Equation \ref{eq:avglambda}), $\overline{\nu} \neq c/\overline{\lambda}$.} between $d\lambda$ and $d\nu$,
$|d\nu| = (c /\lambda^2) |d\lambda|$.
For monochromatic flux densities, $F_{\nu} \neq F_{\lambda}$ and instead we have,
at a particular wavelength or frequency, that $F_{\nu}|d\nu| = F_{\lambda}|d\lambda|$.  This results in the monochromatic transformation equation
\begin{equation}
    F_{\nu} = \frac{\lambda^2}{c} F_{\lambda} \;\;\;\; \rm{erg}\, \rm{s}^{-1}\, \rm{cm}^{-2}\, \rm{Hz}^{-1} \; .
\label{eq:FnuFlam}
\end{equation}
If we define the pivot wavelength (see \citealt{2019A&A...622A.103B}) as
\begin{equation}
    \lambda_{\rm{pivot}} = \sqrt{\frac{\int_{0}^{\infty} \mathcal{T}(\lambda) d\lambda}{\int_{0}^{\infty} \frac{\mathcal{T}(\lambda)}{\lambda^2} d\lambda}} \;\; \rm{cm} \,
\end{equation}
then the flux density per unit frequency and the flux density per unit wavelength observed through a filter with transmission curve $\mathcal{T}$ are related by\footnote{We can prove Equation \ref{eq:FnuFlamPivot} by substituting for the pivot wavelength such that we have $F_{\nu_{\rm{pivot}},obs} = \frac{\int_{0}^{\infty} \mathcal{T}(\lambda) d\lambda}{\int_{0}^{\infty} \mathcal{T}(\lambda) \frac{c}{\lambda^2} d\lambda} \frac{\int_{0}^{\infty} \mathcal{T}(\lambda) F_{\lambda,obs}  d\lambda }{\int_{0}^{\infty} \mathcal{T}(\lambda) d\lambda} = \frac{\int_{0}^{\infty} \mathcal{T}(\nu) F_{\nu,obs}  d\nu }{\int_{0}^{\infty} \mathcal{T}(\nu) d\nu} = F_{\nu_{\rm{pivot}},obs}$ where we have used $|d\nu| = (c /\lambda^2) |d\lambda|$ and $F_{\nu,obs} |d\nu| =  F_{\lambda,obs}|d\lambda|$ to simplify the integrals in the middle steps.}
\begin{equation}
F_{\nu_{\rm{pivot}},obs} = \frac{\lambda_{\rm{pivot}}^2}{c} F_{\lambda_{\rm{pivot}},obs}  \;\;\;\; \rm{erg}\, \rm{s}^{-1}\, \rm{cm}^{-2}\, \rm{Hz}^{-1} \;  
\label{eq:FnuFlamPivot}
\end{equation}
where $\nu_{\rm{pivot}} = c/\lambda_{\rm{pivot}}$.
This definition of the pivot wavelength makes Equation \ref{eq:FnuFlamPivot} have the same form as the monochromatic Equation \ref{eq:FnuFlam}.
The pivot wavelength is used to characterize bolometer filters in Table \ref{tab:dustopacity}.

\section{Dust Opacity}\label{OpacitySection}

Both the absorption coefficient, $\alpha_{\nu}(\vec{r})$, and the emissivity coefficient, $j_{\nu}(\vec{r})$, depend on the dust cross-section, $\sigma_{\nu}$, or the dust mass opacity, $\kappa_{\nu}$.  It is necessary to know how the dust absorption, scattering, and extinction cross-sections and mass opacities vary with frequency or wavelength in different astrophysical environments.
The effective cross-sections for dust grains to absorb or scatter a photon at frequency $\nu$, are written as
\begin{eqnarray}
   \sigma_{\nu, abs} & = &  A_{\rm{cross}} Q_{\nu,abs} \label{eq:grainabscrosssection}\\
   \sigma_{\nu, sca} & = &  A_{\rm{cross}} Q_{\nu,sca} \label{eq:grainscacrosssection}\\
    \sigma_{\nu, ext} & = & \sigma_{\nu, abs} + \sigma_{\nu, sca} =  A_{\rm{cross}} Q_{\nu,ext}  \;\;\;\; \rm{cm^2} \;,
\label{eq:graineffcrosssection}
\end{eqnarray}
where $A_{\rm{cross}}$ is the cross-sectional area of the dust grain (cm$^2$) and $Q_{\nu,abs}$ and $Q_{\nu,sca}$ are unitless dust absorption and scattering coefficients\footnote{$Q_{\nu} = Q_{\lambda}$ for each $\nu = c/\lambda$. Therefore, the notation is interchangable.} that are a numerical correction to the physical grain cross-section to account for the efficiency of absorbing and scattering a photon at frequency $\nu$. 
Since extinction is the combination of absorption and scattering, then we can define $Q_{\nu,ext} = Q_{\nu,abs} + Q_{\nu,sca}$. 
The physical grain cross-section is given by
\begin{equation}
A_{\rm{cross}} = \pi a^2_{\rm{grain}} \;\;\;\; \rm{cm}^2 \;\; ,
\end{equation}
where the grain radius, $a_{\rm{grain}}$, is defined as the radius of a spherical grain with the same grain volume, $V_{\rm{grain}}$,  
\begin{equation}
a_{\rm{grain}} = \left( \frac{3 V_{\rm{grain}}}{4\pi} \right)^{1/3}  \;\;\;\; \rm{cm} \;\;.
\end{equation}
The dust mass opacity of a single dust grain is defined as the effective cross-section of the dust grain divided by the mass of the dust grain, $m_{\rm{grain}}$. 
As a result there are 3 complementary equations for the absorption, scattering, and extinction dust mass opacity
\begin{eqnarray}
\kappa_{\nu,\rm{abs},\rm{grain}} & = & \frac{\sigma_{\nu, abs}}{m_{\rm{grain}}}  =  \frac{A_{\rm{cross}} Q_{\nu,abs}}{m_{\rm{grain}}} \\
\kappa_{\nu,\rm{sca},\rm{grain}} & = & \frac{\sigma_{\nu, sca}}{m_{\rm{grain}}}  =  \frac{A_{\rm{cross}} Q_{\nu,sca}}{m_{\rm{grain}}} \\
\kappa_{\nu,\rm{ext},\rm{grain}} & = & \frac{\sigma_{\nu, ext}}{m_{\rm{grain}}}  =  \frac{A_{\rm{cross}} (Q_{\nu,abs} + Q_{\nu,sca})}{m_{\rm{grain}}} \label{eq:kappagraindefn} \;\;\;\; \rm{cm^2} \; (\rm{g} \; \rm{of} \; \rm{dust})^{-1}\;\;.
\label{eq:kappagrain}
\end{eqnarray}
For the process of thermal dust emission, it is the absorption dust mass opacity, $\kappa_{\nu,\rm{abs},\rm{grain}}$, that is the relevant mass opacity needed to describe how much the grain radiates at each frequency as the grain has the same cross-section for emission as for absorption.

The dust grains in the ISM come in a wide variety of sizes, shapes, and compositions.
How the different grain constituents interact with electromagnetic radiation can be characterized by their respective complex indicies of refraction\footnote{Another measure of the response of materials to electromagnetic radiation is the complex dielectric constant, $\epsilon$, which is related to the complex index of refraction, $m$, by $\epsilon(\nu) = m^2(\nu)$.}.
Generally, the complex index of refraction is measured by laboratory experiments and then used to calculate dust absorption and scattering coefficients\footnote{For example see the Database of Optical Constants for Cosmic Dust,  \url{https://www2.astro.uni-jena.de/Laboratory/OCDB/} 
}.
The subsequent calculation of $Q_{\nu,abs}$ and $Q_{\nu,sca}$ uses techniques such as Mie scattering calculations \citep{1908AnP...330..377M,1983asls.book.....B,ishimaru2017electromagnetic} and the Discreet Dipole Approximation \citep{1973ApJ...186..705P, 2000lsnp.book..131D} to simulate the interaction of an electromagnetic plane wave with a dust grain. 
While there are analytic formulae for $Q_{\nu}$ in the limits of simple geometries and approximations (e.g., spheroidal uniform grains), more general cases, often more appropriate for ISM dust grains, are calculated numerically.
Grains may include mixtures of materials with different indicies of refraction for which techniques like effective medium theory \citep{1991A&A...251..210O,choy1999effective} with approximations such as the Garnett mixing rule or the  Bruggeman mixing rule\footnote{For a mixture of $i$ different materials, the effective index of refraction,  $m_{eff}$, for the Bruggeman mixing rule is found by solving the non-linear equation $\sum_i V_i \frac{(m_i/m_{eff})^2 - 1}{(m_i/m_{eff})^2 + 2} = 0$, where $V_i$ is the volume filling fraction of each material \citep{2015EPJWC.10200005M}.  The corresponding equation for the Garnett mixing rule where the mixture is assumed to have a dominate component with index of refraction $m_j$, is $\sum_i V_i \frac{(m_i/m_{j})^2 - 1}{(m_i/m_{j})^2 + 2} = \frac{(m_{eff}/m_j)^2 - 1}{(m_{eff}/m_j)^2 + 2}$ \citep{2015EPJWC.10200005M}.  See \cite{2018ApJ...869L..45B} for formulae using the complex dielectric function, $\epsilon(\nu)$.} \citep{1904RSPTA.203..385M,1935AnP...416..636B,2015EPJWC.10200005M,2018ApJ...869L..45B} are required to calculate an effective index of refraction. 
Also, the grains may have varying degrees of porosity where a fraction of the total grain volume is vacuum (i.e., \citealt{1989ApJ...341..808M, 2008ApJ...689..260S,2014A&A...568A..42K,2025A&A...698A.200C}).
A complete description of the calculation of $Q_{\nu,abs}$ and $Q_{\nu,sca}$ and the measured optical properties for all of the different materials and all of the different geometries that may be found in the ISM are beyond the scope of this tutorial (but see Chapter 22 of \citealt{Draine2011} for an introduction to this topic and see Section 3 of \citealt{2015EPJWC.10200005M} for a summary of computation techniques). 
Instead, we shall first focus on the basic properties of $Q_{\nu,abs}$ and $Q_{\nu,sca}$ and then on published calculations of the dust mass opacity for populations of grain sizes and materials that are observed in different astrophysical environments.

\begin{figure}[h]
\includegraphics[scale=0.43, trim= 0mm 50mm 0mm 40mm, clip]{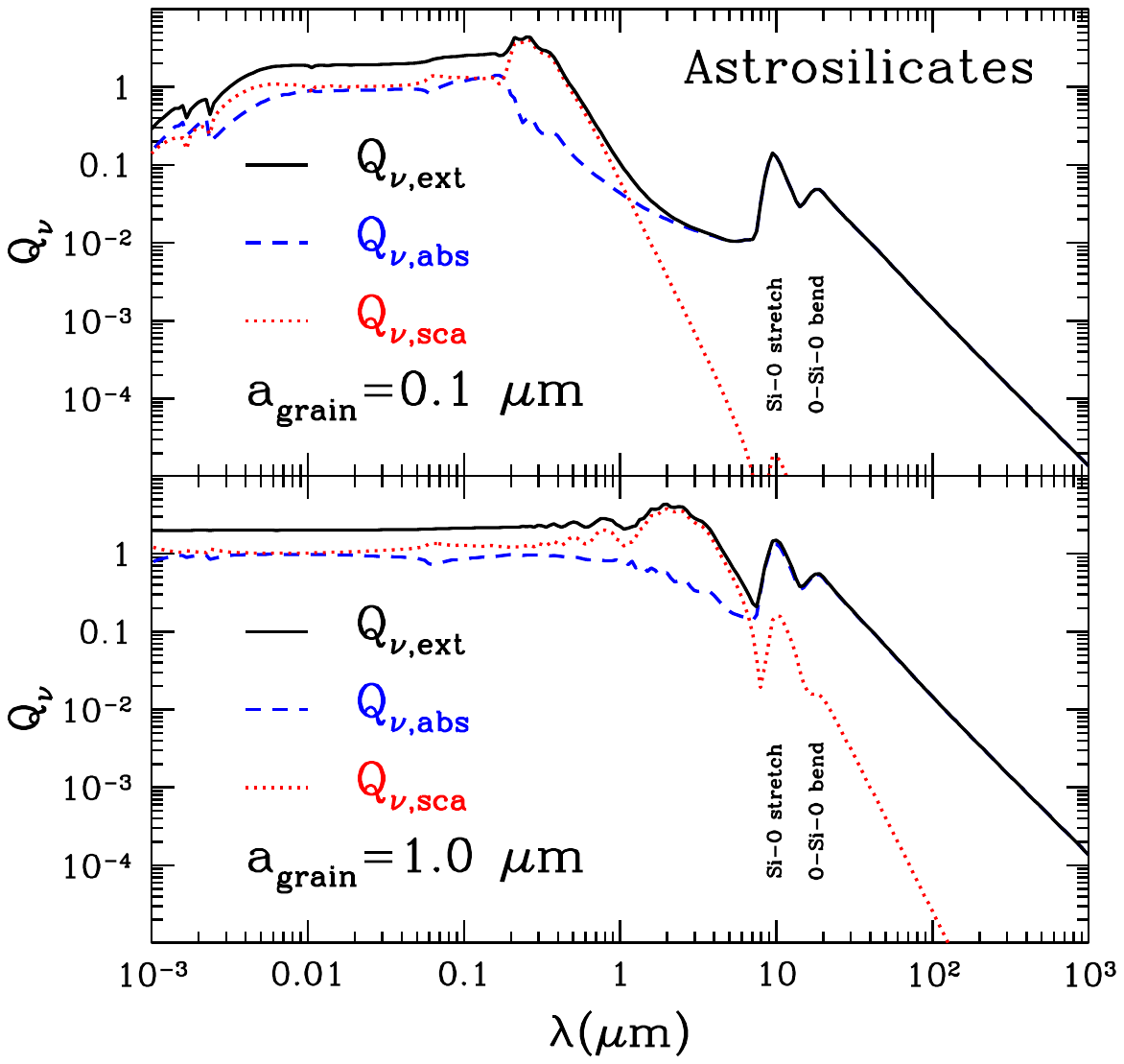}
\includegraphics[scale=0.43, trim= 0mm 50mm 0mm 40mm, clip]{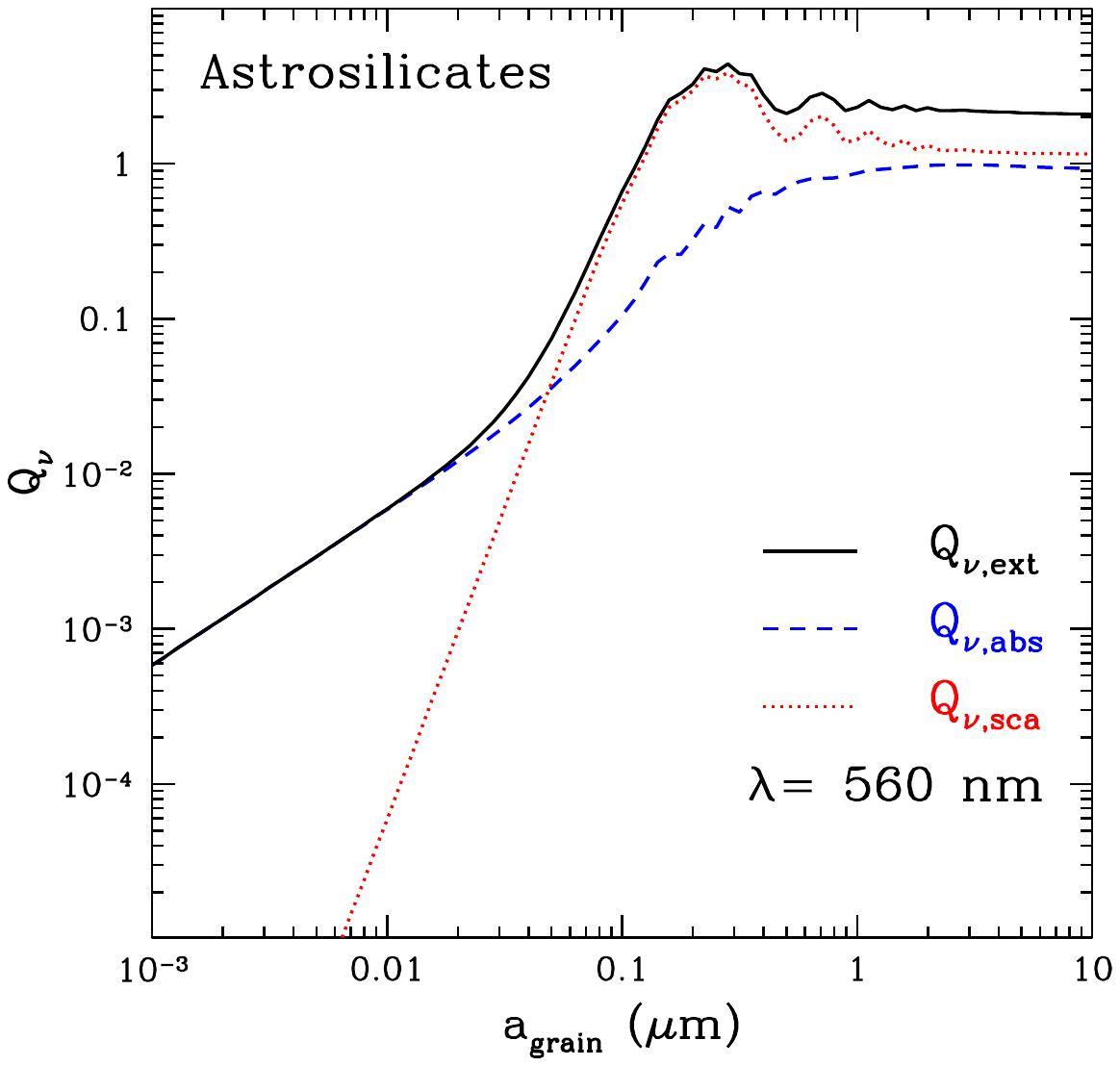}
\caption{Left: $Q_{\nu,ext} = Q_{\nu,abs} + Q_{\nu,sca}$ is plotted from the calculations of \cite{1984ApJ...285...89D} and \cite{1993ApJ...402..441L} for spherical astrosilicate dust grains with radii of (top) $0.1$ $\mu$m and (bottom) $1.0$ $\mu$m (data available from \url{https://www.astro.princeton.edu/~draine/dust/dust.diel.html}).  Right: $Q_{\nu,ext} = Q_{\nu,abs} + Q_{\nu,sca}$ is plotted for different grain sizes at a wavelength of 560 nm (V band).}
\label{fig:QFig}
\end{figure}

Figure~\ref{fig:QFig} shows the astrosilicate grain calculation of \cite{1984ApJ...285...89D} and \cite{1993ApJ...402..441L} to illustrate the basic absorption and scattering properties of ISM dust grains.
Generally, $Q_{\nu,ext}$ tends to be largest within a factor of a few of $\lambda = c/\nu \sim a_{\rm{grain}}$ that corresponds to the effective dust grain radius (see Figure \ref{fig:QFig}).
For grain sizes much larger than the wavelength ($a_{\rm{grain}} \gg \lambda$), both $Q_{\nu,abs}$, and $Q_{\nu,sca}$ tend toward $1$ meaning the grain absorbs light and scatters light with its geometrical cross-section (see upper right panel of Figure \ref{fig:QFig}).
For grain sizes much smaller than the wavelength ($a_{\rm{grain}} \ll \lambda$), both $Q_{\nu,abs}$  and $Q_{\nu,sca}$ tend to decrease with a power-law dependence.
Scattering is typically not important in this limit implying that $Q_{\nu,ext} \approx Q_{\nu,abs}$.

Several spectral features may be  seen in the wavelength dependence of $Q_{\nu,abs}$ and $Q_{\nu,sca}$ for astrosilicates.
At very short wavelengths ($\lambda < 0.005$ $\mu$m), the absorption edges from deeply bound electrons (e.g., those in the K shell) in common grain constituents such as O, Si, and Fe can be seen.
The prominent spectral features in the mid-infrared are the $9.7$ $\mu$m and $18$ $\mu$m siliciate features due to the fundamental vibration frequencies of Si-O stretching bonds and O-Si-O bending bonds,  respectively.
Figure \ref{fig:QFig} only shows the calculations for astrosilicates, but carbonaceous grains are another common constituent of dust grains and have different spectral features (e.g., the C-H stretching mode at 3.4 $\mu$m and PAH mid-infrared features).

\begin{figure}[h]
\includegraphics[scale=0.37,trim= 0mm 50mm 0mm 25mm, clip]{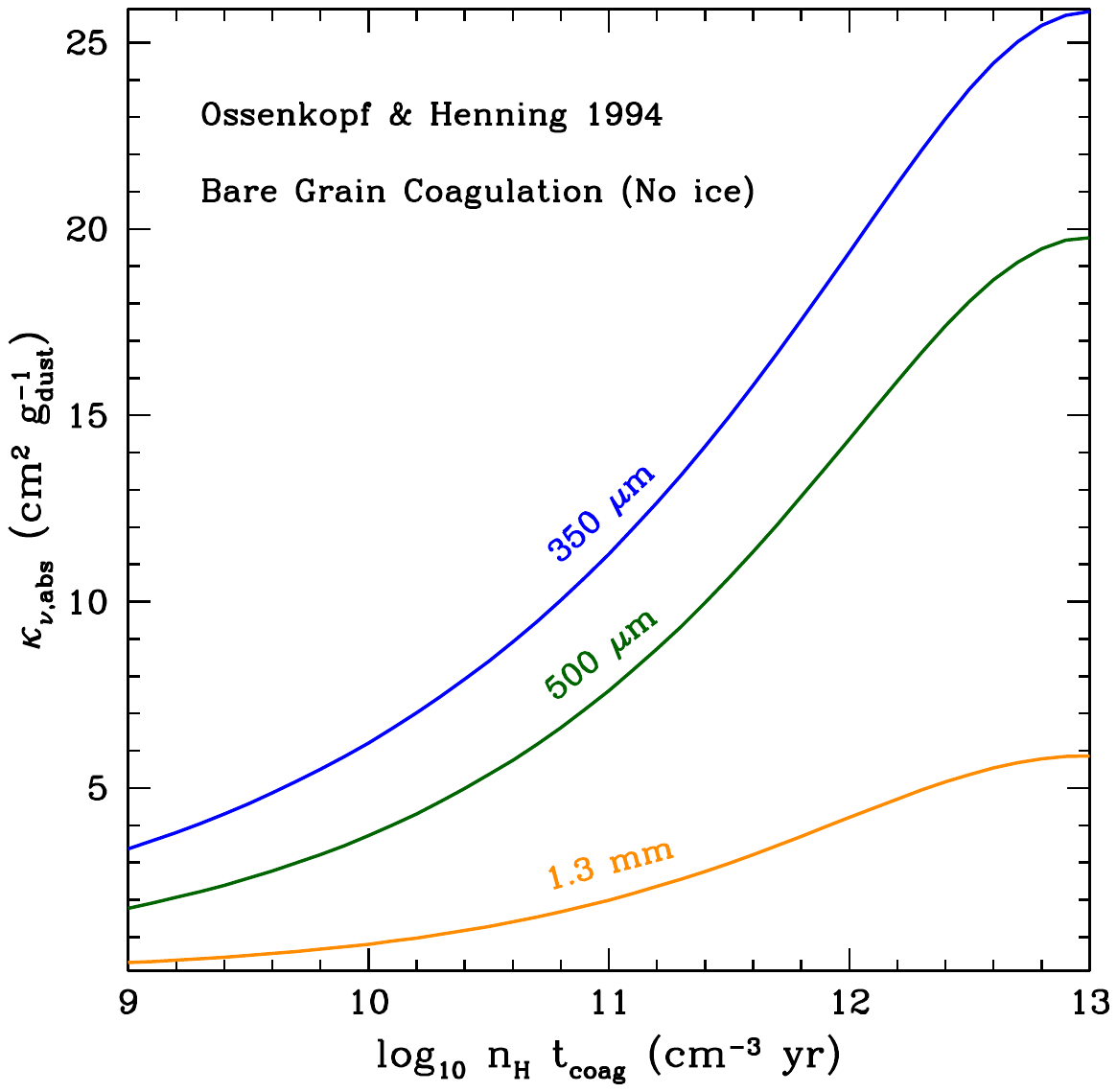}
\includegraphics[scale=0.47,trim= 25mm 35mm 25mm 30mm, clip]{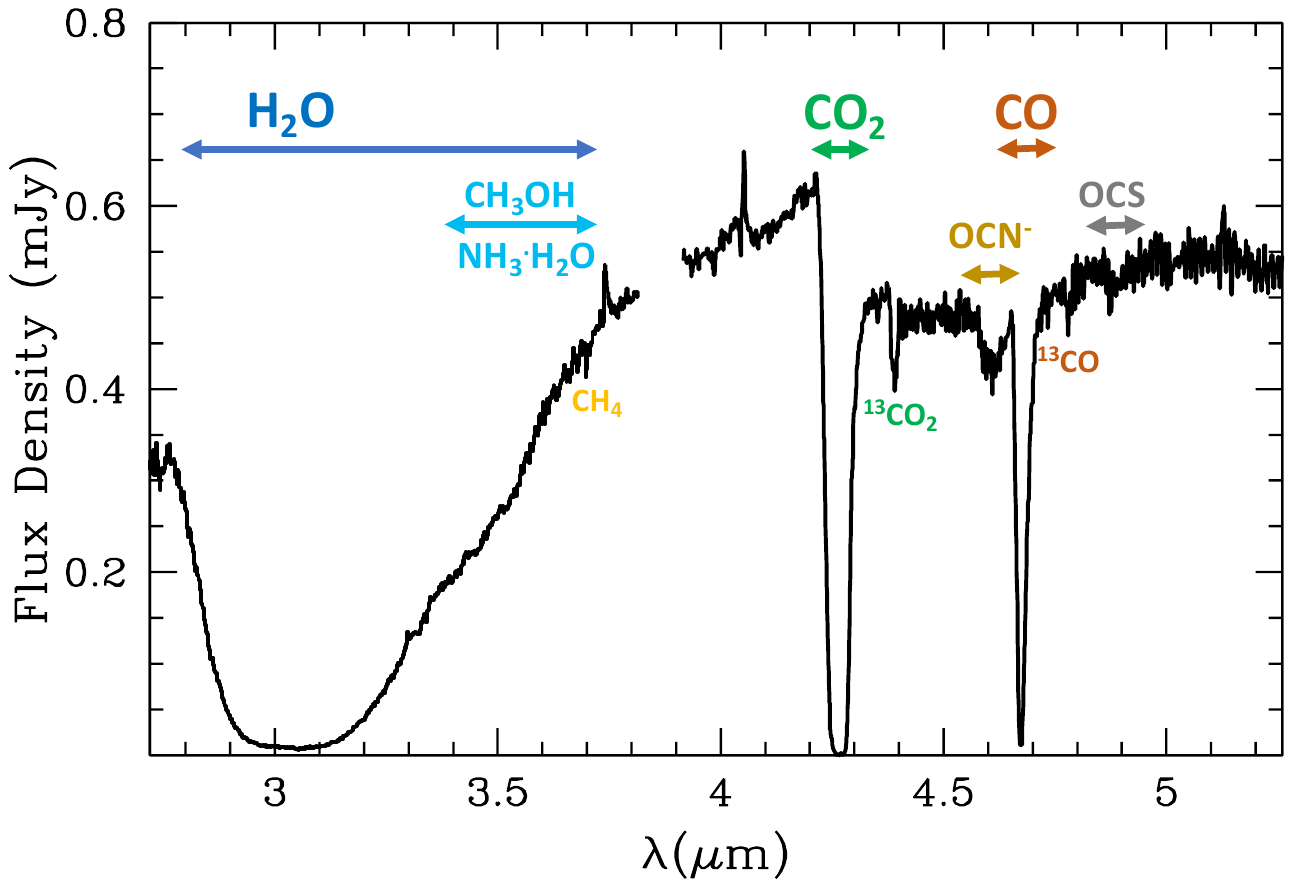}
\caption{Left: The absorption mass opacity from  \cite{1994A&A...291..943O} is plotted as a function of the product of the H atom gas density times the dust grain coagulation timescale. The plotted curves at 350 $\mu$m (blue), 500 $\mu$m (green), and 1.3 mm (orange) are interpolated using Equation 6 of \cite{1994A&A...291..943O}.  Right: Near-infrared spectrum of ice on dust grains toward a background star, NIR38 ($A_{\rm{V}} = 60$ mag), 
in the Chamaeleon I molecular cloud obtained with \textit{JWST} over $2.72 - 5.26$ $\mu$m (\citealt{2023NatAs...7..431M}; data available at \url{https://zenodo.org/records/7501239}).  Strong absorption of background starlight is seen due to H$_2$O, CO$_2$, and CO ices.}
\label{fig:IceFig}
\end{figure}

For each astrophysical environment, there is a distribution of dust grain sizes and masses that must be averaged over to calculate the total $\kappa_{\nu}$ from the individual  $\kappa_{\nu,\rm{grain}}$.
For a grain size distribution, $\mathcal{N}_{d}(a)$ (the number of dust grains of radius a), with mass distribution, $m_{\rm{grain}}(a)$ (g of dust grain with radius a), then the total mass opacity is given by
\begin{equation}
\kappa_{\nu} = \frac{\int_{a_{min}}^{a_{max}} \mathcal{N}_{d}(a) m_{grain}(a) \kappa_{\nu,grain}(a) da}{\int_{a_{min}}^{a_{max}} \mathcal{N}_{d}(a) m_{grain}(a) da} \;\;\;\;  \rm{cm^2} \; (\rm{g} \; \rm{of} \; \rm{dust})^{-1} \;\;,
\label{eq:intsizedistrbution}
\end{equation}
(e.g., see \citealt{2018ApJ...869L..45B}).\footnote{There is an equivalent equation for just the absorption dust opacity, $\kappa_{\nu,abs}$ , with $\kappa_{\nu,grain}$ replaced with $\kappa_{\nu,abs,grain}$ inside the integral. Note that different versions of this equation exist in the literature.  The size distribution is sometimes written as dn/da.  For example, \cite{2024RNAAS...8...68P} write Equation \ref{eq:intsizedistrbution} as $\kappa_{\nu} = \frac{\int_{a_{min}}^{a_{max}} \kappa_{\nu,grain}(a) (dn/da) \pi a^4 \rho_d(a) d(\ln a)}{{\int_{a_{min}}^{a_{max}} (dn/da) \pi a^4 \rho_d(a) d(\ln a)}}$ where they are integrating over a logarithmic grid of grain sizes, $d(\ln a) = (da)/a$.  A logarithmic grid may be preferred since the grain size population may span orders of magnitudes.}
It is $\kappa_{\nu}$ or $\kappa_{\nu,abs}$, averaged by Equation \ref{eq:intsizedistrbution}, that is typically tabulated in the literature.
The size distribution of dust grains in the diffuse ISM of the Milky Way is known as the MRN distribution (named for Mathis, Rumpl, and Nordsieck) where $\mathcal{N}_{d}(a) \propto a^{-3.5}$ for $0.1 \leq a \leq 0.25$  $\mu$m \citep{1977ApJ...217..425M}.
In dense environments such as cores in molecular clouds or in protoplanetary disks, coagulation of dust grains result in growth of the maximum grain sizes and variations in the grain size distribution (e.g., \citealt{1993A&A...280..617O, 2001ApJ...553..321D,2009A&A...502..845O, 2015A&A...579A..15K, 2018ApJ...869L..45B}).
This growth ultimately affects both the absolute value of the dust opacity at each frequency and the shape of the opacity curve (\ie\ the far-infrared-millimeter power-law exponent which we discuss below).
The left panel of Figure \ref{fig:IceFig} shows an example of how  $\kappa_{\nu,abs}$ varies from $350$ $\mu$m to 1.3 mm for the coagulation calculation of \cite{1994A&A...291..943O}.
The opacity is plotted as a function of the product of the H atom gas density times the dust grain coagulation timescale, $t_{\rm{coag}}$, where the grain size distribution starts with an MRN distribution prior to coagulation.  
In the \cite{1994A&A...291..943O} simulations, the opacity is approximately constant for $n_{\rm{H}} t_{\rm{coag}} \leq 10^{9}$ cm$^{-3}$ yr (no coagulation) and $n_{\rm{H}} t_{\rm{coag}} \geq 10^{13}$ cm$^{-3}$ yr.

The dust also accretes and forms icy mantles of predominantly H$_2$O, CO, and CO$_2$ ice (see Figure \ref{fig:IceFig}). 
H$_2$O ice forms from the hydrogenation of O atoms that stick to dust grain surfaces and is observed along lines-of-sight above a threshold of $A_{\rm{V}} \gtrsim 3$ mag that varies in different environments (e.g., \citealt{2001ApJ...547..872W, 2011ApJ...731....9C}). 
Other volatile molecules, such as CO, freeze out of the gas phase at higher densities and cold temperatures \citep[e.g., CO at $n \geq 10^4$ cm$^{-3}$ and $T_K \lesssim 20$ K;][]{2002A&A...389L...6B}.
Reactions with O atoms form CO$_2$ ice \citep{2011ApJ...735...15G}. 
The sucessive hydrogenation of CO drives a rich surface chemistry  \citep[e.g., CO + H $\rightarrow$ HCO + H $\rightarrow$ H$_2$CO + H $\rightarrow$ CH$_3$O + H $\rightarrow$ CH$_3$OH;][]{2002ApJ...571L.173W}, along with energetic processing (\ie\ from cosmic rays, uv radiation,  etc.) to form more complex molecules \citep{2022ApJ...941L..13Y, 2024A&A...683A.124R}. 
The opacity from these ices affect $\kappa_{\nu,abs}$ and must be accounted for in dense environments.

Some dust opacity models are comprised of multiple dust grain constituents and populations that are individually averaged over their size and mass distributions by Equation \ref{eq:intsizedistrbution}.
For example, the \cite{2024A&A...684A..34Y} THEMIS 2.0 diffuse ISM dust opacity models are composed of three distinct populations of grains with different compositions (see Appendix \ref{AppendixOpacity} for more details).
For each grain population, $i$, the gas mass to dust mass ratio, $(M_{\rm{H}}/M_{\rm{dust},i})$, is reported. 
The combined mass opacity for all grain populations is then the average of $\kappa_{\nu, i}$ weighted by the dust mass to gas mass fraction of each grain population,
\begin{equation}
    \kappa_{\nu} = \frac{\sum_{i} \left( \frac{M_{\rm{dust},i}}{M_{\rm{H}}} \right)  \kappa_{\nu, i}} {\sum_{i} \frac{M_{\rm{dust},i}}{M_{\rm{H}}}} \;\;\;\;  \rm{cm^2} \; (\rm{g} \; \rm{of} \; \rm{dust})^{-1} \;\;.
\label{eq:massfracweightedkappa}
\end{equation}

\begin{figure}[h]
\includegraphics[scale=0.9,trim= 3mm 50mm 0mm 40mm, clip]{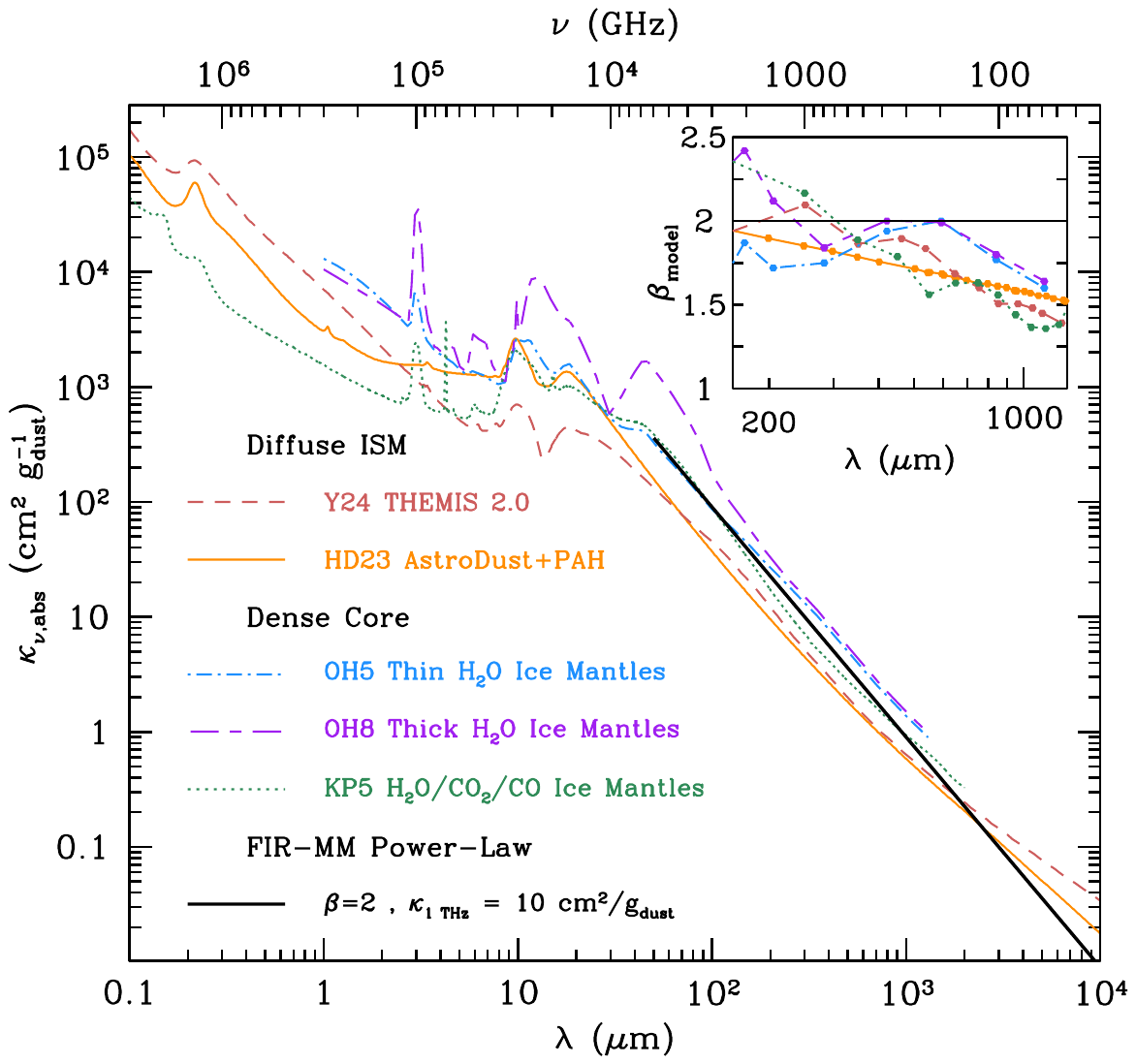}
\caption{Example dust absorption mass opacity for models of the dust in different environments within the Milky Way.  
The inset panel shows $\beta_{model}$, averaged over $\sim 100$ $\mu$m, plotted between $160$ $\mu$m and 1.3 mm for each model. 
The diffuse ISM models include: 
Y24 THEMIS 2.0 (\citealt{2024A&A...684A..34Y}, red dashed line) and HD23 AstroDust+PAH from (\citealt{2023ApJ...948...55H}; orange, solid line)   
that match extinction, emission, and polarization constraints from the diffuse ISM. 
The dense core models include: OH5 and OH8 opacities from \cite{1994A&A...291..943O} for grains that have coagulated for $10^5$ years at a gas density of $10^6$ cm$^{-3}$ with thin water ice mantles (OH5; blue, dot-dashed line) or with thick water ice mantles (OH8; purple, long-short dashed line),  and KP5 opacities from \cite{2024RNAAS...8...68P} that are a good fit to mid-infrared and submillimeter observations of protostellar cores (green, dotted line).
The OH5, OH8, and KP5 models were not extrapolated beyond their published wavelength ranges.
A single power-law with $\beta = 2$ is shown from $50$ $\mu$m to 1 cm (black, thick solid line) that has a value of $10$ cm$^2$/g of dust at 1 THz ($299.8$ $\mu$m). 
Detailed descriptions of each dust opacity model may be found in Appendix \ref{AppendixOpacity}.}
\label{fig:Opacity1}
\end{figure}

The $\kappa_{\nu,abs}$ for selected representative models of dust grain populations are shown in Figure \ref{fig:Opacity1} for the diffuse ISM of the Milky Way (we adopt the acronyms HD23 for \citealt{2023ApJ...948...55H} and Y24 for \citealt{2024A&A...684A..34Y}) and in dense starless, prestellar, and Class 0 and I protostellar cores (we adopt the acronyms OHx, where x correspond to columns in Table 2 of \citealt{1994A&A...291..943O}, and KP5 for \citealt{2024RNAAS...8...68P}).
In general, $\kappa_{\nu,abs}$ decreases by more than seven orders of magnitude from ultraviolet to centimeter wavelengths. 
The variation between models typically span one order of magnitude or more at wavelengths between $1$ $\mu$m and 1.3 mm.
The $\kappa_{\nu,abs}$ is larger at all wavelengths $\gtrsim 30$ $\mu$m for the dense core coagulated grain models with ice mantles compared to the diffuse ISM dust opacities.
Quantitative values for each dust model at commonly observed wavelengths are given in Table \ref{tab:dustopacity} in Appendix \ref{AppendixOpacity}.
More detailed descriptions of each dust opacity model are also given in Appendix \ref{AppendixOpacity}.

\begin{figure}[h]
\includegraphics[scale=0.57]{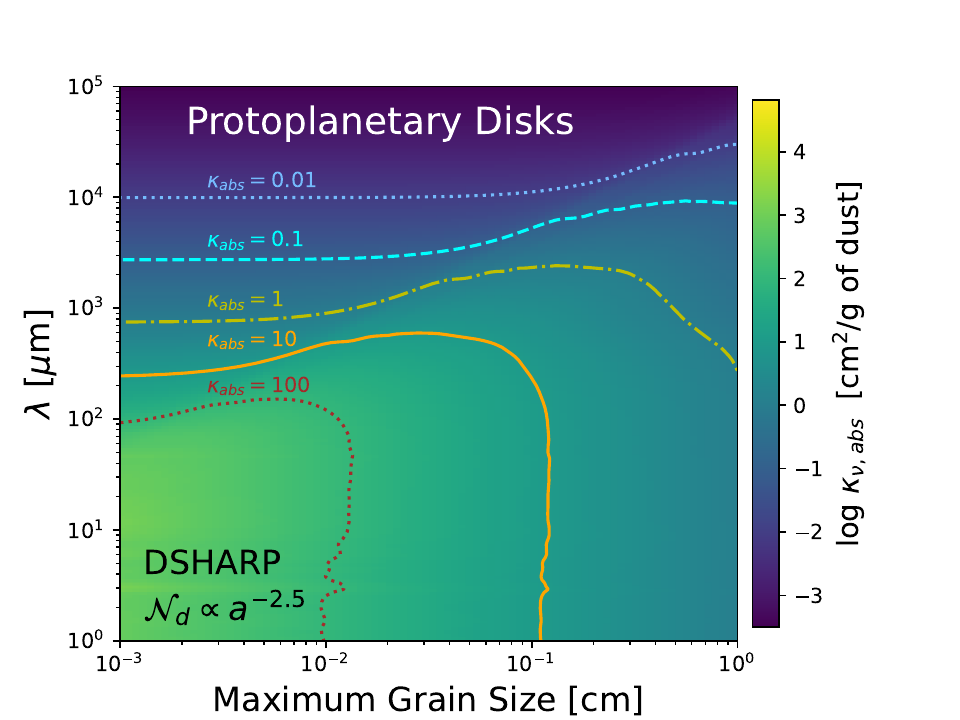}
\includegraphics[scale=0.57]{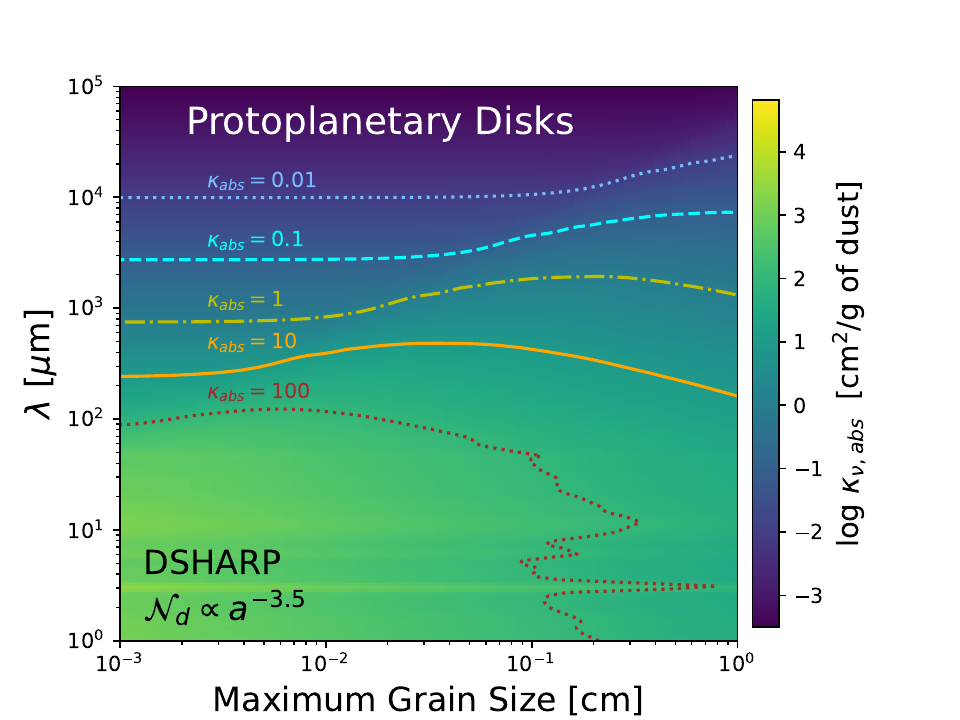}
\vspace{0cm}
\caption{Plots of $\kappa_{\nu,abs}$ for the DSHARP dust mass opacities \citep{2018ApJ...869L..45B} appropriate for protoplanetary disks over wavelengths from $1$ $\mu$m to 10 cm.  The opacities are integrated over the size distribution up to a maximum grain size shown on the x-axis.  Contours are plotted for $\kappa_{\nu,abs} = 0.01, 0.1, 1, 10, 100$ cm$^2$/gram of dust.  The left panel shows an $\mathcal{N}_d(a) \propto a^{-2.5}$ size distribution while the right panel shows an $\mathcal{N}_d(a) \propto a^{-3.5}$ size distribution.}
\label{fig:OpacityDSHARP}
\end{figure}

\begin{figure}[h]
\includegraphics[scale=1.0]{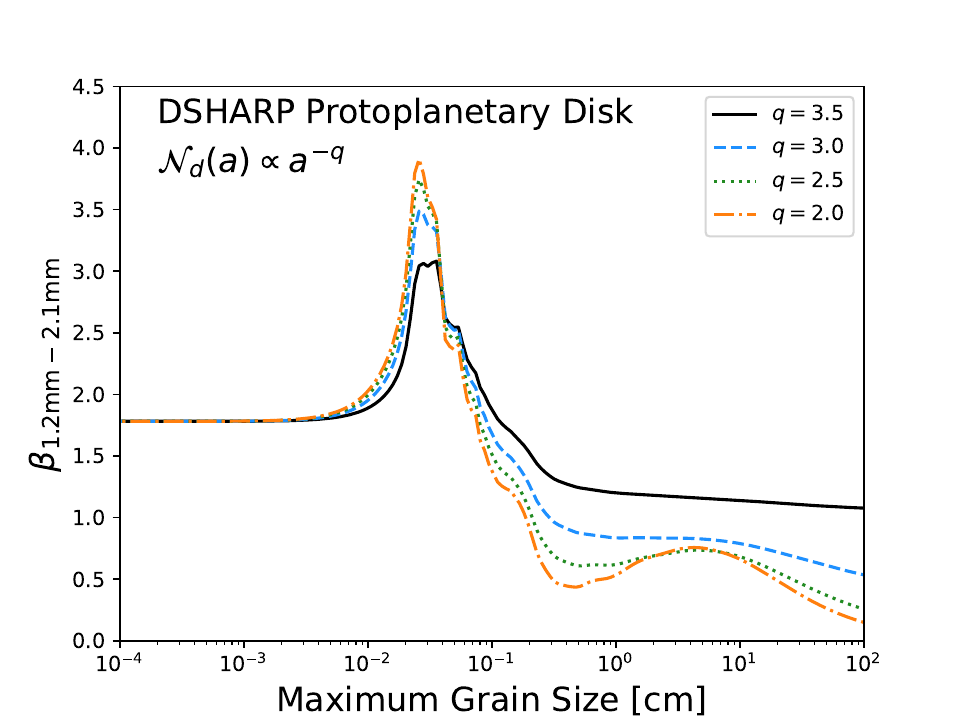}
\vspace{0cm}
\caption{The dust opacity spectral index, $\beta$ between 1.2 mm and 2.1mm is plotted versus the maximum grain size (cm) for the DSHARP dust mass opacities presented in Figure 4 of \cite{2018ApJ...869L..45B} for grains with $\mathcal{N}_d(a) \propto a^{-3.5}$ (black solid line), $a^{-3}$ (blue dashed line), $a^{-2.5}$ (green dotted line), and $a^{-2}$ (orange dash-dotted line).}
\label{fig:Opacity2}
\end{figure}

At frequencies that correspond to far-infrared and longer wavelengths, where most of the emission from cold dust originates, the dust mass opacity is often approximated by a single power-law of the form
\begin{equation}
    \kappa_{\nu,abs}(\vec{r}) = \kappa_{\nu_0}(\vec{r}) \left( \frac{\nu}{\nu_0(\vec{r})} \right)^{\beta(\vec{r})} \;\;\;\;  \rm{cm^2} \; (\rm{g} \; \rm{of} \; \rm{dust})^{-1} \;\;,
\label{eq:kappa_nu3}
\end{equation}
where $\kappa_{\nu_0}$ is the opacity at a nominal frequency $\nu_0$ and $\beta$ is the dust mass opacity spectral index.
For a dust opacity model, we can calculate the $\beta$ from the spectral index from frequency points for which $\kappa_{\nu,abs}$ is calculated
\begin{equation}
    \beta_{\rm{model}} = \frac{\ln \left(\frac{\kappa_{\nu_1}}{\kappa_{\nu2}} \right)}{\ln \left( \frac{\nu_1}{\nu_2} \right)}  \;\;\;\; .
\end{equation}
For each of the models shown in Figure \ref{fig:Opacity1}, the inset shows the variation of $\beta_{model}$ at wavelengths between $160$ $\mu$m and 1.3 mm.
While a single power-law $\beta$ is often assumed for simplicity in calculations, it is clear from the inset in Figure \ref{fig:Opacity1} that a single power-law fit is insufficient over the entire wavelength range for the models shown.
All of the models, with the exception of OH5, have a generally decreasing $\beta_{model}$ between $160$ $\mu$m and 1.3 mm.

In very dense regions, such as protoplanetary disks, grains can coagulate to very large sizes that become comparable to the millimeter wavelengths of the observations (see Figure \ref{fig:OpacityDSHARP}).
The DSHARP models of \cite{2018ApJ...869L..45B} provide a demonstration of the effects of such large grain size distributions.  
In Figure \ref{fig:Opacity2} the dust mass  spectral index, $\beta$, between $1.2$ mm and $2.1$ mm wavelengths is shown as a function of the maximum grain size in a power-law dust size distribution of the form $\mathcal{N}_{d}(a) \propto a^{-q}$.
When the observing wavelengths are much longer than the maximum grain size, $\beta_{1.2\rm{mm}-2.1\rm{mm}}$ is near $1.8$.
A noticeable increase in $\beta_{1.2\rm{mm}-2.1\rm{mm}}$ occurs when the maximum grain size is about a factor of $20$ less than the observed wavelengths. 
When the maximum grain size becomes progressively larger than the observed wavelengths, $\beta_{1.2\rm{mm}-3.3\rm{mm}}$ decreases below the ISM values with the steepness of the decline depending on the power-law index, $q$, of the dust grain size distribution.
Shallower grain size distributions have a steeper decline in $\beta_{1.2\rm{mm}-3.3\rm{mm}}$ with increasing maximum grain size. 
For example, when $q \leq 3.0$, then $\beta_{1.2\rm{mm}-3.3\rm{mm}} < 1$ for maximum grain sizes greater than the observed wavelengths.
For lower values of $q$ there are more dust grains closer to the maximum grain size than for a steeper values of $q$ and those larger grains have a more substantial effect on the observed opacities and therefore $\beta_{1.2\rm{mm}-3.3\rm{mm}}$ (see Equation \ref{eq:intsizedistrbution}).
There is a substantial literature studying this decrease in $\beta$ due to dust grain growth in protoplanetary disks (i.e., \citealt{1991ApJ...381..250B,2004A&A...416..179N,2005ApJ...631.1134A,2011A&A...529A.105G,2012ApJ...760L..17P,2015ApJ...813...41P,2017ApJ...849...63R,2020A&A...640A..19T,2025A&A...700A.188C}).
In Section \ref{sec:SEDModels} and Appendix \ref{AppendixSpectralIndex} we discuss how SED modeling and spectral indicies may be used to constrain $\beta$ from observations.

We have ignored scattering in this tutorial because a proper accounting of its effect on the monochromatic specific intensity and polarization of light requires 3D radiative transfer \citep{2013ARA&A..51...63S}. 
Scattering can have important effects on the observed flux density and polarization at long (e.g. millimeter) wavelengths of objects, such as protoplanetary disks, that have significant grain growth and optical depth  (e.g. 
\citealt{2015ApJ...809...78K,2016MNRAS.456.2794Y,2019ApJ...877L..18Z,2023Natur.623..705S, 2025ApJ...981...12H,2025ApJ...985..148L}).
Dust self-scattering in optically thick protoplanetary disks typically reduces the observed intensity to be less than the Planck function ($I_{\nu,obs} < B_{\nu}(T_d)$; see \cite{1986rpa..book.....R}) with
the exact amount of reduction depending on the dust scattering and absorption properties\footnote{The efficiency with which dust scatters at a given frequency is often characterized by the albedo, $\omega_{\nu}$, which takes the form of $\omega_{\nu} = \kappa_{\nu,sca}(1 - g)/[\kappa_{\nu,abs} + \kappa_{\nu,sca}(1 - g)]$ for anisotropic scattering where $g$ is the expectation value of the cosine of the scattering angle \citep{2018ApJ...869L..45B}. } 
and the specific geometry and physical details of the radiative transfer problem (\ie\ see Figure 9 of \citealt{2018ApJ...869L..45B} and the discussion in \citealt{2019ApJ...877L..18Z}).
We discuss numerical radiative transfer calculations that include scattering with 3D modeling codes in Section \ref{sec:CodesforRadTrans}.

One additional possible effect on the properties of cold dust opacities is that $\kappa_{\nu}$ may vary with the dust grain temperature.
At cold dust temperatures ($T_d \sim 10$ K), the solid state quantum effects of a disordered charge distribution in the presence of two-level system (TLS) defects in amorphous solids composing the grains may result in grain opacities that are also a function of the grain temperature, $\kappa_{\nu}(T_d)$ \citep{2007A&A...468..171M,2011A&A...534A.118P}.
These solid state quantum effects, if present in cold interstellar grains, complicate SED modeling that attempts to constrain simultaneously dust temperature and the dust mass opacity spectral index.

\section{Thermal Dust Continuum Modeling} \label{ModelingSection}

Thermal dust continuum models involve fitting the observed SED of the object and, if the object is resolved, modeling the observed on-sky intensity distribution of emission.
In this section we derive some of the basic equations for dust continuum modeling.
We discuss the basic principles and the degeneracies inherent in such modeling.
Finally we discuss radiative transfer modeling that self-consistently calculates the dust temperature distribution.
The calculation of SED metrics such as the bolometric temperature and the peak frequency are provided in Appendix \ref{AppendixTbol} and Appendix \ref{SEDPeakAppendix} respectively. 
A detailed discussion of measuring and interpreting SED spectral indicies can be found in Appendix \ref{AppendixSpectralIndex}.

\subsection{SED Modeling}\label{sec:SEDModels}

The observed SED of a source is the set of flux densities observed at the average frequencies, $\overline{\nu}$ (see Equation \ref{eq:avgnu}), of the continuum filters.
The complete equation for the flux density of dust continuum emission is found by substituting Equation \ref{eq:Inu} into Equation \ref{eq:FilterFlux} (assuming the small angle approximation),
\begin{eqnarray}
    F_{\overline{\nu},obs} & = & \frac{\int_{\nu = 0}^{\infty} \int_{\phi = 0}^{2\pi} \int_{\theta = 0}^{\theta_{ap}} I_{\nu,bg}(\theta,\phi) e^{-\tau_{\nu}(\theta,\phi)} P_n(\theta,\phi) \mathcal{T}(\nu) \theta d \theta d \phi d\nu }{\int_{0}^{\infty} \mathcal{T}(\nu) d\nu} \nonumber \\ & + & \frac{\int_{\nu = 0}^{\infty} \int_{\phi = 0}^{2\pi} \int_{\theta = 0}^{\theta_{ap}} \int_{\tau_{\nu}^{\prime} = 0}^{\tau_{\nu}(\theta,\phi)} B_{\nu}[T_d(\theta,\phi,\tau_{\nu}^{\prime})] e^{-(\tau_{\nu}(\theta,\phi) - \tau_{\nu}^{\prime})}   P_n(\theta,\phi) \mathcal{T}(\nu) d \tau_{\nu}^{\prime} \theta d \theta d \phi d\nu }{\int_{0}^{\infty} \mathcal{T}(\nu) d\nu} \;\;\;\; \rm{erg}\, \rm{s}^{-1}\, \rm{cm}^{-2}\, \rm{Hz}^{-1} \; , 
\label{eq:FULLSED}
\end{eqnarray}
where the integral of $\tau_{\nu}^{\prime}$ is along each unique line-of-sight, A, that corresponds to coordinates ($\theta,\phi)$ on the sky within the limits of integration $\theta \leq \theta_{ap}$.
If the observed flux density is optically thin, then we can simplify the integral over $\tau_{\nu}^{\prime}$ to an integral over $z$ whose integral limit depends on the source geometry (see Section \ref{SphGeoApp}),
\begin{eqnarray}
    F_{\overline{\nu},obs} & = & \frac{\int_{\nu = 0}^{\infty} \int_{\phi = 0}^{2\pi} \int_{\theta = 0}^{\theta_{ap}} I_{\nu,bg}(\theta,\phi) [1 - \tau_{\nu}(\theta,\phi)] P_n(\theta,\phi) \mathcal{T}(\nu) \theta d \theta d \phi d\nu }{\int_{0}^{\infty} \mathcal{T}(\nu) d\nu} \nonumber \\ & + & \frac{\int_{\nu = 0}^{\infty} \int_{\phi = 0}^{2\pi} \int_{\theta = 0}^{\theta_{ap}} \int_{z}  B_{\nu}[T_d(\theta,\phi,z)] \rho_d(\theta,\phi,z) \kappa_{\nu}(\theta,\phi,z)  P_n(\theta,\phi) \mathcal{T}(\nu) dz \theta d \theta d \phi d\nu }{\int_{0}^{\infty} \mathcal{T}(\nu) d\nu} \;\;\;\; \rm{erg}\, \rm{s}^{-1}\, \rm{cm}^{-2}\, \rm{Hz}^{-1} \; , 
\label{eq:FULLSEDTHIN}    
\end{eqnarray}
The solution of Equation \ref{eq:FULLSED} or \ref{eq:FULLSEDTHIN} is only possible with prior knowledge of the three-dimensional temperature, density, and dust opacity structure of the source and the source geometry that maps $(x,y,z)$ coordinates inside the source to $(\theta,\phi,\tau_{\nu}^{\prime})$ or $(\theta,\phi,z)$ observer coordinates.
In practice, some or all of these quantities are unknown and the observer will attempt to fit the SED to constrain them.
In this section we discuss the general methods of SED fitting and the effects of the breakdown of assumptions such as isothermality.
We caution from the outset that SED modeling generally involves significant degeneracies and that radiative transfer modeling, with well benchmarked and publicly-available radiative transfer codes, is preferred when practical.

\begin{figure}[h]
\includegraphics[scale=0.8, trim= 0mm 50mm 0mm 40mm, clip]{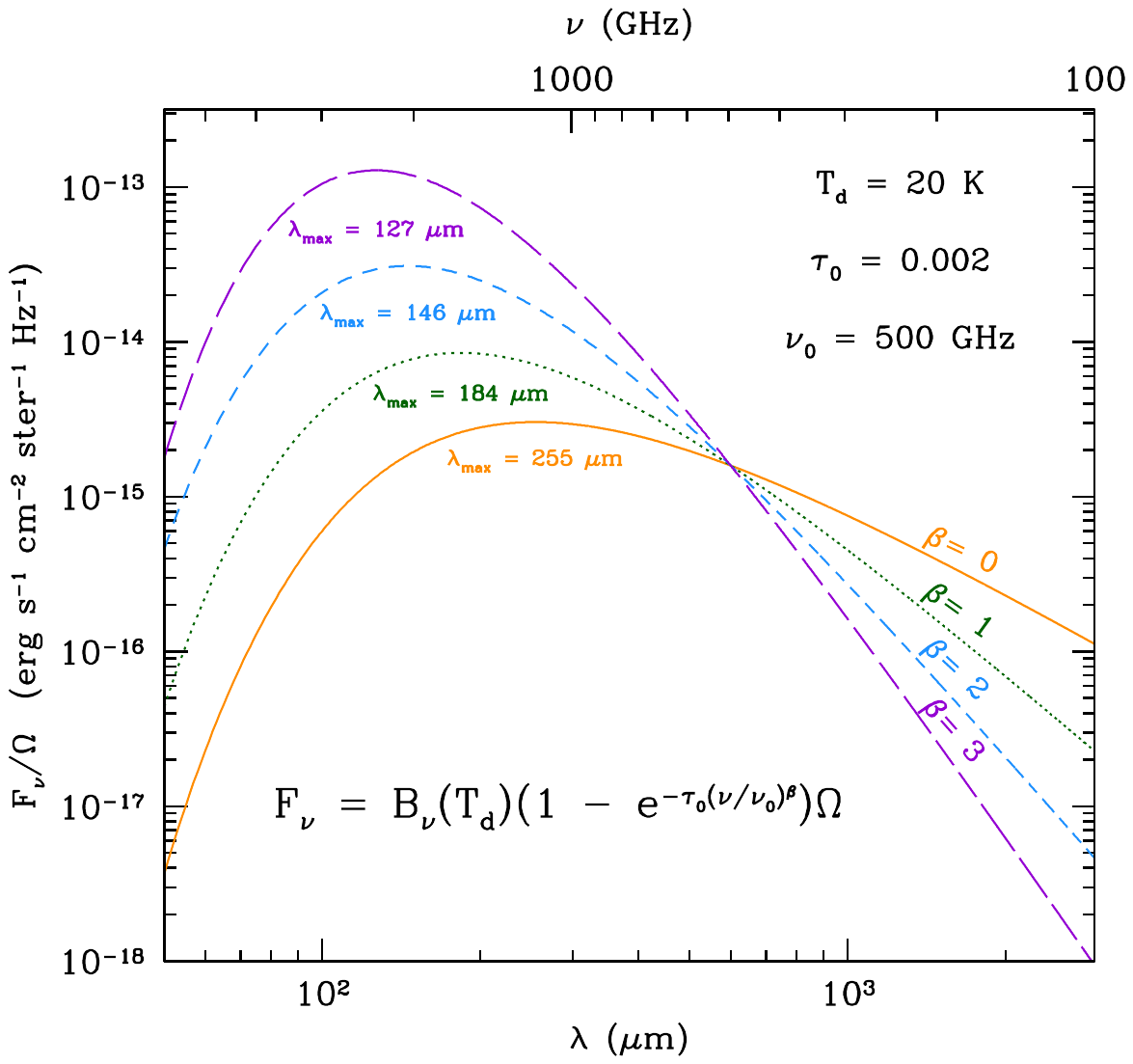}
\caption{Modified 10 K Blackbody SEDs for dust opacity indicies of $\beta = 0$ to $3$.  The optical depth was set to 0.002 at 500 GHz for all models.  The average intensity of the modified blackbody is plotted on the y-axis (cgs units).  This figure was inspired from Figure 1 of \cite{2009ApJ...696.2234S}.}
\label{fig:SEDBetaFig}
\end{figure}

The starting point for many SED studies is not Equations \ref{eq:FULLSED} or \ref{eq:FULLSEDTHIN} due to their complexity, but is instead the extreme assumptions that the source is isothermal and the same opacity law applies throughout the source.
As we have seen, dust opacities are often characterized by a power-law over a range of frequencies (Equation \ref{eq:kappa_nu3}) which permits us to parameterize the optical depth as
\begin{equation}
    \overline{\tau_{\nu}} = \tau_0 \left( \frac{\nu}{\nu_0} \right)^{\beta} \;\;\;\; ,
    \label{eq:taunubar}
\end{equation}
where $\tau_0$ is the optical depth at frequency $\nu_0$.
The isothermal SED Equation \ref{eq:FluxSED} then becomes
\begin{equation}
       \lim_{I_{\nu,\rm{bg}} \rightarrow 0} \; \lim_{T_d(\vec{r}) = T_d} \; \lim_{\kappa_{\nu}(\vec{r}) = \kappa_{\nu}}  \; F_{\nu,obs} =  B_{\nu}(T_d) \left[ 1 - e^{-\tau_0 \left( \frac{\nu}{\nu_0} \right)^{\beta}}  \right] \Omega_{ap} \;\;\;\; \rm{erg}\, \rm{s}^{-1}\, \rm{cm}^{-2}\, \rm{Hz}^{-1} \;\; .
\label{eq:sedtaupow}
\end{equation}
Equation \ref{eq:sedtaupow} is an isothermal, constant opacity parametrization of Equation \ref{eq:FULLSEDTHIN} in the limit that there is perfect coupling with the telescope beam ($P_n \approx 1$) and that the continuum filter transmission curve is a Dirac delta function at the observed frequency ($\mathcal{T}(\nu) \approx \delta(\overline{\nu})$).
This equation is sometimes referred to as a ``Modified Blackbody" where the word ``modified" refers to the effect of the opacity at each frequency.
Figure \ref{fig:SEDBetaFig} shows examples of modified blackbody SEDs with different opacity power-law indices.
In the optically thin limit, Equation \ref{eq:sedtaupow} reduces to
\begin{equation}
\lim_{I_{\nu,\rm{bg}} \rightarrow 0} \; \lim_{T_d(\vec{r}) = T_d}  \; \lim_{\kappa_{\nu}(\vec{r}) = \kappa_{\nu}} \; \lim_{\tau \ll 1} \; F_{\nu,obs} =  B_{\nu}(T_d) \kappa_0 \left( \frac{\nu}{\nu_0} \right)^{\beta} \overline{\Sigma_d} \Omega_{ap} \;\;\;\; \rm{erg}\, \rm{s}^{-1}\, \rm{cm}^{-2}\,  \rm{Hz}^{-1} \;\; ,
\label{eq:sedtaupowthin}
\end{equation}
where we have used Equation \ref{eq:taukappasigma} to substitute for the optical depth.
$\overline{\Sigma_d}$ can be converted to $\overline{\Sigma_g}$ or $\overline{N_g}$ using Equations \ref{eq:sigmag} and \ref{eq:sigmad}.
This is also sometimes written by substituting $\overline{\Sigma_d} \Omega_{ap} = M_d/D^2$ using Equation \ref{eq:Md_thin}.

The spectral index, $\alpha_s$\footnote{$\alpha$ is the traditional letter used to denote the spectral index in the literature.  It should not be confused with $\alpha_{\nu}$, the absorption coefficient and $\alpha$ the right ascension of the object coordinates.  We have added the subscript, $s$, to differentiate spectral index from the other $\alpha$ variables.}, is calculated from the SED as
\begin{equation}
    \alpha_s = \frac{\partial \ln{F_{\nu}}}{\partial \ln{\nu}} \;\;\;\; .
\label{eq:specind}    
\end{equation}
At long wavelengths on the Rayleigh-Jeans side of the observed SED (right side of the peak in Figure \ref{fig:SEDBetaFig}), thermal dust continuum emission may often be approximated as a power-law  in part, due to the power-law nature of the dust opacities, with spectral index $\alpha_s$,
\begin{equation}
F_{\nu,obs} = F_{\nu, f} \left( \frac{\nu}{\nu_f} \right)^{\alpha_s} \;\;\;\; \rm{erg}\, \rm{s}^{-1}\, \rm{cm}^{-2}\, \rm{Hz}^{-1} \; , 
\label{eq:SEDPL}
\end{equation}
where $F_{\nu, f}$ is the flux density at a fiducial frequency $\nu_f$.
If we take the logarithm of Equation \ref{eq:SEDPL}, then we find
\begin{equation}
    \ln{F_{\nu,obs}} = \ln{F_{\nu, f}} + \alpha_s \ln{\nu} - \ln{\nu_f} \;\;\;\; 
\end{equation}
and the spectral index defined by Equation \ref{eq:specind} is single valued.
If the SED is not a single power-law over some range of frequencies, then Equation \ref{eq:specind} still defines a spectral index, but that spectral index is a function of frequency, $\alpha_s(\nu)$.
The spectral index of a modified blackbody (Equation \ref{eq:sedtaupow}) is given by
\begin{equation}
\alpha_s(\nu, \overline{\tau_{\nu}}, T_d)  =  3 + \beta\frac{\overline{\tau_{\nu}}}{e^{\overline{\tau_{\nu}}} - 1} - \frac{\frac{h\nu}{kT_d}}{1 - e^{-h\nu/kT_d}} \;\;\;\; .
\label{eq:sedspecindexmain}
\end{equation}
In Appendix \ref{AppendixSpectralIndex}, we derive this formula and discuss how $\beta$ can be determined from the observed spectral index between two frequencies of the SED.

The peak wavelength of a modified blackbody SED\footnote{A peak wavelength given by $\lambda_{\rm{pk}} = c / \nu_{\rm{pk}}$, as in Equation \ref{eq:WienLawbetalam} and shown in Figure \ref{fig:SEDBetaFig}, is only appropriate for flux density units that are per unit frequency (Hz$^{-1}$) like Jy.  
It is not the same $\lambda_{\rm{pk}}$ that would be found from a derivation of Wien's Displacement Law using $B_{\lambda}(T_d)$.  
In that case, the observed flux densities would have to be quoted in units that were per unit wavelength interval (cm$^{-1}$).
Care must be taken when quoting the peak frequency or peak wavelength to use the correct form of the Planck function and therefore the correct version of Wien's Law given the units of the flux density measurements (use $B_{\nu}(T_d)$ for per unit frequency and $B_{\lambda}(T_d)$ for per unit wavelength). This is because $d\nu \neq d\lambda$, and therefore $B_{\nu}(T_d) \neq B_{\lambda}(T_d)$ for a given $\nu$ that corresponds to a given $\lambda$ ($\nu = c/\lambda$).    
Instead, the nonlinear differential relationships, $|d\nu| = (c /\lambda^2) |d\lambda|$ or $|d\lambda| = (c/\nu^2) |d\nu|$, mean that $B_{\nu}$ and $B_{\lambda}$ are two separate and different Planck functions, one per unit frequency and the other per unit wavelength, that are related to each other by the relationship $B_{\nu}(T_d) |d\nu| = B_{\lambda}(T_d) |d\lambda|$.  Appendix \ref{SEDPeakAppendix} derives a generalized Wien's Law using B$_{\nu}$ since most flux density observations of dusty SEDs are quoted in units that are per unit frequency.}, or a generalized Wien's Law for different choices of $\beta$, is given by 
\begin{equation}
      \lambda_{\rm{pk}} = \frac{c}{\nu_{\rm{pk}}} = \frac{h c}{k T_d} \left[ (3 + \beta\epsilon) + W_0\left(- \frac{3 + \beta\epsilon}{e^{3 + \beta\epsilon}}\right)  \right]^{-1} \; \rm{cm} \;\;.
\label{eq:WienLawbetalam}
\end{equation}
where $W_0$ is the principle branch of the Lambert W-function \citep{Corless1996OnTL} and $\epsilon = \overline{\tau_{\nu}}/(e^{\overline{\tau_{\nu}}} - 1)$.
This equation is derived in Appendix \ref{SEDPeakAppendix}.

The modified blackbody curves in Figure \ref{fig:SEDBetaFig} illustrate the general trends that smaller opacity power-law indicies, $\beta$, result in the peak of the SED occurring at longer wavelengths and that the spectral index of the SED is smaller at long wavelengths.
The opposite is true for larger values of $\beta$, as the SED peak occurs at shorter wavelengths and the spectral index of the SED is larger at long wavelengths.
For example, at $T_d = 20$ K, $\lambda_{max} = 255$ $\mu$m for $\beta = 0$ and $\lambda_{max} = 127$ $\mu$m for $\beta = 3$ while the spectral index measured at a wavelength of 2 mm ($149.9$ GHz) is $\alpha_s = 1.81$ for $\beta = 0$ and $\alpha_s = 4.81$ for $\beta = 3$.
Note that the spectral index for $\beta = 0$ is not equal to $2$ as would be expected for a blackbody in the Rayleigh-Jeans limit ($h\nu/kT_d \ll 1$).  
This is because we are not strictly in the Rayleigh-Jean limit at 2 mm for a $T_d = 20$ K modified blackbody ($h(149.9 \,\rm{GHz})/k(20 \,\rm{K}) \approx 0.36 \not\ll 1$)\footnote{For $T_d = 20$ K, you would have to measure the spectral index at a wavelengths of $\geq 3.7$ mm in order to have $2.0 \geq \alpha_s \geq 1.9$, within less than $5$\% of the strict Rayleigh-Jean value.}.
The assumption of the Rayleigh-Jeans limit is often misused in dust continuum studies (see Appendix \ref{AppendixSpectralIndex} for further examples).

\begin{figure}[h!]
\includegraphics[scale=0.5, trim= 50mm 0mm 60mm 0mm, clip]{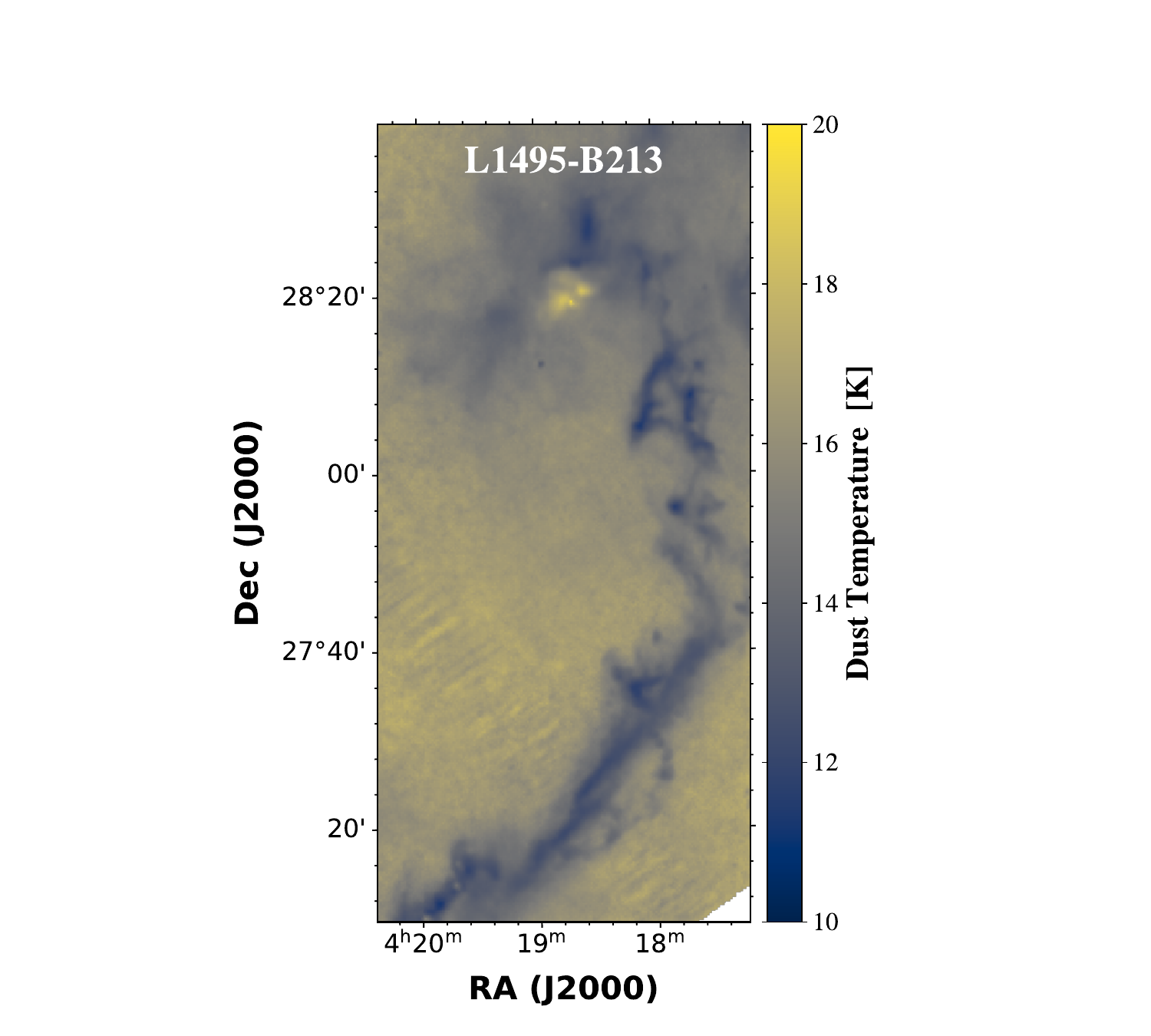}
\includegraphics[scale=0.5]{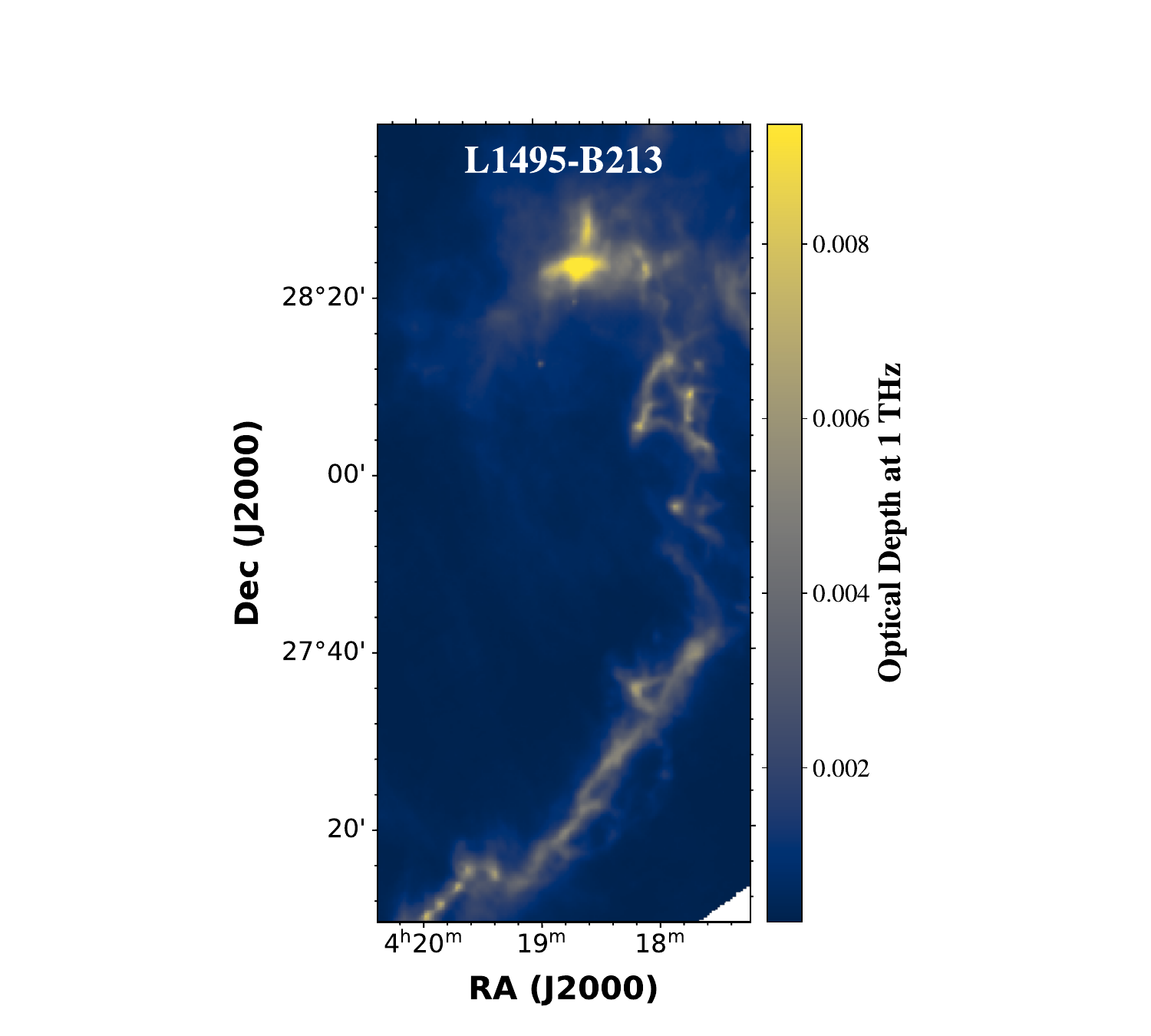}
\caption{The SED modeling of \cite{2022ApJ...941..135S} simultaneously constrain the dust temperature (LEFT) and the optical depth at 1 THz (RIGHT) toward the L1495-B213 regions of the Taurus Molecular Cloud.  Notice that the right panel looks identical to Figure \ref{fig:H2Column} since Equation \ref{eq:tauNg} was used to convert this optical depth map to the H$_2$ column density map.}
\label{fig:SEDModelOutputs}
\end{figure}

Traditional modified blackbody fitting involves $\chi^2$ fitting Equation \ref{eq:sedtaupow} to the observed flux densities to constrain $T_d$, $\beta$, and $\tau_0$.
The optical depth at a characteristic frequency, $\tau_0$, can be converted into $\overline{\Sigma_d}$, $\overline{\Sigma_g}$, or $\overline{N_g}$ with a choice of the dust opacity at $\nu_0$ (see Equations \ref{eq:taukappasigma}, \ref{eq:sigmag} and \ref{eq:tauNg}).
If either $T_d$ or $\beta$ are assumed to have a fixed value, then the other quantities can be determined from $\chi^2$ fitting the observations compared to model modified blackbodies.
For example, \cite{2022ApJ...941..135S} model the SED in common resolution pixels of \textit{Herschel} maps at $160, 250, 350$, and $500$ $\mu$m of the Herschel Gould Belt Survey \citep{2010A&A...518L.102A} by assuming $\beta = 2.0$ and that $\kappa_{\nu} = 10$ cm$^2$ per gram of dust at 1 THz\footnote{\cite{2022ApJ...941..135S} also explore the systematic effects on their fits of different choices of $\beta$ including using a lower resolution map of $\beta$ determined from SED modeling of \textit{Planck} observations (see Section 11.1.1 of their paper).}.
From $\chi^2$ fitting, they determine $T_d$ and the optical depth at 1 THz (Figure \ref{fig:SEDModelOutputs}), which they convert to the H$_2$ column density (see Figure \ref{fig:H2Column}), in every pixel.
\cite{2022ApJ...941..135S} perform a thorough analysis of the effects of uncertainties on SED modeling with fixed $\beta$.

\begin{figure}[h]
\includegraphics[scale=0.5, trim= 0mm 50mm 0mm 30mm, clip]{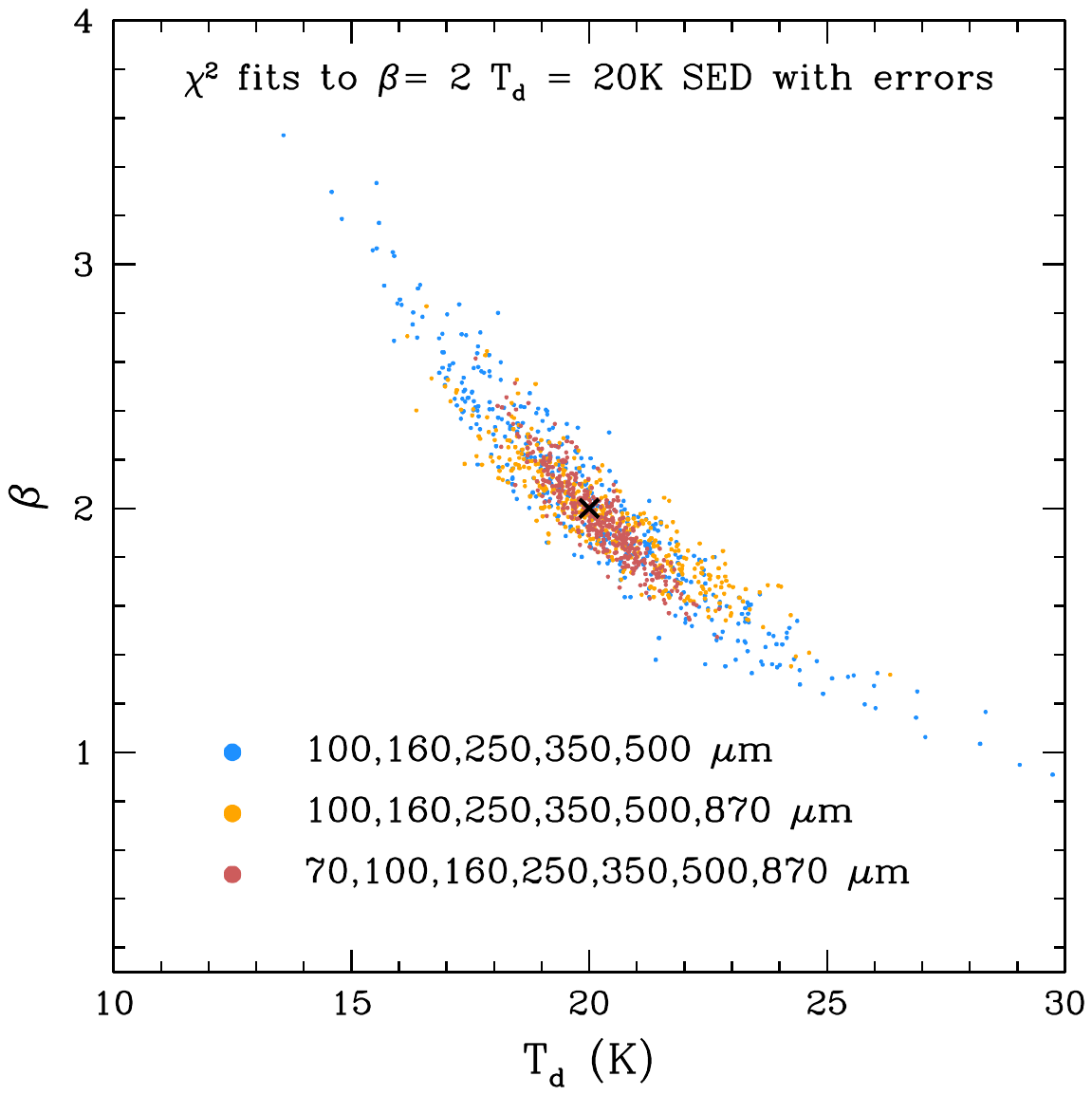}
\caption{The combination of dust opacity spectral index $\beta$ and dust temperatures are plotted that are best-fits to $\chi^2$ fitting of modified blackbody SEDs of dust around HII regions with photometric errors. The points are from the simulation data in Figure 11 of \cite{2012A&A...542A..10A} and are SED fits to simulated photometry with errors of a modified blackbody model with $\beta = 2$ and $T_d = 20$ K (indicated by the black cross).  Blue points are fits to SEDs observed at 100, 160, 250, 350, 500 $\mu$m.  Orange points are fits to the SEDs observed at 100, 160, 250, 350, 500, 870 $\mu$m.  The red points are fits to SEDs observed at 70, 100, 160, 250, 350, 500, 870 $\mu$m.}
\label{fig:DustBetaTd}
\end{figure}

If both $T_d$ and $\beta$ are free variables in the fit, then $\chi^2$ fitting produces an anti-correlation and degeneracy between these variables \citep{2009ApJ...696..676S,2012A&A...541A..33J,2015A&A...584A..94J,2021ApJ...919...30D}.
A good example of this degeneracy is illustrated in Figure \ref{fig:DustBetaTd} which is a modification of a figure in \cite{2012A&A...542A..10A}.
They generated modified blackbody models with $T_d = 20$ K and $\beta = 2$ from which they simulated observations with errors by drawing from Gaussian distributions that accounted for typical statistical and systematic calibration errors at the \textit{Herschel Space Observatory} wavelengths of $70$, $100$, $160$, $250$, $350$, and $500$ $\mu$m plus ground-based observations at $870$ $\mu$m.
The points in Figure \ref{fig:DustBetaTd} show the best-fitted $\chi^2$ for each simulated observation.
Such simulations result in a ``banana shape" in $\chi^2$ values in a plot of $\beta$ vs. $T_d$ (see Figure \ref{fig:DustBetaTd}), for which the extent of the shape depends on the size of the observational errors and the number of different wavelengths sampling the SED (including the number of wavelengths bracketing the peak of the SED).
As the number of observed wavelengths on each side of the SED peak increases, the extent of the degeneracy decreases. 
Even for these well-sampled SEDs (and it is not possible to find better wavelength coverage with existing facilities), there are considerable uncertainties in simultaneously determining $T_d$ and $\beta$.
Low signal-to-noise photometry can lead to multiple minima in $\chi^2$ fitting (see \citealt{2012A&A...541A..33J}).
There is an extensive literature analyzing the degeneracy between $T_d$ and $\beta$ and utilizing other statistical tools, such as 
Bayesian inference, for this fitting problem to reduce the ``banana-shaped" degeneracy between $T_d$ and $\beta$ in the presence of observational and systematic errors \citep{2012ApJ...752...55K,2013A&A...556A..63J,2019MNRAS.489.4389L,2023A&A...673A.145J}.

The SED fitting methods described above tend to assume isothermality and constant dust opacity and are therefore fitting only a single modified blackbody SED. 
Real sources likely have dust temperature and dust opacity variations.
We can make a simple model to understand the effects of source temperature and opacity variations on the SED.
Assume we have a source with two temperature components, a cold dust component at $T_{d,c}$ and a hot dust component at temperature $T_{d,h}$ that have dust opacities $\kappa_{\nu,c}$ and $\kappa_{\nu,h}$ respectively.
In the optically thin limit\footnote{If the SED becomes optically thick at some frequency, then $L_{\nu}$ at the optically thick frequencies will just be equal to $B_{\nu}(T_{d,\tau_{\nu} = 1})$ evaluated at the dust temperature of the optically thick surface or ``photosphere" of the object, which may differ at different optically thick frequencies.}, the ratio of monochromatic luminosities observed at the same frequency of the two components is given by (see Equation \ref{eq:Lnu_mono_thin_isotherm})
\begin{equation}
   \lim_{\tau \ll 1} \; \frac{L_{\nu,c}}{L_{\nu,h}} = \frac{B_{\nu}(T_{d,c}) M_{d,c} \kappa_{\nu,c}}{B_{\nu}(T_{d,h}) M_{d,h} \kappa_{\nu,h}} \;\; .
\label{eq:Lratiokap} 
\end{equation}
where $M_{d,c}$ and $M_{d,h}$ are the cold and hot dust masses respectively. 
Since $T_{d,c} < T_{d,h}$, then $B_{\nu}(T_{d,c}) < B_{\nu}(T_{d,h})$ at all frequencies\footnote{The family of Planck function curves at different temperatures do not cross.}.
If the dust opacity is constant in the source, then the colder component will contribute less to the SED at a given frequency unless it is sufficiently more massive than the warm component to overcome the ratio of Planck functions.
If the opacities are not constant, then the colder component will contribute less to the SED at a given frequency unless the product of $M_d \kappa_{\nu}$ is sufficiently larger for the cold component than the warm component to overcome the ratio of Planck functions.
Real astrophysical sources are unlikely to have only two isothermal temperature components, but instead are more likely to have smooth gradients in dust temperature.
The simple model discussed here, however, has some utility in that the relative contribution to the SED of each temperature component is weighted by the mass and the opacity of dust radiating at that temperature.
Some SED fitting techniques such as Point Process Mapping (PPMAP) attempt to account for source temperature and opacity variation.
PPMAP uses Bayesian inference to fit multiple modified blackbodies with different $T_d$ and $\beta$ while also preserving the resolution of the data at different wavelengths \citep{2015MNRAS.454.4282M,2017MNRAS.471.2730M,2019MNRAS.489..962H,2019MNRAS.489.5436W,2021MNRAS.504.6157H,2025A&A...694A..24M}.

For simplicity in our pedagogical model, we have used $L_{\nu}$, the total monochromatic luminosity of the source. 
Observations, however, are made with a telescope with a power pattern, $P_n(\theta, \phi)$,  resulting in the observed flux density being measured within the solid angle of the observed aperture, $\Omega_{ap}$ (see Section \ref{Caveats}).
The solid angle of the aperture (and the telescope response pattern within that aperture) effectively acts as a mass filter in resolved sources.
This is apparent from Equations \ref{eq:sedtaupowthin} and \ref{eq:Md_thin} with the product of $\overline{\Sigma_d} \Omega_{ap} = \overline{\Sigma_d} A_{ap}/D^2 = M_d/D^2$, where $M_d$ is the dust mass observed within the solid angle of the aperture.
For example, a dense ($>10^5$ cm$^{-3}$) starless core has a small, central, shielded cold interior and a warmer exterior with a much larger volume and mass that is heated by the surrounding ISRF \citep{2001ApJ...557..193E,2001A&A...376..650Z,2005ApJ...632..982S,2013A&A...551A..98L,2016A&A...592A..61L,2023MNRAS.521.4579S}. 
A small telescope beam can couple better to the small, cold temperature component 
by restricting, through the small beam solid angle, how much of the mass of the warmer component it observes (see  \citealt{2015A&A...574L...5P} for a discussion of this problem which is never fully ameliorated for dense starless cores).

\subsubsection{Example: Galaxy SED Fitting with Multiple Temperature Components}

A good example of modified blackbody fitting where temperature gradients should be considered is the dust emission spectrum of galaxies. 
This topic has an extensive literature, especially for fitting the SEDs of high redshift galaxies  (e.g., \citealt{2006ApJ...650..592K,casey12a,2012ApJ...760....6M,casey14a,2019ApJ...887..144J,2020MNRAS.494.3828D,2021ApJ...919...30D,2022ApJ...934...56B,2023A&A...678A..27I,2023MNRAS.523.3119W,2024ApJ...961..226L,2025MNRAS.540.1560B}).
Single temperature modified blackbody fits to galaxy SEDs, while used commonly in the literature, can often fail dramatically near the Wien side of the thermal IR SED of galaxies if extremely hot clumps of dust are present  \citep[e.g., near an AGN;][]{kovacs10a}.  
\cite{casey14a} summarize galaxy SED fitting methods. 
We highlight a couple of these methods to illustrate the techniques.

One approach is to approximate the mid-infrared component from hot dust as a power-law, and fit the normalizations of the modified blackbody component and the power-law together \citep{casey12a},
\begin{equation}
    F_{\nu,obs}\left(T\right) = N_{\rm MBB}\frac{\left(1-e^{-\tau_\nu}\right)\nu^3}{e^{h\nu/kT}-1} + N_{\rm PL}\nu^{-\alpha_{\rm PL}} e^{-\left(\frac{\nu_c}{\nu}\right)^2}
\end{equation}
where $N_{\rm MBB}$ and $N_{\rm PL}$ are the relative weights of the modified blackbody and power-law components, respectively, $\alpha_{\rm PL}$ is the power-law slope, and $\nu_c$ is the frequency at which $dF_{\nu,obs}/d\nu = -\alpha_{\rm PL}$.
Variations of this two component SED model have been used in several studies (e.g., see \citealt{casey12a,2022ApJ...930..142D,2023ApJ...951...48M,2024MNRAS.530.4887W,2024ApJ...965..123L}).

Another approach is to assume a dust temperature distribution, such that the aggregate observed SED is comprised of the composite emission of dust with a range of temperatures. 
We can modify Equation \ref{eq:fluxred7}  by replacing the difference in Planck functions at a single dust temperature by an integral over an emitted  temperature distribution, at redshift $z_{red}$, that has a normalized probability distribution $\mathcal{P}(T_{d_e})$\footnote{$\int_{0}^{\infty}  \mathcal{P}(T_{d_e}) dT_{d_e} = 1$}.
This results in an observed background subtracted flux density, $\Delta F_{\nu,obs}  =   F_{\nu,obs} - B_{\nu_o}(T_{\rm{CMB}_o})\Omega(z_{\rm{red}})$, of
\begin{equation}
   \lim_{T_{d_e}(\vec{r}) = T_{d_e} \, \forall \, T_{d_e}}  \;   \Delta F_{\nu,obs} =   \frac{\mu_{\rm{lens}} (1 + z_{\rm{red}})A_{src}}{D_L^2} \int_{T_{d_e}}  \left[B_{\nu_e}(T_{d_e}) - B_{\nu_e}(T_{\rm{CMB}_e}) \right] \mathcal{P}(T_{d_e}) (1 - e^{-\overline{\tau_{\nu_e}}} ) dT_{d_e}   \;\;\;\; \rm{erg}\, \rm{s}^{-1}\, \rm{cm}^{-2}\, \rm{Hz}^{-1} \; ,
\label{eq:fluxredTddist}
\end{equation}
where the effect of the CMB on the emitted dust temperature at different redshifts is calculated in \citet{2013ApJ...766...13D} as:
\begin{equation}
    T_{d_e} = \left\{ T_{d_o}^{(4+\beta)}  +  T_{\rm{CMB}_o}^{(4+\beta)} [ (1 + z_{\rm{red}})^{(4+\beta)} - 1]   \right\}^{\frac{1}{(4+\beta)}} \;\; \rm{K} \; .
\label{eq:Tdhighz}
\end{equation}
In the optically thin limit, Equation \ref{eq:fluxredTddist} becomes
\begin{equation}
   \lim_{T_{d_e}(\vec{r}) = T_{d_e} \,  \forall \, T_{d_e}}  \; \lim_{\tau \ll 1} \;   \Delta F_{\nu_o} =   \frac{\mu_{\rm{lens}} (1 + z_{\rm{red}})M_d\kappa_{\nu}}{D_L^2} \int_{T_{d_e}}  \left[B_{\nu_e}(T_{d_e}) - B_{\nu_e}(T_{\rm{CMB}_e}) \right] \mathcal{P}(T_{d_e}) dT_{d_e}   \;\;\;\; \rm{erg}\, \rm{s}^{-1}\, \rm{cm}^{-2}\, \rm{Hz}^{-1} \; .
\label{eq:fluxredTddistthin}
\end{equation}
The integrals over the dust temperature distribution are computed numerically and depend on the choice of the shape of the probability distribution.
For instance,  \cite{2025MNRAS.540.3693S} use a skewed-Gaussian distribution of dust temperatures.
Another choice is a power-law distribution of dust temperatures above a minimum dust temperature \citep{kovacs10a, casey14a} .

\subsection{Dust Continuum Radiative Transfer Modeling}

Dust continuum radiative transfer modeling self-consistently calculates the dust temperature distribution, $T_d(\vec{r})$ given an input density distribution, the dust opacity distribution, and the internal and external sources of heating of the source.
This is achieved by calculating the radiative balance between grain heating, due to the absorption of radiation, and grain cooling, due to thermal emission,
\begin{equation}
    \left(\frac{dE}{dt}\right)_{abs} = \left(\frac{dE}{dt}\right)_{emit} \;\;\;\; \rm{erg} \; \rm{s}^{-1} \;\;.
\label{eq:EnergyBalance}
\end{equation}
Analytical expressions for $T_d(\vec{r})$ may be calculated for special circumstances, but in general, the calculations must be done numerically using a radiative transfer code.
We show an example of the analytical calculation of the dust temperature for a central luminous source and then discuss the current state-of-the-art dust continuum radiative transfer codes.
A separate analytical derivation of the dust temperature distribution for grain heating by the interstellar radiation field is given in Appendix \ref{AppendixISRF}.

\subsubsection{Optically Thin Dust Temperature Profile with a Central Source}\label{sec:TdCentralSource}

Large dust grains have enough internal vibrational modes that they are able to distribute the energy quickly from an absorbed short wavelength photon. 
If the rate of energy impinging on the dust grain is constant in time, then a large grain can reach an equilibrium temperature, $T_d$.
We calculate the equilibrium dust temperature of a dust grain by balancing the rate at which energy is absorbed with the rate of energy emitted by the grain (Equation \ref{eq:EnergyBalance}).

The rate at which a grain absorbs energy is proportional to the incident flux density onto the dust grain, $f_{\nu}$, and the effective cross-section for absorption, $\sigma_{\nu, abs}$, of a photon of frequency $\nu$,
\begin{equation}
    \left(\frac{dE}{dt}\right)_{abs} = \int_{\nu = 0}^{\infty} f_{\nu} \sigma_{\nu, abs} d\nu \;\;\;\; \rm{erg} \; \rm{s}^{-1} \;\;.
\label{eq:dEdt_abs}
\end{equation}
In general, the incident flux density onto the dust grain is the monochromatic specific intensity at the position of the dust grain, $I_{\nu}(\vec{r},\hat{r})$, integrated over angular factors in all directions, $\hat{r}$, using a spherical coordinate system $(\theta^{\prime},\phi^{\prime})$ centered on the grain,
\begin{equation}
    f_{\nu} = \int_{\phi^{\prime}=0}^{2\pi} \int_{\theta^{\prime}=0}^{\pi} I_{\nu}(\vec{r},\theta^{\prime},\phi^{\prime}) \cos{(\theta^{\prime})} \sin{(\theta^{\prime})} d\theta^{\prime} d\phi^{\prime} \;\;\;\; \rm{erg} \; \rm{s}^{-1}\, \rm{cm}^{-2}\, \rm{Hz}^{-1}  \;\;.
    \label{eq:IncidentFluxGeneral}
\end{equation}
If the main source of incident flux density is a central source with luminosity, $L_{\nu, *}$, that is a distance, $r$, away (e.g., the luminous source is at the center of the cloud in Figure \ref{fig:Geo1}), and if the path between the source and the dust grain is optically thin to the short wavelength radiation that the dust grain can absorb, then the incident flux density is given by\footnote{Equation \ref{eq:IncidentFlux} assumes an isotropic emitting source with the luminosity spread over the surface area of a sphere with radius r and no absorption between the object and the dust grain located at distance r.  It is possible to derive Equation \ref{eq:IncidentFlux} from the more general Equation \ref{eq:IncidentFluxGeneral}. For an isotropic emitting object, the flux density emitting from the surface of the object is $f_{\nu}(R_{*}) = \int_{\phi^{\prime\prime}=0}^{2\pi} \int_{\theta^{\prime\prime}=0}^{\pi/2} I_{\nu, *} \cos{(\theta^{\prime\prime})} \sin{(\theta^{\prime\prime})} d\theta^{\prime\prime} d\phi^{\prime\prime} = \pi I_{\nu, *} = L_{\nu, *} / 4 \pi R_{*}^2$ where we have oriented a new spherical coordinate system $(\theta^{\prime\prime},\phi^{\prime\prime})$ with its origin on the surface of the object, with the $\theta^{\prime\prime} = 0$ pole of the spherical coordinate system pointing toward the center of the object, and integrating over the hemisphere of this coordinate system with $I_{\nu} \neq 0$ (looking toward the object). 
Solving for $I_{\nu, *} = L_{\nu, *}/4 \pi^2 R_{*}^2$, plugging into Equation \ref{eq:IncidentFluxGeneral}, and assuming that the dust grain is located at a distance r that is far away from the source ($r \gg R_{*}$) then we find that $f_{\nu}(r) = \int_{\phi^{\prime}=0}^{2\pi} \int_{\theta^{\prime}=0}^{R_{*}/r} (L_{\nu, *}/4 \pi^2 R_{*}^2)  \theta^{\prime} d\theta^{\prime} d\phi^{\prime} =  (L_{\nu, *}/4 \pi^2 R_{*}^2) \pi R_{*}^2/r^2 = L_{\nu, *}/4 \pi r^2$ where the $\theta^{\prime} = 0$ direction points from the dust grain toward the center of the luminous object which subtends and angular radius of $R_{*}/r$.}
\begin{equation}
    f_{\nu, *} = \frac{L_{\nu, *}}{4 \pi r^2} \;\;\;\; \rm{erg} \; \rm{s}^{-1}\, \rm{cm}^{-2}\, \rm{Hz}^{-1}  \;\;.
\label{eq:IncidentFlux}
\end{equation}
An example is a dust grain in an optically thin debris disk orbiting a young star.
Substituting for the effective cross section for absorption, $\sigma_{\nu, abs} = A_{\rm{cross}} Q_{\nu,abs} =  \kappa_{\nu,abs} m_{\rm{grain}}$, in Equation~\ref{eq:dEdt_abs}, we find that
\begin{equation}
    \left(\frac{dE}{dt}\right)_{abs} = \int_{\nu = 0}^{\infty}  \frac{L_{\nu, *}}{4 \pi r^2} \, \kappa_{\nu,abs} \, m_{\rm{grain}} \, d\nu \;\;\;\; \rm{erg} \; \rm{s}^{-1} \;\;.
\label{eq:dEdt_abs2}
\end{equation}
It is important to note that $\kappa_{\nu, abs}$ is the mass dust opacity due to absorption (not extinction = absorption + scattering; see Equation \ref{eq:graineffcrosssection}).
We also note that this term should technically be $\kappa_{\nu, abs, grain}$ for a single dust grain and therefore also a function of the grain size.  However, if we use the $\kappa_{\nu, abs}$ averaged over the grain size and grain mass distributions (see Equation \ref{eq:intsizedistrbution}), then our result will apply to the average equilibrium dust temperatures of the population of grains at each radius.
If you instead want to calculate a single grain temperature, then replace $\kappa_{\nu, abs}$ with 
$\kappa_{\nu, abs, grain}$ in all equations in this section.

It is convenient to define the spectral luminosity-weighted mean opacity, 
\begin{eqnarray}
    \langle \kappa_{\nu,abs} \rangle_{L_{*}} & = & \frac{\int_{\nu = 0}^{\infty} L_{\nu, *} \kappa_{\nu, abs} d\nu}{\int_{\nu = 0}^{\infty} L_{\nu, *} d\nu}  \nonumber \\ 
    & = & \frac{\int_{\nu = 0}^{\infty} L_{\nu, *} \kappa_{\nu, abs} d\nu}{L_{*}}\;\;\;\;  \rm{cm^2} \; (\rm{g} \; \rm{of} \; \rm{dust})^{-1}  \;\; .
\label{eq:kappa_mean}
\end{eqnarray}
The energy absorption rate (Equation~\ref{eq:dEdt_abs2}) can be written as
\begin{equation}
    \left(\frac{dE}{dt}\right)_{abs} = \frac{L_{*}}{4 \pi r^2} \, m_{\rm{grain}} \, \langle \kappa_{\nu,abs} \rangle_{L_{*}} \;\;\;\; \rm{erg} \; \rm{s}^{-1} \;\; .
\label{eq:dEdt_abs3}
\end{equation}

The rate of emission from a dust grain can be calculated from the spectral luminosity of a dust grain assuming it is optically thin to its own emission (see Equation~\ref{eq:Lnu_mono_thin_isotherm} above)
\begin{eqnarray}
    L_{\nu, grain} & = & 4 \pi j_{\nu} V_{\rm{grain}} \nonumber \\ 
    L_{\nu, grain} & = & 4 \pi \alpha_{\nu} B_{\nu}(T_d) V_{\rm{grain}} \;\;\;\; \rm{erg} \; \rm{s}^{-1} \, \rm{Hz}^{-1} \;\; ,
\label{eq:Lnu_grain}
\end{eqnarray}
where $V_{\rm{grain}}$ is the volume of the dust grain.
We can write $\alpha_{\nu} = n_d \sigma_{\nu, abs}$, the product of the number density of grains, $n_d$ (cm$^{-3}$), and the effective cross-section of the dust grains for absorption, $\sigma_{\nu, abs} = m_{\rm{grain}} \kappa_{\nu,abs}$ (see Equation \ref{eq:kappagrain}).
Recognizing that the product of $n_d V_{grain} = 1$ for the volume of a single dust grain under consideration, then the total rate of energy emitted is given by
\begin{eqnarray}
    \left(\frac{dE}{dt}\right)_{emit} & = & \int_{\nu = 0}^{\infty} L_{\nu, grain} d\nu \nonumber \\
    & = & \int_{\nu = 0}^{\infty} 4 \pi m_{\rm{grain}} \kappa_{\nu,abs} B_{\nu}(T_d) d \nu \;\;\;\; \rm{erg} \; \rm{s}^{-1} \;\; .
\label{eq:dEdt_emit}
\end{eqnarray}
Following \cite{1993duun.book.....E}  and \cite{Draine2011}, it is convenient to define the Planck mean opacity at dust temperature $T_d$ as
\begin{equation}
    \langle \kappa_{\nu,abs} \rangle_{T_d} = \frac{\int_{\nu = 0}^{\infty} B_{\nu}(T_d) \kappa_{\nu,abs} d \nu}{ \int_{\nu = 0}^{\infty} B_{\nu}(T_d) d \nu} \;\;\;\;  \rm{cm^2} \; (\rm{g} \; \rm{of} \; \rm{dust})^{-1}  \;\; .
\label{eq:kappa_nu2}
\end{equation}
The total energy rate of emission (Equation~\ref{eq:dEdt_emit}) becomes
\begin{eqnarray}
    \left(\frac{dE}{dt}\right)_{emit} & = & 4 m_{\rm{grain}}  \langle \kappa_{\nu,abs} \rangle_{T_d} \int_{\nu = 0}^{\infty} \pi B_{\nu}(T_d) d \nu \nonumber \\
    & = & 4 m_{\rm{grain}} \, \langle \kappa_{\nu,abs} \rangle_{T_d} \,  \sigma_{\rm{SB}} T_d^4 \;\;\;\; \rm{erg} \; \rm{s}^{-1} \;\; ,
\label{eq:dEdt_emit2}
\end{eqnarray}
where $\sigma_{\rm{SB}} = 5.6704 \times 10^{-5}$ erg s$^{-1}$ cm$^{-2}$ K$^{-4}$ is the Stephan-Boltzmann constant.

The dust equilibrium temperature is calculated by equating the rate of energy absorption (Equation~\ref{eq:dEdt_abs3}) and the rate of energy emission (Equation~\ref{eq:dEdt_emit2}) using Equation~\ref{eq:EnergyBalance}
\begin{equation}
      \frac{L_{*}}{4 \pi r^2} \, \langle \kappa_{\nu,abs} \rangle_{L_{*}} = 4  \langle \kappa_{\nu,abs} \rangle_{T_d} \, \sigma_{\rm{SB}} T_d^4 \;\;\;\; \rm{erg} \; \rm{s}^{-1} \;\; .
\label{eq:EnergyBalance2}
\end{equation}
Notice that the mass of the dust grain has cancelled out.  This means that the weighted averages of $\kappa_{\nu}$ in Equation \ref{eq:EnergyBalance2} can be replaced with weighted averages of $Q_{\nu,abs}$ if you only have a tabulation of the latter.
There are two terms on the right hand side of Equation~\ref{eq:EnergyBalance2} that depend on $T_d$.  
We can make a reasonable analytic estimate of how $\langle \kappa_{\nu, abs} \rangle_{T_d}$ depends on $T_d$ by assuming that the dust opacity varies with frequency as a single power-law with exponent $\beta$ over all frequencies of interest for emission at all radii by modifying Equation \ref{eq:kappa_nu3},
\begin{equation}
    \kappa_{\nu,abs} = \kappa_{\nu_0} \left( \frac{\nu}{\nu_0} \right)^{\beta} \;\;\;\;  \rm{cm^2} \; (\rm{g} \; \rm{of} \; \rm{dust})^{-1}  \;\; ,
\label{eq:kappa_nu4}
\end{equation}
where $\kappa_{\nu_0}$ is the opacity at a nominal frequency $\nu_0$.
In this case, $\langle \kappa_{\nu,abs} \rangle_{T_d}$ (Equation~\ref{eq:kappa_nu2}) can be calculated exactly as
\begin{equation}
   \langle \kappa_{\nu,abs} \rangle_{T_d} = \frac{ \frac{2 h \kappa_{\nu_0}}{c^2 \nu_0^{\beta}} \int_{\nu=0}^{\infty} \frac{\nu^{(3+\beta)}}{\exp{(h\nu/kT_d)} - 1} \, d\nu }{\frac{\sigma_{\rm{SB}} T_d^4}{\pi}} \;\;\;\;  \rm{cm^2} \; (\rm{g} \; \rm{of} \; \rm{dust})^{-1}  \;\;,
\label{eq:kappa_nu6}
\end{equation}
where we have collected together terms that depend on frequency in the numerator and we have integrated the Planck function in the denominator.
The integral in the numerator is a frequency moment of the Planck function which has the solution\footnote{In general, $\int_0^{\infty} \frac{w^s}{\exp{(w)} - 1} dw = \Gamma(s+1) \zeta(s+1)$ for $s > 0$.  We transform our integral into this form with the variable transformations $s = 3+\beta$, $w = h\nu/kT_d$, $\nu = (kT_d/h)w$, and $d\nu = (kT_d/h) dw$ such that the integral in Equation \ref{eq:PlanckIntegralSoln} becomes   $(kT_d/h)^{s+1}\int_0^{\infty} \frac{w^s}{\exp{(w)} - 1} dw$.  Multiplying the integrand by 1 in the form of ($e^{-w}/e^{-w}$) we find for the integral that  $\int_0^{\infty}  w^s e^{-w} \frac{1}{1 - \exp{(-w)}} dw$.  The fraction inside the integral can be re-written as an infinite geometric series, $\frac{1}{1 - \exp{(-w)}} = \sum_{n=0}^{n=\infty} e^{-nw} $.  Bringing the coefficient of $e^{-w}$ from the integrand into the sum increases the starting value by 1 so that we have $\int_0^{\infty}  w^s \sum_{n=1}^{n=\infty} e^{-nw} dw$. If $s > 0$, then each term in the sum results in a convergent integral, therefore we can reverse the order of the sum and integration and look at each individual term in the sum as the generic integral,  $\int_0^{\infty}  w^s e^{-nw} dw$.  Making the variable transformation $u =nw$ such that $w = u/n$ and $dw = du/n$, we find that $n^{-(s+1)} \int_0^{\infty}  u^s e^{-u} du = n^{-(s+1)} \Gamma(s+1)$ since the integral is, by definition, a Gamma function. Plugging the result from the integral back into the sum we find that (pulling all constants outside the sum), 
$(kT_d/h)^{s+1} \Gamma(s+1) \sum_{n=1}^{n=\infty} n^{-(s+1)}$.  The sum is a Riemann Zeta function, $\sum_{n=1}^{n=\infty} n^{-(s+1)} = \zeta(s+1)$.}
\begin{equation}
    \int_0^{\infty} \frac{\nu^{(3+\beta)}}{\exp(h
\nu/kT_d) - 1} d\nu = \left( \frac{kT_d}{h} \right)^{(4+\beta)} \Gamma(4+\beta) \zeta(4+\beta) \;\;\;\; ,
\label{eq:PlanckIntegralSoln}
\end{equation}
where $\Gamma$ and $\zeta$ are the Gamma Function and Riemann Zeta Function respectively and will just be numbers.
The limits of integration are formally $\nu \in (0,\infty)$ but, in practice, for cold dust grain temperatures, the exponential decay of the Planck function on the Wien-side of the SED (left of the peak in Figure \ref{fig:SEDBetaFig}) makes those frequencies negligible when calculating the integral such that the single power-law approximation to $\kappa_{\nu,abs}$ is still approximately valid (see Section \ref{OpacitySection} for a discussion of the validity of the single-power law assumption).
Substituting Equation~\ref{eq:PlanckIntegralSoln} into Equation~\ref{eq:kappa_nu6} gives us the final expression for $\langle \kappa_{\nu,abs} \rangle_{T_d}$,
\begin{equation}
    \langle \kappa_{\nu,abs} \rangle_{T_d} = \frac{2 \pi \Gamma(4 + \beta) \zeta(4 + \beta) \kappa_{\nu_0} (k T_d)^{4+\beta}}{c^2 \nu_0^{\beta} h^{3 + \beta} \sigma_{\rm{SB}} T_d^4} \;\;\;\;  \rm{cm^2} \; (\rm{g} \; \rm{of} \; \rm{dust})^{-1} \;\;.
\label{eq:kappa_nu5}
\end{equation}
Ultimately, $\langle \kappa_{\nu,abs} \rangle_{T_d} \propto T_d^{\beta}$.
From Equation~\ref{eq:EnergyBalance2}, we see that the right hand side has a total dependence of $T_d^{4+\beta}$.
Since no other terms depend on $T_d$ in Equation~\ref{eq:EnergyBalance2}, then we can substitute Equation~\ref{eq:kappa_nu5} into Equation~\ref{eq:EnergyBalance2} and solve for the equilibrium dust temperature,
\begin{equation}
    T_d = \frac{h \nu_0}{k} \left( \frac{\langle \kappa_{\nu,abs} \rangle_{L_{*}} \, c^2 }{32 \pi^2 \Gamma(4 + \beta) \zeta(4 + \beta) \kappa_{\nu_0} h \nu_0^{4} }  \, \frac{L_{*}}{r^2} \right)^{\frac{1}{(4+\beta)}} \;\;\;\; K \;\;.
\label{eq:TdLcen}
\end{equation}

For optically thin dust absorption from a central luminous source, Equation \ref{eq:TdLcen} indicates that $T_d \propto (L_{*} / r^2)^{1/(4+\beta)}$. 
For a large dust grain with $\beta = 0$, this proportionality reduces to  $T_d \propto (L_{*}^{1/4} r^{-1/2})$.
This is the same functional dependence on luminosity and radius for the equilibrium temperature of a fast rotating blackbody planet.
For dust grains that have coagulated and have $\beta = 1$, then $T_d \propto (L_{*}^{1/5} r^{-2/5})$.
For ISM-like dust grain with $\beta = 2$, then $T_d \propto (L_{*}^{1/6} r^{-1/3})$.
As the dust opacity index $\beta$ increases, the radial dependence of the equilibrium dust temperature becomes steeper and the luminosity dependence becomes shallower.
The dependence of grain temperature on luminosity is relatively weak.
As an example, for dust with $\beta = 2$, the $L_{*}^{1/6}$ dependence means that, at a fixed radius from the source, the luminosity would have to increase by a factor of 64 to increase the dust equilibrium temperature by a factor of 2.
The derivation for the equilibrium dust temperature of dust exposed to the interstellar radiation field (ISRF) is in Appendix \ref{AppendixISRF} where a similar result is obtained with $T_d \propto u_{\rm{ISRF}}^{1/(4+\beta)}$ (see Equation \ref{eq:TdISRF}) for the total energy density of the ISRF, $u_{\rm{ISRF}}$ (erg cm$^{-3}$).

All of these relations were derived in the limit that the absorption of short wavelength radiation by the dust grain was optically thin.
In the limit that this absorption is optically thick over some range of radii, a radiative transfer calculation must be performed to determine $T_d(\vec{r})$.
We discuss these calculations in the next section.

\subsubsection{Numerical Calculations with Radiative Transfer Codes}\label{sec:CodesforRadTrans}

While the analytical techniques for SED modeling developed above can be useful for interpreting the observations of dust continuum emission, there are significant degeneracies and severe assumptions (such as isothermality or constant dust opacities). 
Using radiative transfer modeling with publically-available codes, that have been benchmarked against standard problems, may be a better option. 
Furthermore, numerical calculations on a computer can include processes such as scattering and polarization that require more than one dimension. 
In this section, we briefly describe Monte Carlo radiative transfer techniques in 3D.

There are many 3D dust continuum radiative transfer codes available in the literature.
We will discuss aspects of three of the currently most popular codes: RADMC-3D \citep{2012ascl.soft02015D}\footnote{\url{https://www.ita.uni-heidelberg.de/~dullemond/software/radmc-3d/index.php}}, HYPERION \citep{2011A&A...536A..79R}\footnote{\url{http://docs.hyperion-rt.org/en/stable/index.html}} and MCFOST  \citep{2006A&A...459..797P,2009A&A...498..967P}\footnote{\url{https://github.com/cpinte/mcfost}}.
The calculations can be performed in 1, 2, or 3 dimensions in a variety of coordinate systems.
These systems are divided into individual cells which can vary in volume.
When doing radiative transfer in multiple dimensions, we introduce a new radiative transfer quantity, the mean intensity, which is measured at a position $\vec{r}$ in the problem by calculating the average integrated intensity coming from all directions,
\begin{equation}
    J_{\nu}(\vec{r}) = \frac{1}{4\pi} \oint_{4\pi \, \rm{ster}} I_{\nu}(\vec{r},\hat{r}^{\prime}) d \Omega  \;\;\;\; \rm{erg}\, \rm{s}^{-1}\, \rm{cm}^{-2}\, \rm{ster}^{-1}\, \rm{Hz}^{-1} \;\; ,
\label{eq:MeanInt}
\end{equation}
(see Figure \ref{fig:Geo1}).
Ultimately, the energy balance in the $i^{\rm{th}}$ cell is found by solving Equation \ref{eq:EnergyBalance} with Equation \ref{eq:dEdt_emit} and revising Equation \ref{eq:dEdt_abs2} to use the mean intensity (canceling common factors of $4\pi$ and average grain mass on both sides of the equation)
\begin{eqnarray}
\left(\frac{dE}{dt}\right)_{abs} & = & \left(\frac{dE}{dt}\right)_{emit} \nonumber \\
    \int_{0}^{\infty} \kappa_{\nu,abs,i} \,J_{\nu,i}\, d\nu \; & = & \;\int_{0}^{\infty} \kappa_{\nu,abs,i}\, B_{\nu}(T_{d,i})\, d\nu \;\;\;\; \rm{erg} \; \rm{s}^{-1} \;\; ,
\label{eq:3dradtran}
\end{eqnarray}
(e.g., \citealt{2013ARA&A..51...63S}).
Calculations are based on a Monte Carlo radiative transfer technique developed in \cite{2001ApJ...554..615B} with modifications for calculation of the mean intensity from \cite{1999A&A...344..282L}. 
Instead of solving the equations of radiative transfer directly, Monte Carlo radiative transfer techniques sample probability distributions to propagate photon packages through a grid of cells in the simulation.
The user inputs the sources of luminosity, the density grid, and the dust grain properties.
The code converts the total luminosity into a series of photon packages that represent multiple photons.
A typical Monte Carlo calculation may have as many as $10^4 - 10^6$ photon packages.
Individual photon packages are followed through the calculation and as a photon package traverses through cells in the coordinate grid, it may be absorbed or scattered given probabilities.
If it is scattered, the photon package changes direction and its polarization.
If it is absorbed, then a new photon package is emitted in a new direction with a new wavelength.
As the photon package transverses the grid, the equilibrium dust temperature is calculated in each cell through which the photon package traverses.
This simulation is continued, one photon package at a time, until all photon packages random walk through the grid and escape.
The codes implement modified random walk techniques to deal with very optically thick regions where the photon packages can become stuck with many absorptions and scatterings (see \citealt{2009A&A...497..155M,2010A&A...520A..70R}).
The codes may also be run in parallel on multiple nodes which permits the calculation of large grids ($> 10^6$) of radiative transfer models with variable input parameters (e.g. \citealt{2019AJ....157..144B,2022ApJ...929...76S,2023MNRAS.521.4579S}).

After all photon packages have escaped the grid, the equilibrium dust temperatures in each cell, along with the input density and dust parameters, are used to calculate SEDs and on-sky intensity images with ray tracing techniques.
The calculated outputs of RADMC-3D, HYPERION, and MCFOST are generally similar.
For instance, modeling of the edge-on protoplanetary disk, HH 48 NE, with RADMC-3D and HYPERION find that the model SEDs agree over all wavelengths to better than $20$\% with a standard deviation of only $5$\% (see Appendix A of \citealt{2023A&A...677A..18S}).
In another example, benchmarking of different hydrodynamic simulations of a protoplanetary disk with an embedded planet find that radiative transfer calculation performed with RADMC-3D and MCFOST agree to within 3\% \citep{2025ApJ...984L..12B}.

\begin{figure}[h]
\includegraphics[scale=0.48]{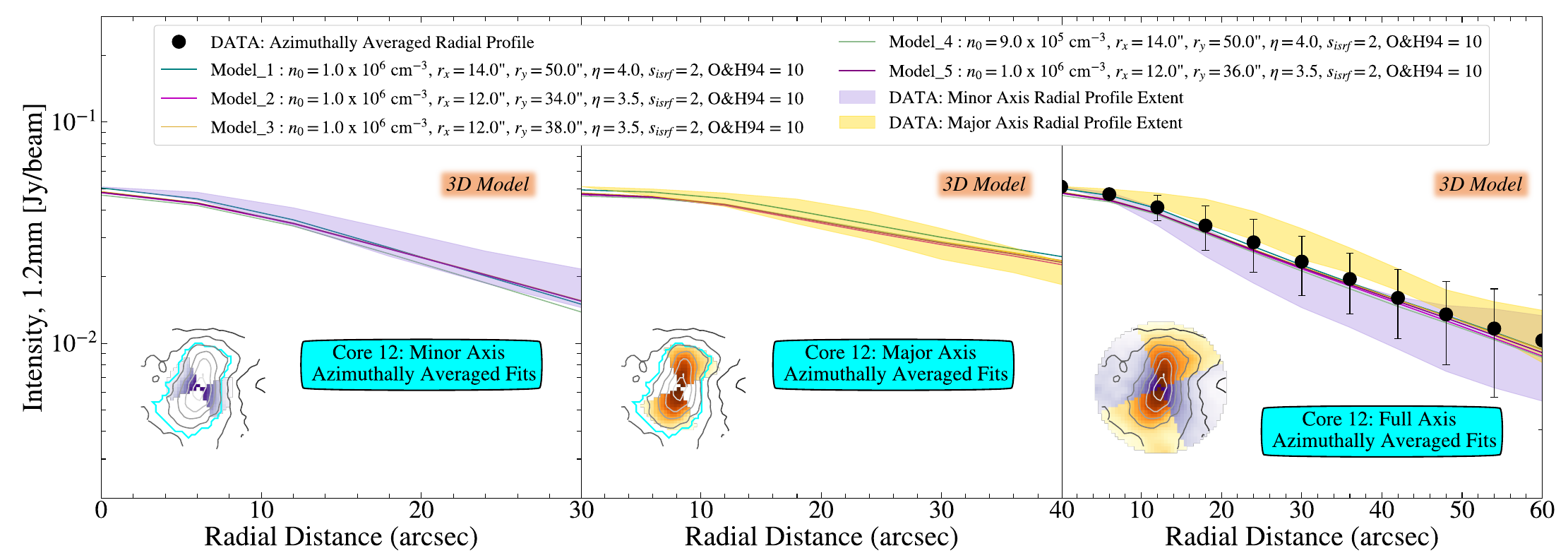}
\caption{The five best-fitted Plummer spheroid models for the starless core Seo 12 in \cite{2023MNRAS.521.4579S} are shown above (see paper for model details).  The 1.2mm continuum image of Seo 12, a starless core in the Taurus Molecular Cloud, is shown in the lower left of each panel.  The shaded regions show the sectors over which the intensity profile is azimuthally averaged.  The left panel shows the azimuthally-averaged profile centered on the minor axis of the core (purple shaded region indicates $\pm\,1\,\sigma$ uncertainty).  The middle panel shows the azimuthally-averaged profile centered on the major axis of the core (yellow shaded region).  The subset of 3D radiative transfer models that fit within both the purple and yellow shaded regions and that fit the $1.2$ mm peak intensity are plotted as solid line with different colors (see key at top of figure for model properties).  These models also fit the full azimuthally-averaged data (right panel; black points).  Note that the shaded purple and yellow regions mostly do not overlap.}
\label{fig:3DRadTran}
\end{figure}

If the source under study is resolved, such as a dense starless or protostellar core, then simultaneously analyzing intensity profiles and the SED can help break SED model degeneracies.
The general technique is to compare a grid of model on-sky intensity distributions with the observed on-sky intensity distribution, as well as a grid of model SEDs with the observed SED, using either $\chi^2$ minimization or Bayesian inference techniques to select the subset of best-fitted models. 
Interferometric observations may also be compared by computing the Fourier transform of the model intensity distribution to produce visibilities (see Appendix \ref{AppendixVisibilities}; e.g.,  \citealt{2009ApJ...700.1502A,2012ApJ...755...23E,2016A&A...588A.143S}). 
Recall from Equation \ref{eq:Inu_nobg_thin} that the monochromatic specific intensity dends on the product of $\rho_d(\vec{r})$, $T_d(\vec{r})$, and $\kappa_{\nu}(\vec{r})$ along the line-of-sight.
For the modeling of protostellar core envelopes, the shape of the intensity profile, $I_{\nu,src}(\theta,\phi)$, is more sensitive to the shape of the density structure than the SED.
Essentially the density changes of several orders of magnitude sampled across the intensity spatial profile help break the degeneracy between $\rho_d(\vec{r})$ and $T_d(\vec{r})$ \footnote{Most radiative transfer modeling of dense cores assumes $\kappa_{\nu}$ is spatially constant and therefore does not properly break the degeneracy between $\rho_d(\vec{r})$, $T_d(\vec{r})$, and $\kappa_{\nu}(\vec{r})$}.
For 1D radiative transfer, azimuthally-averaged profiles of the observed intensity distributions are compared with the model intensity profile (\ie\   \citealt{2002ApJ...575..337S,2002A&A...389..908J}).
For 3D radiative transfer, the comparison of models with observations requires techniques to capture the non-azimuthal aspect of the observations.
As example of azimuthal-averaging over sectors centered on a starless core's major and minor axes from \cite{2023MNRAS.521.4579S} is shown in Figure \ref{fig:3DRadTran}.
Ultimately, the techniques for 3D model selection will depend on the geometry of the problem and the details of the observations.

\section{Summary}

In this tutorial, we have summarized the fundamental equations governing thermal dust emission with radiative transfer in one dimension. 
The most important equations are summarized in Table \ref{tab:eqsummary} with links in the table back to the sections where the equations are derived.
Where practical, we have started from first principles and shown the limits and assumptions needed to derive the equations.
Many commonly used equations in the literature have severe assumptions such as isothermality or spatially-constant dust opacities which may not be appropriate in real sources (caveat emptor!).
A good mantra embraced by the authors of this tutorial is ``when in doubt, model it out" using one of the benchmarked, publicly-available multi-dimensional radiative transfer codes.

\begin{acknowledgments}
Y. Shirley is extremely grateful to the National Research Council of Canada's Herzberg Astronomy and Astrophysics Research Centre for hosting my sabbatical during which the majority of this tutorial was written.
Y. Shirley was partially supported by an Astronomy and Astrophysics Grant AST-2205474 from the National Science Foundation.
Y. Shirley thanks Adam Block for taking and providing the optical image of the central Taurus Molecular Cloud.
Y. Shirley thanks Ayushi Singh for sharing the H$_2$ column density map of the Taurus Molecular cloud and thanks Loren Anderson for sharing the data from their simulation of the $\beta-T_d$ degeneracy in SED fitting. 
Y. Shirley also thanks Bruce Draine, George Rieke, Neal Evans, Sarah Sadavoy, Samuel Fielder, Breanna Crompvoets, and Lucille Steffes for discussions that improved this paper.
This research has made use of the NASA/IPAC Infrared Science Archive, which is funded by the National Aeronautics and Space Administration and operated by the California Institute of Technology.
This research has made use of the SVO Filter Profile Service ``Carlos Rodrigo", funded by MCIN/AEI/10.13039/501100011033/ through grant PID2020-112949GB-I00.
The cartoon image of the radio telescope in Figures \ref{fig:Geo1}, \ref{fig:Geo2}, and \ref{fig:GeoSph} was modified from an image obtained online that has a free simplified Pixabay license.
\end{acknowledgments}

\vspace{5mm}
\facilities{ALMA}


\software{astropy \citep{Astropy2013,Astropy2018}
          }



\appendix

\section{Summary of Useful Equations}\label{AppendixEquationSummary}

We summarize the most important equations derived in this article in Table \ref{tab:eqsummary}.
The limits of no scattering and $\frac{dI_{\nu}(\vec{r},\hat{z})}{dt} = 0$ are explicit to every equation derived in this tutorial. 
We also ignore the effects of telescope beam patterns and filters discussed in Section \ref{Caveats} for all equations that use flux density in Table \ref{tab:eqsummary}.  
The terminology ``Constant $T_d$, $\kappa_{\nu}$," etc. means that the variable is spatially constant. 
The variable in question, however, may still be a function of other quantities such as frequency.

{\catcode`\&=11
\gdef\Hocuk2017AandA{\cite{2017A&A...604A..58H}}}

\begin{longrotatetable}
\begin{deluxetable*}{llll}
\tablecaption{Summary of Useful Equations\label{tab:eqsummary}}
\tablewidth{0pt}
\tablehead{
\colhead{Equation} &
\colhead{Eq. \#(s)} &
\colhead{CGS Units} &
\colhead{Assumptions \& Notes} }
\startdata
\multicolumn{4}{c}{\textbf{Monochromatic Specific Intensity}} \\
\hline
$I_{\nu,\rm{src}} = \int_{\rm{A}} B_{\nu}[T_d(\vec{r})] \kappa_{\nu}(\vec{r}) \rho_d(\vec{r}) dz $ & \ref{eq:Inu_nobg_thin} & $\rm{erg}\, \rm{s}^{-1}\, \rm{cm}^{-2}\, \rm{ster}^{-1}\, \rm{Hz}^{-1}$ & $\tau_{\nu} \ll 1$\\
$I_{\nu,\rm{src}}(\theta) = 2 \int_{r = D\theta}^R B_{\nu}[T_d(r)] \kappa_{\nu}(r) \rho_d(r) \frac{r}{\sqrt{r^2 - D^2 \theta^2}} dr $ & \ref{eq:Inu_Abel} & $\rm{erg}\, \rm{s}^{-1}\, \rm{cm}^{-2}\, \rm{ster}^{-1}\, \rm{Hz}^{-1} $ & $\tau_{\nu} \ll 1$; Spherical geometry \\
$I_{\nu,\rm{src}} = B_{\nu}(T_d) \kappa_{\nu} \Sigma_d$ & \ref{eq:Inu_nobg_thin_iso_constk} & $\rm{erg}\, \rm{s}^{-1}\, \rm{cm}^{-2}\, \rm{ster}^{-1}\, \rm{Hz}^{-1}$ & Const $T_d$, $\kappa_{\nu}$; $\tau_{\nu} \ll 1$\\
$I_{\nu,obs}  =  \frac{I_{\nu_e}}{(1 + z_{\rm{red}})^3}$ & \ref{eq:Inucubed} & $ \rm{erg}\, \rm{s}^{-1}\, \rm{cm}^{-2}\, \rm{ster}^{-1}\, \rm{Hz}^{-1}$ & Redshift = $z_{\rm{red}}$; o = observed frame; e = emitted frame \\
\hline
\multicolumn{4}{c}{\textbf{Optical Depth}} \\
\hline
$\tau_{\nu} = \int_{\rm{A}} \kappa_{\nu}(\vec{r}) \rho_d(\vec{r}) dz$ & \ref{eq:taunu}, \ref{eq:taunu_limit} & (unitless) & \nodata \\
$\overline{\tau_{\nu}} = \kappa_{\nu} \overline{\Sigma_d} = \kappa_{\nu} \frac{\mu_g m_{\rm{H}}}{R_{gd}}  \overline{N_g}$ & \ref{eq:angleavgtau}, \ref{eq:taukappasigma}, \ref{eq:tauNg} & (unitless) & Const $\kappa_{\nu}$, $\mu_g$, $R_{gd}$; $\tau_{\nu} \ll 1$ \\
\hline
\multicolumn{4}{c}{\textbf{Solid Angle}} \\
\hline
$\Omega_{ap} = \int_{\phi = 0}^{2\pi} \int_{\theta=0}^{\theta_{ap}} \theta d\theta d\phi = \frac{A_{\rm{ap}}}{D^2}$ & \ref{eq:ApertureSolidAngle}, \ref{eq:solidang} & ster & $\theta_{ap}$ small;  See Section \ref{Caveats} for examples \\
$\Omega(z_{\rm{red}}) = \frac{A_{\rm{src}} (1 + z_{\rm{red}} )^4}{D_L^2}$ & \ref{eq:SolidAng_redshift} & ster & Valid for any redshift, $z_{\rm{red}}$ \\
\hline
\multicolumn{4}{c}{\textbf{Flux Density}} \\
\hline
$F_{\nu,obs} = [I_{\nu,bg}e^{-\overline{\tau_{\nu}}} + B_{\nu}(T_d)(1 - e^{-\overline{\tau_{\nu}}}) ]\Omega_{ap}$ & \ref{eq:FluxSEDwithbg} & $\rm{erg}\, \rm{s}^{-1}\, \rm{cm}^{-2}\, \rm{Hz}^{-1}$ & Const $T_d$, $I_{\nu,bg}$  \\
$F_{\nu,obs} =  B_{\nu}(T_d) \left[ 1 - e^{-\kappa_0 \left( \frac{\nu}{\nu_0} \right)^{\beta}\overline{\Sigma_d}}  \right] \Omega_{ap}$ & \ref{eq:sedtaupow} & $ \rm{erg}\, \rm{s}^{-1}\, \rm{cm}^{-2}\, \rm{Hz}^{-1}$ & Const $T_d$, $\kappa_{\nu}$; $I_{\nu,bg} = 0$ ; Valid when $\kappa_{\nu} \propto \nu^{\beta}$ \\
$\Delta F_{\nu,obs} =  F_{\nu,obs} - \mu_{\rm{lens}} \frac{B_{\nu_e}(T_{\rm{CMB}_e})\Omega(z_{\rm{red}})}{(1 + z_{\rm{red}})^3}$ & \ref{eq:DelFnuz} &  $\rm{erg}\, \rm{s}^{-1}\, \rm{cm}^{-2}\, \rm{Hz}^{-1} $ & Redshift = $z_{\rm{red}}$; o = observed frame; e = emitted frame \\
\hline
\multicolumn{4}{c}{\textbf{Dust Mass Surface Density}} \\
\hline
$\overline{\Sigma_d}\Omega_{ap} =   \int_{\phi = 0}^{2\pi} \int_{\theta=0}^{\theta_{ap}} \Sigma_d(\theta,\phi) \theta d\theta d\phi$ & \ref{eq:MassSurfDenAvg} &  $ \rm{g} \; \rm{of} \; \rm{dust}\; \rm{cm}^{-2} $ & $\theta_{ap}$ small \\  
$\overline{\Sigma_d} = \frac{F_{\nu} - I_{\nu,bg}\Omega_{ap}}{[B_{\nu}(T_d) - I_{\nu,bg}] \kappa_{\nu} \Omega_{ap}}$ & \ref{eq:MassSurfDenAvg_final}, \ref{eq:DeltaF_Ibg} & $\rm{g} \; \rm{of} \; \rm{dust}\; \rm{cm}^{-2}$  & Const $T_d$, $\kappa_{\nu}$, $I_{\nu,bg}$ ;  $\tau_{\nu} \ll 1$ \\
$\overline{\Sigma_d} = \frac{F_{\nu}}{B_{\nu}(T_d) \kappa_{\nu} \Omega_{ap}}$ & \ref{eq:MassSurfDenAvg_nobg_final} &  $ \rm{g} \; \rm{of} \; \rm{dust}\; \rm{cm}^{-2}$ & Const $T_d$, $\kappa_{\nu}$; $I_{\nu,bg} = 0$;  $\tau_{\nu} \ll 1$\\
\hline
\multicolumn{4}{c}{\textbf{Gas Mass Surface Density}} \\
\hline
$\Sigma_{g} = R_{gd} \Sigma_d$ , $\overline{\Sigma_{g}} = R_{gd} \overline{\Sigma_d}$ & \ref{eq:sigmag} & $ \rm{g} \; \rm{of} \; \rm{gas} \; \rm{cm^{-2}}$ & Const $R_{gd}$;  $R_{gd} \approx 100$ in Milky Way ISM\\
$\Sigma_g^{thin} < 4.0 \;\; \left( \frac{\tau_{\nu,\rm{thick}}}{0.2} \right) \left( \frac{R_{gd}}{100} \right) \left( \frac{5 \, \rm{cm}^2 \, \rm{g}^{-1} }{\kappa_{\nu}} \right)$ & \ref{eq:NH2thin} &  $\rm{g} \; \rm{of} \; \rm{gas}\; \rm{cm}^{-2} $ & Limit for $\Sigma_g$ where $\tau_{\nu} < 0.2$ and is negligible \\
\hline
\multicolumn{4}{c}{\textbf{Gas Column Density}} \\
\hline
$\overline{N_{g}} = \frac{R_{gd} (F_{\nu,obs} - I_{\nu,bg}\Omega_{ap})}{\mu_{g} m_{\rm{H}} [B_{\nu}(T_d) - I_{\nu,bg}] \kappa_{\nu} \Omega_{ap}}$ & \ref{eq:GasColumnDen_final}, \ref{eq:DeltaF_Ibg} & number of gas particles $\rm{cm}^{-2}$ & Const $T_d$, $\kappa_{\nu}$, $I_{\nu,bg}$, $R_{gd}$, $\mu_{g}$ ;  $\tau_{\nu} \ll 1$ \\
$\overline{N_{g}} = \frac{R_{gd} F_{\nu,src}}{\mu_{g} m_{\rm{H}} B_{\nu}(T_d) \kappa_{\nu} \Omega_{ap}}$ & \ref{eq:GasColumnDen_nobg_final} & number of gas particles $\rm{cm}^{-2}$ & Const $T_d$, $\kappa_{\nu}$, $R_{gd}$, $\mu_{g}$ ; $I_{\nu,bg} = 0$; $\tau_{\nu} \ll 1$ \\
$N_{\rm{H}_2}^{thin}  <  8.51 \times 10^{23} \;\; \left( \frac{\tau_{\nu,\rm{thick}}}{0.2} \right) \left( \frac{R_{gd}}{100} \right)  \left( \frac{5 \, \rm{cm}^2 \, \rm{g}^{-1} }{\kappa_{\nu}} \right)$ & \ref{eq:NH2thin}& $\rm{H}_2 \; \rm{molecules} \; \rm{cm}^{-2}$ & Limit for $N_{\rm{H}_2}$ where $\tau_{\nu} < 0.2$ and is negligible\\ 
\hline
\multicolumn{4}{c}{\textbf{Mean Atomic/Molecular Weight}} \\
\hline
$\mu_g = 1.404 \;\; \rm{if} \; N_g = N_{\rm{H}}$ & \ref{eq:muhfinal} & (unitless) & Proto-solar metallicity - see Section \ref{sec:RadTrans} for details\\ 
$\mu_g = 2.809 \;\; \rm{if}\; N_{g} = N_{\rm{H}_2}$ & \ref{eq:muh2final} & (unitless) & Proto-solar metallicity - see Section \ref{sec:RadTrans} for details\\ 
$\mu_g = 2.351 \; \;\rm{if} \; N_{g} = N_{\rm{all} \, \rm{gas} \, \rm{particles}}$ & \ref{eq:mupfinal} & (unitless) & Proto-solar metallicity - see Section \ref{sec:RadTrans} for details\\ 
\hline
\multicolumn{4}{c}{\textbf{Dust Mass}} \\
\hline
$M_d = \frac{(F_{\nu,obs} - I_{\nu,bg}\Omega_{ap}) D^2}{[B_{\nu}(T_d) - I_{\nu,bg}] \kappa_{\nu}}$ & \ref{eq:Md_thin_final}, \ref{eq:DeltaF_Ibg} & $\rm{g} \; \rm{of} \; \rm{dust}$ &  Const $T_d$, $\kappa_{\nu}$, $I_{\nu,bg}$;  $\tau_{\nu} \ll 1$\\
$M_d = \frac{F_{\nu,src} D^2}{B_{\nu}(T_d) \kappa_{\nu}}$ & \ref{eq:Md_thin_nobg_final} & $ \rm{g} \; \rm{of} \; \rm{dust} $ & Const $T_d$, $\kappa_{\nu}$; $I_{\nu,bg} = 0$;  $\tau_{\nu} \ll 1$\\
$M_d = \frac{1}{\mu_{\rm{lens}} (1 + z_{\rm{red}})} \frac{\Delta F_{\nu_o} D_L^2}{[B_{\nu_e}(T_{d_e}) -  B_{\nu_e}(T_{\rm{CMB}_e})] \kappa_{\nu_e}}$ & \ref{eq:Md_thin_red_cmb} & $ \rm{g} \; \rm{of} \; \rm{dust} $ & Const $T_d$, $\kappa_{\nu}$, $I_{\nu,bg}$;  $\tau_{\nu} \ll 1$; Use when $z_{\rm{red}}$ important\\
\hline
\multicolumn{4}{c}{\textbf{Total (Gas+Dust) Mass}} \\
\hline
$M_{\rm{tot}} = (R_{gd} + 1)M_d$ & \ref{eq:Md_thin_final_tot} & g &  Const $T_d$, $\kappa_{\nu}$, $I_{\nu,bg}$, $R_{gd}$;  $\tau_{\nu} \ll 1$\\
\hline
\multicolumn{4}{c}{\textbf{Monochromatic Luminosity}} \\
\hline
$L_{\nu} =  4 \pi  B_{\nu}(T_d) \kappa_{\nu} M_d$ & \ref{eq:Lnu_mono_thin_isotherm} & $\rm{erg}\, \rm{s}^{-1}\, \rm{Hz}^{-1}$ & Const $T_d$, $\kappa_{\nu}$; $\tau_{\nu} \ll 1$ \\
\hline
\multicolumn{4}{c}{\textbf{SED Peak Frequency}} \\
\hline
$\nu_{\rm{pk}} = \frac{k T_d}{h} \left[ (3 + \beta\epsilon) + W_0\left(- \frac{3 + \beta\epsilon}{e^{3 + \beta\epsilon}}\right) \right]$ \;, $\epsilon = \frac{\overline{\tau_{\nu}}}{e^{\overline{\tau_{\nu}}} - 1}$ & \ref{eq:WienLawbeta} & $ \rm{Hz}$ & Const $T_d$, $\kappa_{\nu}$; Valid when $\kappa_{\nu} \propto \nu^{\beta}$; $W_0$ Lambert W funct. \\ 
\hline
\multicolumn{4}{c}{\textbf{Spectral Index}} \\
\hline
$\alpha_s  = \frac{\partial \ln{F_{\nu}}}{\partial \ln{\nu}} =  3 + \beta\frac{\overline{\tau_{\nu}}}{e^{\overline{\tau_{\nu}}} - 1} - \frac{\frac{h\nu}{kT_d}}{1 - e^{-h\nu/kT_d}}$ & \ref{eq:sedspecindex} & (unitless) & Const $T_d$, $\kappa_{\nu}$;  $I_{\nu,bg} = 0$; Valid when $\kappa_{\nu} \propto \nu^{\beta}$ \\
$\alpha_{s,\nu_1/\nu_2} = \frac{\ln \left( \frac{F_{\nu_1}}{F_{\nu_2}} \right) }{\ln \left( \frac{\nu_1}{\nu_2} \right)} = 3 + \beta - \frac{\ln \left(\frac{e^{h\nu_1/kT_d} - 1}{e^{h\nu_2/kT_d} - 1}\right)}{\ln \left( \frac{\nu_1}{\nu_2} \right)}$ &\ref{eq:specind2freq}, \ref{eq:gammaempirical} &(unitless) & Const $T_d$, $\kappa_{\nu}$;  $I_{\nu,bg} = 0$; Valid when $\kappa_{\nu} \propto \nu^{\beta}$ \\
\hline
\multicolumn{4}{c}{\textbf{Equilibrium Dust Temperature}} \\
\hline
$T_d = \frac{h\nu}{k} \left[ \ln \left( 1 + \frac{2h\nu^3}{c^2 I_{\nu,src}}\right) \right]^{-1} = \frac{h\nu}{k} \left[ \ln \left( 1 + \frac{2h\nu^3 \Omega_{ap}}{c^2 F_{\nu,src}}\right) \right]^{-1}$ & \ref{eq:TdthickInu}, \ref{eq:FluxSEDThick} & K & Const $T_d$, $\kappa_{\nu}$;  $I_{\nu,bg} = 0$; $\tau_{\nu} \gg 1$ \\
$T_d = \frac{h \nu_0}{k} \left( \frac{\langle \kappa_{\nu,abs} \rangle_{L_{*}} \, c^2 }{32 \pi^2 \Gamma(4 + \beta) \zeta(4 + \beta) \kappa_{\nu_0} h \nu_0^{4} }  \, \frac{L_{*}}{r^2} \right)^{\frac{1}{(4+\beta)}}$ & \ref{eq:TdLcen} &  K & Const $\kappa_{\nu}$ ; $\kappa_{\nu} \propto \nu^{\beta}$ at freq $\nu$; $\tau_{\nu} \ll 1$; Grain at $r$ from $L_{*}$ \\
$T_d = \frac{h \nu_0}{k} \left( \frac{\langle \kappa_{\nu,abs} \rangle_{u} \, c^3 }{8 \pi \Gamma(4 + \beta) \zeta(4 + \beta) \kappa_{\nu_0} h \nu_0^{4} } \; u_{\rm{ISRF}}  \right)^{\frac{1}{(4+\beta)}} $ & \ref{eq:TdISRF} &  K & Const $\kappa_{\nu}$ ; $\kappa_{\nu} \propto \nu^{\beta}$ at freq $\nu$; $\tau_{\nu} \ll 1$; Exposed to ISRF \\
$T_d \approx [11 + 5.7\tanh (0.61 - \log_{10}(A_{\rm{V}}))] \chi_{\rm{UV}}^{\frac{1}{5.9}}$ & \ref{eq:TdAvEmp} & K & Derived in \Hocuk2017AandA 
\\
$T_{d_e} = \left[ T_{d_o}^{(4+\beta)}  +  T_{\rm{CMB}_o}^{(4+\beta)} [ (1 + z_{\rm{red}})^{(4+\beta)} - 1]   \right]^{\frac{1}{(4+\beta)}}$ & \ref{eq:Tdhighz} & K & Derived in \cite{2013ApJ...766...13D}
\\
\enddata
\end{deluxetable*}
\end{longrotatetable}

\section{Calculation of the Helium and Metals Contributions to $\mu_p$}\label{sec:AppendixMuZ}

In this Appendix, we show the details of the calculation of the terms in the denominator of Equation \ref{eq:mupfinal} which we repeat here
\begin{equation}
    \mu_p  =  \frac{\frac{M_g}{m_{\rm{H}}\mathcal{N}(H)}}{\frac{\mathcal{N}(\rm{H}_2)}{\mathcal{N}(\rm{H})} + \frac{\mathcal{N}(\rm{He})}{\mathcal{N}(\rm{H})} + \sum_i \frac{\mathcal{N}(Z_i)}{\mathcal{N}(\rm{H})} } \;\; .
    \label{eq:mupfinal2}
\end{equation}

The first term in the denominator is $1/2$ which comes from assuming that all H atoms are in molecules and that  $\mathcal{N}(\rm{H}_2)/\mathcal{N}(\rm{H}) = 1/2$.

For the Helium term, $3.971525 m_{\rm{H}} \mathcal{N}(\rm{He}) \approx M(\rm{He})$, where $3.971525$ is the atomic weight of helium, accounting for stable isotope abundances, divided by the atomic weight of $^1$H ($3.971525 = 4.002602/1.00782503$). 
We calculate that $\mathcal{N}(\rm{He})/\mathcal{N}(\rm{H}) \approx (1/3.971525)[M(\rm{He})/M(\rm{H})] \approx 0.096354$ for a protosolar mass ratio of helium to hydrogen of $M(\rm{He})/M(\rm{H}) = [M(\rm{He})/M_g]/[M(\rm{H})/M_g] = 0.2725/0.7121 = 0.38267$ \citep[][Table 4]{2021A&A...653A.141A}.
We note that the $^4$He and $^3$He protosolar abundances are also listed in Table B.1 of \cite{2021A&A...653A.141A}; however, the reported logarithmic abundance is only reported to two decimal places (a third decimal place is needed since $^4$He is so abundant with respect to hydrogen).
Our calculation above is more precise.

For the third term in the denominator of Equation~\ref{eq:mupfinal2}, we start by considering the two most abundant metals in the Universe: oxygen and carbon.
A significant fraction of C and O atoms can be locked up in dust grains, therefore we need to know their gas phase abundance in the ISM which is a fraction of their total abundance.
We use column 3 of Table 2 of \cite{2021ApJ...906...73H} which summarizes the gas phase abundance of the most abundant constituents of dust grains (C, O, Mg, Al, Si, S, Ca, Fe, and Ni) in the ISM.
Oxygen is the most abundant gas phase metal at 434 ppm (ppm = parts per million meaning that $\mathcal{N}(\rm{O})/\mathcal{N}(\rm{H}) = 434 \times 10^{-6}$).
Carbon is the next most abundant at 198 ppm.
The abundances of the remaining elements that are important dust grain constituents are significantly smaller: sulfur at 9.6 ppm, magnesium at 7.1 ppm, silicon at 6.6 ppm, and iron at 0.88 ppm \citep{2021ApJ...906...73H}.
All other metals have values of 0.1 ppm or less.

After oxygen and carbon, the next most abundant metals which remain in the gas phase are neon and nitrogen (see Table 23.1 of \citealt{Draine2011} for N abundances; Ne is an inert gas). 
They are not included in the \cite{2021ApJ...906...73H} table because they are not major constituents of dust grains.
Therefore, we use Table B.1 of \cite{2021A&A...653A.141A} which lists the protosolar abundances of stable isotopes in the Sun.
The abundances are quoted as logarithmic abundances, $\log_{10} \epsilon_{\rm{Z}} = \log_{10}[\mathcal{N}(\rm{Z})/\mathcal{N}(\rm{H})] + 12$, such that the value for $^1$H is $\log_{10} \epsilon_{^1\rm{H}} = 12.0$.
Neon has three isotopes with logarithmic abundances of $\log_{10} \epsilon_{^{20}\rm{Ne}} = 8.09$, $\log_{10} \epsilon_{^{21}\rm{Ne}} = 5.51$, and $\log_{10} \epsilon_{^{22}\rm{Ne}} = 6.97$.
The total Neon abundance is $\mathcal{N}(\rm{Ne})/\mathcal{N}(\rm{H}) = [10^{8.09} + 10^{5.51} + 10^{6.97}]/10^{12} \approx 1.3268\times10^{-4}$ or 132.68 ppm.
For Nitrogen, the two stable isotopes add up to $\mathcal{N}(\rm{N})/\mathcal{N}(\rm{H}) = [10^{7.89} + 10^{5.25}]/10^{12} \approx 7.780\times10^{-5}$ or 77.80 ppm.
The next most abundant element that is not already included in our analysis is argon which has an abundance of 2.78 ppm.
Other gas phase metals that do not deplete into dust grains will be less abundant than argon.

There is a large drop in abundance of gas phase metals after O, C, Ne, and N.
If we calculate $\sum_i \frac{\mathcal{N}(Z_i)}{\mathcal{N}(\rm{H})} = \frac{\mathcal{N}(\rm{O}) + \mathcal{N}(\rm{C}) + \mathcal{N}(\rm{Ne}) + \mathcal{N}(\rm{N})}{\mathcal{N}(\rm{H})} = 842.49$ ppm then $\mu_p = 2.351$, where we have used $\mu_{\rm{H}} = \frac{M_g}{m_{\rm{H} \mathcal{N}(\rm{H})}} = 1.404$ (Equations~\ref{eq:muh} and \ref{eq:muhfinal}).
If we include all of the metals in Table 2 of \cite{2021ApJ...906...73H} plus Ne, N, and Ar,  then we find that $\sum_i \frac{\mathcal{N}(Z_i)}{\mathcal{N}(\rm{H})} = \frac{\mathcal{N}(\rm{O}) + \mathcal{N}(\rm{C}) + \mathcal{N}(\rm{Ne})  + \mathcal{N}(\rm{N})
+  \mathcal{N}(\rm{S})
+ \mathcal{N}(\rm{Mg}) + \mathcal{N}(\rm{Si}) + \mathcal{N}(\rm{Ar}) + \mathcal{N}(\rm{Fe}) + \mathcal{N}(\rm{Al})+ \mathcal{N}(\rm{Ca}) + \mathcal{N}(\rm{Ni})}{\mathcal{N}(\rm{H})} = 869.69$ ppm, but we still calculate $\mu_p = 2.351$.
The inclusion of these metals less abundant than nitrogen only increased the total metals sum by 27.2 ppm.
The inclusion of even rarer metals will not change the precision of $\mu_{p}$ in the third decimal.
Neither would the inclusion of HD because the local $\mathcal{N}(\rm{D})/\mathcal{N}(\rm{H}) \approx 15$ ppm \citep{2003SSRv..106...49L}.
Therefore, we only need to keep track of the gas phase abundances of O, C, Ne, and N to calculate $\mu_p$ to three decimal places.
Lastly, we mention that if we include the uncertainty in the abundance of metals in the dust grains from column 4 of Table 2 of \cite{2021ApJ...906...73H} then we find that $\mu_p = 2.3514 \pm 0.0006$ (using all 9 metal abundances in Table 2); however, this error does not include uncertainties in the abundances of N, Ne, Ar, and the \cite{2021A&A...653A.141A} protosolar ratios.  
We have also not considered the formation of other molecules such as CO which, if it is assumed that we are in an environment where every free C atom in the gas forms CO, would result in $\mu_p = 2.352$.

\section{Abel Transformations for Spherical Geometry}\label{AbelApp}

The integrals in Equation \ref{eq:Inu_Abel} and Equation \ref{eq:Inu_isotherm_Abel} for spherical geometry are Abel integral transforms if the outer radius of the sphere goes to infinity ($\lim R \rightarrow \infty$) which we re-write here as
\begin{equation}
      \lim_{I_{\nu,\rm{bg}} \rightarrow 0} \; \lim_{\tau_{\nu} \ll 1}  I_{\nu,src}(\theta) = 2 \int_{r = D\theta}^{\infty}  B_{\nu}[T_d(r)] \kappa_{\nu}(r) \rho_d(r) \frac{r}{\sqrt{r^2 - D^2 \theta^2}} dr  \;\;\;\; \rm{erg}\, \rm{s}^{-1}\, \rm{cm}^{-2}\, \rm{ster}^{-1}\, \rm{Hz}^{-1} \;.
\label{eq:AppendixAbelIntegral}
\end{equation}
Therefore, the non-geometric part of the integrand (i.e., the factors prior to the fraction inside the integral of Equation \ref{eq:Inu_Abel} or Equation \ref{eq:AppendixAbelIntegral}) can be solved using an inverse Abel transform (see  \citealt{1986ftia.book.....B})
\begin{equation}
   \lim_{R \rightarrow \infty}  B_{\nu}[T_d(r)] \kappa_{\nu}(r) \rho_d(r) = - \frac{1}{2 \pi} \int_{b = r}^{\infty} \frac{dI_{\nu,src}}{db} \frac{1}{\sqrt{b^2 - r^2}} db \;\;. 
\label{eq:Inu_InvAbel}
\end{equation}
Note that any contribution from $I_{\nu,\rm{bg}}$ is negligible in an inverse Abel transform if the background does not spatially vary (its derivative is zero).
Similarly, Equation \ref{eq:Inu_isotherm_Abel} is the Abel integral transform of the dust density profile and therefore we can solve for the density profile using the inverse Abel transform
\begin{equation}
    \lim_{T_d(\vec{r}) = T_d}   \lim_{\kappa_{\nu}(\vec{r})=\kappa_{\nu}} \lim_{R \rightarrow \infty} \rho_d(r) = - \frac{1}{2 \pi B_{\nu}(T_d)\kappa_{\nu}} \int_{b = r}^{\infty} \frac{dI_{\nu,src}}{db} \frac{1}{\sqrt{b^2 - r^2}} db  \;\;.
\end{equation}
In practice when applying the inverse Abel transform, noise in the observed $dI_{\nu,src}/db$ strongly affects the integral \citep{2014A&A...562A.138R}.
One approach is to fit $I_{\nu,src}(b)$ with an analytical function whose derivative can be calculated exactly \citep{2019A&A...623A.118C}.

\section{Interferometric Visibilities}\label{AppendixVisibilities} 

We briefly discuss interferometric observations for a homogeneous, co-planar array, meaning all antennas have the same single-dish power patterns, $P_n$, and occupy the same elevation.
There are many excellent books and articles describing the fundamentals of interferometry (e.g. \citealt{1999ASPC..180.....T,2017isra.book.....T}; and the online proceedings of the IRAM Millimeter Interferometry Summer School 2 by A. Dutrey at: \url{https://web-archives.iram.fr/IS/IS2002/html_2/book.html}).

The distribution of emission on the sky can be considered the sum of an infinite series of sinusoids with various amplitudes and phases, analogous to the way any one-dimensional function can be reproduced by an infinite Fourier series.  
In the two-dimensional case of sky emission, however, the sinusoids have ``spatial frequencies," and these are projected onto axes called u and v that reflect their relative east-west and north-south directionality on the sky, respectively. 
Within interferometers, pairs of antennas sample the sinusoids that comprise the on-sky distribution of emission at specific spatial frequencies with u and v components corresponding to the east-west and north-south distances between antennas projected on the sky\footnote{The u-v plane is defined as a plane intersecting the center of the telescope array with normal pointing in the direction that the interferometer is pointing (which corresponds to $\theta = 0$ in Figure \ref{fig:Geo2} and line of sight C in Figure \ref{fig:Geo1}).}.   
In general, widely separated antennas sample high spatial frequencies and vice versa.  Each short integration measures an amplitude and phase at a specific spatial frequency.  
By tracking a source over time, the rotation of the Earth will lengthen or shorten the projected distances between pairs of antennas as the target rises or sets, respectively, allowing more of the so-called ``u-v plane" to be sampled.  
The resulting data are termed ``visibilities" at specific spatial frequencies - these visibilities have amplitudes and phases that are calibrated by looking at standard sources during the observations.  

Images are recovered via a two-dimensional Fourier transform of the ensemble of calibrated visibilities obtained from all pairs of antennas.  
The visibilities can be weighted to maximize sensitivity or resolution.  
Note that the sampling of the u-v plane is finite and un-sampled locations on the u-v plane will lead to aliasing in the image. 
Much of the resulting image artifacts that can be handled via deconvolution (depending on the extent of u-v coverage, of course) but there is a hard limit to u-v plane coverage set by the minimum distance between neighboring antennas, i.e., their diameter.  
This particular hole in u-v plane coverage corresponds to the lowest spatial frequencies, meaning that emission on large angular scales cannot be recovered from imaging.  
To recover this emission, however, data from arrays of smaller antennas or large, single-dish telescopes can be added to the ensemble.
The u and v distances are typically measured in units of wavelengths (\ie, if u = 1000 times the wavelength of the observations, then u = 1 klam where klam is the abbreviation for kilo-lambda).

One quantity used to compare observations with models of thermal dust continuum emission is the monochromatic interferometric visibility amplitudes, $V_{\nu}$, which is calculated from the two-dimensional Fourier transform of the specific intensity and the single telescope power pattern,
\begin{equation}
    V_{\nu}(d_{uv}) = \int_{\phi = 0}^{2\pi} \int_{\theta = 0}^{\infty} I_{\nu,obs}(\theta, \phi) P_n(\theta, \phi) e^{-2\pi i d_{uv} \theta cos(\phi - \psi)} \theta  d \theta d \phi \;\;\;\; \rm{erg}\, \rm{s}^{-1}\, \rm{cm}^{-2}\, \rm{Hz}^{-1} \;\;,
\label{eq:Visibilities}
\end{equation}
where $d_{uv} = \sqrt{u^2 + v^2}$ is a quantity known as the uv-distance, $\tan(\psi) = v/u$ is a phase angle in the u-v plane, and we have assumed that the small angle approximation in $\theta$ applies (see \citealt{1998PhDT........18L} for a derivation).
Note that the limits of integration for $\theta$ from $0$ to infinity indicate that we are now integrating in a planar polar coordinate system due to the assumption of the small angle approximation.

If the source intensity distribution is azimuthally-symmetric ($I_{\nu}$ does not depend on $\phi$), then the limits of integration of $\phi$ from $0$ to $2\pi$ mean that the phase angle $\psi$ can be ignored and the integral reduces to the Hankel transform of zeroth order \citep{1998PhDT........18L} of the intensity times the azimuthally-averaged single telescope beam profile,
\begin{equation}
    V_{\nu}(d_{uv}) = 2 \pi \int_{\theta = 0}^{\infty} I_{\nu,obs}(\theta) P_n(\theta) J_0(2 \pi  d_{uv} \theta) \theta  d \theta \;\;\;\; \rm{erg}\, \rm{s}^{-1}\, \rm{cm}^{-2}\, \rm{Hz}^{-1} \;\;,
\label{eq:HankelTransform}
\end{equation}
where $J_0$ is a zeroth-order Bessel function of the first kind\footnote{The Hankel transform is also called a Fourier-Bessel transform and is the two-dimensional Fourier transform of a azimuthally-symmetric function.  The zeroth-order Bessel function obeys the Hansen-Bessel formula, $2 \pi J_0(t) = \int_0^{2\pi} e^{i t cos(\phi)} d\phi$}.
As an example, let's assume that the intensity is approximated by an azimuthally-symmetric Gaussian function with FWHM of $b_{\rm{fwhm}} = \sigma_{b} \sqrt{8 \ln{2}}$,
\begin{equation}
    I_{\nu,src}(b) = I_{pk} e^{-\frac{1}{2}\frac{b^2}{\sigma_{b}^2}} \;\; ,
\end{equation}
then the Hankel transform results in \citep{papoulis1981systems}
\begin{equation}
    V_{\nu}(d_{uv}) = I_{pk} \sigma_{b}^2 \exp \left[ - \frac{1}{2} \frac{(2\pi d_{uv})^2}{\left(\frac{1}{\sigma_{b}^2}  \right)^2}\right]  \;\;\;\; .
\end{equation}
The Hankel transform of a Gaussian function is also a Gaussian function, but with a FWHM that is proportional to $1/b_{\rm{fwhm}}$, the inverse of the source intensity FWHM.
This illustrates the inverse nature of Fourier transforms where a broad intensity distribution on the sky results in a narrow visibility distribution in $d_{uv}$ and vice versa.
If instead the sky intensity distribution is an azimuthally-symmetric power-law, $I_{\nu,src}(b) \propto b^{-m}$ (i.e., \citealt{1998apsf.book.....H,2000ApJS..131..249S}), 
then Equation \ref{eq:HankelTransform}  transform to visibility amplitudes as $V_{\nu}(d_{uv}) \propto d_{uv}^{m-2}$ \citep{1998PhDT........18L}\footnote{If $m \in (\frac{1}{2},2)$ then Equation \ref{eq:HankelTransform} can be solved using Equation 6.561.14 of \cite{1965tisp.book.....G}}.

\section{Calculating the Average Optical Depth for Various Geometries}\label{AppendixAvgTau}

In this appendix we explore the calculation of the average optical depth, $\overline{\tau_{\nu}}$, defined in Equation \ref{eq:angleavgtau} which we repeat here
\begin{equation}
    1 - e^{-\overline{\tau_{\nu}}} = \frac{\int_{\Omega_{src}} (1 - e^{-\tau_{\nu}(\theta,\phi)}) d\Omega}{\Omega_{src}} \;\; ,
\end{equation}
for a slab geometry and for spherical geometry.

\subsection{Slab Geometry}

If the object is an isothermal, constant opacity uniform density slab with the same thickness, $L$, that is emitting over a solid angle\footnote{This geometry could also be a cylinder or a disk that is viewed face-on, meaning the cylinder or disk presents a circular cross-section to the observer.}, then every line-of-sight within the solid angle has a constant optical depth of 
\begin{eqnarray}
    \tau_{\nu}(\theta,\phi) & = & \int_{0}^{L} \kappa_{\nu} \rho_d dz \\
                            & = & \kappa_{\nu} \rho_d L \\
                            & = & \tau_c \;\;\;\; ,
\end{eqnarray}
where $L$ is the path length through the slab and we have defined the variable $\tau_c$ to be the constant optical depth.
As a result, the $(1 - e^{-\tau_{\nu}(\theta,\phi)})$ term that is the integrand of Equation \ref{eq:angleavgtau} is a constant and can be brought out of the integral resulting in $\overline{\tau_{\nu}} = \tau_c$.
The constant density uniform thickness slab is the geometric limit that is typically assumed, but not often acknowledged, when analyzing the SED with Equation \ref{eq:FluxSED}.
Most astrophysical sources, however, have more complex geometries with physical variables that change along each line-of-sight.

\subsection{Spherical Geometry}

\begin{figure}[h!]
\includegraphics[scale=0.6]{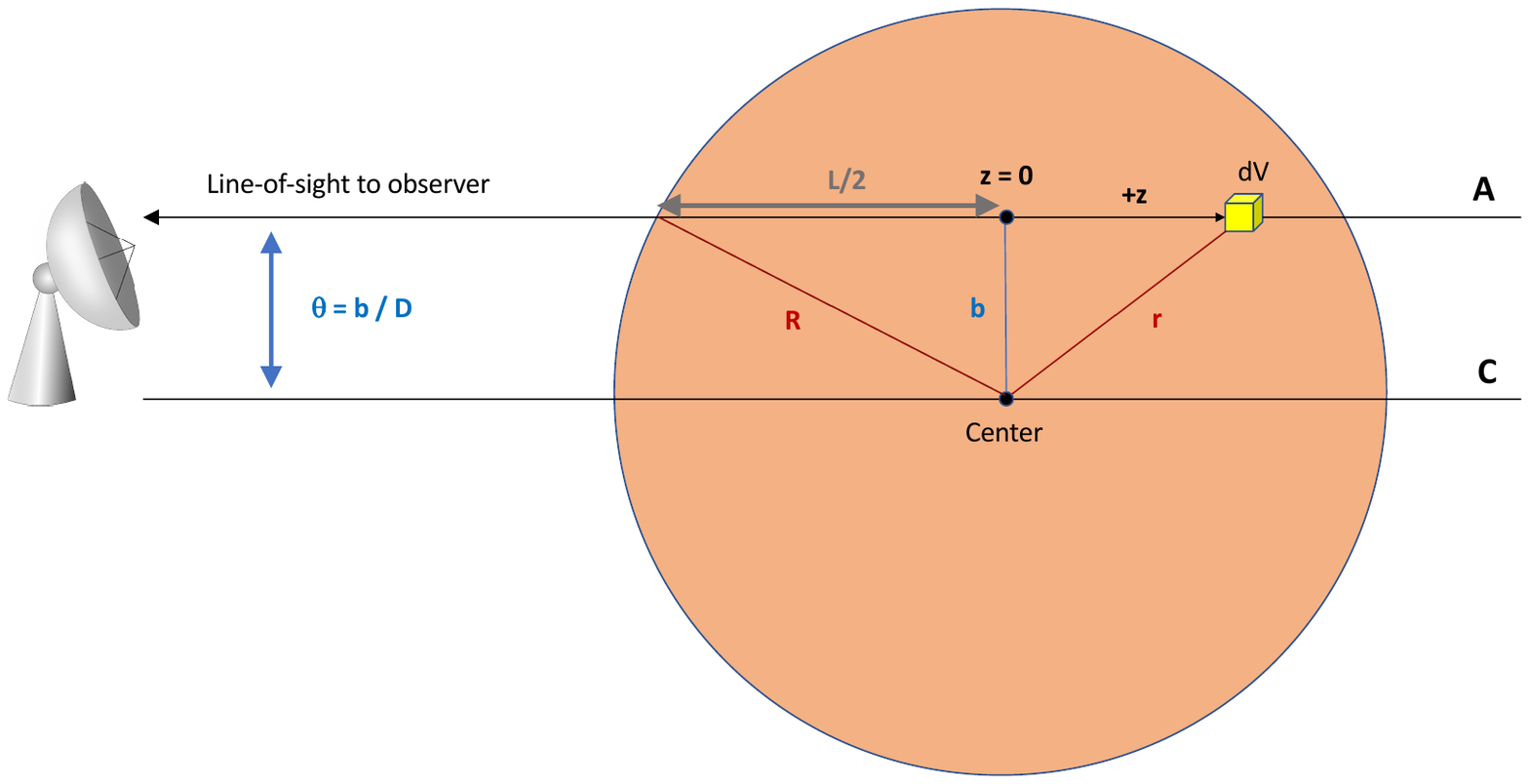}
\vspace{-2cm}
\caption{Geometry of a spherical dust cloud with total path length $L$ through the source at angular impact parameter $\theta$.}
\label{fig:GeoSphTau}
\end{figure}

Next we consider the spherical geometry shown in Figure \ref{fig:GeoSphTau}. 
We need to first derive an expression for the optical depth as a function of viewing angle $\tau_{\nu}(\theta,\phi)$.
For simplicity, we start by assuming that we have an isothermal, constant opacity, constant density sphere with radius R emitting thermal dust emission.
We do not make any assumptions, initally, about optical depth limits.
Since the sphere has a constant density, then the problem of finding an equation for $\tau_{\nu}(\theta,\phi)$ reduces to finding the total path length, $L(\theta)$ through the sphere for each line-of-sight A at an angular impact parameter $\theta$.
The distance from $z = 0$ to the edge of the sphere is half of the total path length, L/2, therefore the total path length is given by
\begin{equation}
    L(\theta) = 2 \sqrt{R^2 - D^2\theta^2} \;\; \rm{cm}\; .
\end{equation}
For a constant opacity and constant density sphere, the total optical depth for each line-of-sight is then given by
\begin{equation}
    \tau_{\nu}(\theta) = \kappa_{\nu} \rho_d L(\theta) \;\; .
\end{equation}
We can now substitute this expression in the definition of $\overline{\tau_{\nu}}$ (Equation \ref{eq:angleavgtau}) to find that
\begin{equation}
    1 - e^{-\overline{\tau_{\nu}}} = \frac{ \int_{\phi = 0}^{2\pi} \int_{\theta=0}^{R/D} 1 - e^{-2 \kappa_{\nu} \rho_d  \sqrt{R^2 - D^2\theta^2}}  \theta d\theta d\phi}{\Omega_{sph}} \;\;,
\end{equation}
when averaged over the entire sphere with solid angle $\Omega_{sph} = \pi R^2/D^2$.
Since $\tau_{\nu}(\theta)$ does not depend on $\phi$ due to spherical symmetry, then the integral of $d\phi$ is $2\pi$.
We can then solve the integral of $d\theta$ by making the substitution $u = -2 \kappa_{\nu} \rho_d  \sqrt{R^2 - D^2\theta^2}$.  The derivative of this expression is $du = 2 \kappa_{\nu} \rho_d D^2 \theta d\theta /  \sqrt{R^2 - D^2\theta^2} = - 4 \kappa_{\nu}^2 \rho_d^2 D^2 \theta d\theta/u$, where we have used the definition of $u$ to eliminate the square root in the denominator. 
We can solve this differential expression for $\theta d\theta = - u du / (4 \kappa_{\nu}^2 \rho_d^2 D^2)$.
Substituting we find
\begin{equation}
    1 - e^{-\overline{\tau_{\nu}}} = - \frac{1}{2 \kappa_{\nu}^2 \rho_d^2 R^2} \int_{u=-2\kappa_{\nu} \rho_d R}^{0} (1 - e^{u})  u du \;\; .
\end{equation}
The lower limit to the integral is related to the total optical depth observed at $\theta = 0$ for the line-of-sight through the center of the sphere which we will define as the new variable $\tau_c = 2 \kappa_{\nu} \rho_d R$.
Re-writing our integral in terms of $\tau_c$ and propagating the minus sign into the integrand to reverse the order of the terms, we find that
\begin{equation}
    1 - e^{-\overline{\tau_{\nu}}} = \frac{2}{\tau_c^2} \int_{u=-\tau_c}^{0} (e^{u} - 1)  u du \;\; .
\label{eq:uintegralfortau}
\end{equation}
The solution to the integral is
\begin{equation}
    1 - e^{-\overline{\tau_{\nu}}} = \frac{2}{\tau_c^2} \left[ \frac{\tau_c^2}{2} + e^{-\tau_c}(\tau_c + 1) - 1 \right] \;\; .
    \label{eq:oneminusexptau}
\end{equation}
At this point, we should check if this expression is reasonable by looking at its solution in different limits.
If the $\lim (\overline{\tau_{\nu}}, \tau_c) \rightarrow \infty$, then both the left-hand side and right-hand side simplify to $1$ as expected.
In the optically thin limit ($(\overline{\tau_{\nu}}, \tau_c) \ll 1$), then we can use the MacLaurin expansion of $e^{-\overline{\tau_{\nu}}} \approx 1 - \overline{\tau_{\nu}}$ and $e^{-\tau_c} \approx 1 - \tau_c + \tau_c^2/2 - \tau_c^3/6$.
Note that we have to expand $\tau_c$ to powers of $\tau_c^3$ because the division by $\tau_c^2$ in the denominator will then leave terms linear in $\tau_c$.
Substituing and simplifying both expressions we find that
\begin{equation}
    \lim_{(\overline{\tau_{\nu}},\tau_{c}) \ll 1} \; \overline{\tau_{\nu}} = \frac{2}{3} \tau_c \;\; .
\end{equation}
This is indeed the same expression that would be obtained if the limit was applied in Equation \ref{eq:uintegralfortau} ($e^u \approx  1 + u$) and the integral became
\begin{equation}
     \lim_{(\overline{\tau_{\nu}},\tau_{c}) \ll 1} \;\overline{\tau_{\nu}} = \frac{2}{\tau_c^2} \int_{u=-\tau_c}^{0}  u^2 du = \frac{2\tau_c^3}{3\tau_c^2} = \frac{2}{3} \tau_c \;\;.
\end{equation}
This expression also implies that, in the optically thin limit for a constant density sphere, $\overline{\Sigma}_d = (2/3) \Sigma_c$.

When we are not in the optically thin or optically thick limits, then the final expression for $\overline{\tau_{\nu}}$ is given by
\begin{equation}
    \overline{\tau_{\nu}} = \ln{\frac{\tau_c^2}{2[1 - e^{-\tau_c}(\tau_c + 1)]}}  \;\; .
\end{equation}
The ratio of $\overline{\tau_{\nu}}/\tau_c$ is a slowly varying, decreasing function for increasing optical depths.
For instance, for $\tau_c = 2.21$, $\overline{\tau_{\nu}} \approx 0.6 \tau_c$ which is still $90$\% of the optically thin value.
At $\tau_c = 5.45$, $\overline{\tau_{\nu}} \approx 0.5 \tau_c$ and at $\tau_c = 13.56$, $\overline{\tau_{\nu}} \approx 0.33 \tau_c$ or approximately half of the optically thin value.

The total flux density is given by Equation~\ref{eq:FluxSED} with $\Omega_{ap} = \frac{\pi R^2}{D^2}$ and inserting Equation~\ref{eq:oneminusexptau} for $1 - e^{-\overline{\tau_{\nu}}}$
\begin{equation}
    F_{\nu} = \frac{2\pi R^2}{\tau_c^2 D^2} B_{\nu}(T_d) \left[ \frac{\tau_c^2}{2} + e^{-\tau_c}(\tau_c + 1) - 1 \right] \;\; .
\end{equation}
In the optically thick limit ($\tau_c \gg 1$), this expression reduces to
\begin{equation}
    \lim_{(\overline{\tau_{\nu}},\tau_{c}) \gg 1} F_{\nu} = \pi B_{\nu}(T_d)\frac{R^2}{D^2} \;\;\;\; ,
\end{equation}
which is just the emergent flux density from the surface (at radius R) scaled by the inverse square law as expected.
In the optically thin limit ($\tau_c \ll 1$), we find that
\begin{eqnarray}
    \lim_{(\overline{\tau_{\nu}},\tau_{c}) \ll 1} F_{\nu} & = & B_{\nu}(T_d) \overline{\tau_{\nu}} \Omega_{sph} \nonumber \\ 
    & = & \frac{2}{3} B_{\nu}(T_d) \tau_c \Omega_{sph} \nonumber \\ 
    & = & \frac{2}{3} B_{\nu}(T_d) (2\kappa_{\nu} \rho_d R) \left(\frac{\pi R^2}{D^2}\right) \nonumber \\
    & = &  \frac{4}{3} \pi R^3 \frac{B_{\nu}(T_d) \kappa_{\nu} \rho_d}{D^2} \nonumber \\ 
    & = & V_{sph}\frac{j_{\nu}}{D^2} \;\; ,
\end{eqnarray}
which is the expected expression for the observed flux density from an optically thin emitting object with constant emissivity coefficient $j_{\nu}$ (e.g., combine Equation \ref{eq:Lnuflux} and Equation \ref{eq:Lnu_mono1}).

The procedure developed in this section may be applied to other geometries by finding the expression for $\tau_{\nu}(\theta,\phi)$ with the line-of-sight total path length as a function of viewing angles. 
The utility of that approach, however, is likely limited for real astrophysical sources having variations in dust temperature, density, and opacity.

\section{Details of Dust Opacity Models}\label{AppendixOpacity}

In this section, we give details of the dust opacity models highlighted in this tutorial. 
We adopt the acronyms HD23 for \citealt{2023ApJ...948...55H}, Y24 for \citealt{2024A&A...684A..34Y}, OHx where x correspond to columns in Table 2 of \citealt{1994A&A...291..943O}, and KP5 for \citealt{2024RNAAS...8...68P}.
As described below, OHSC corresponds to \citealt{1994A&A...291..943O} for starless cores that is interpolated from the published values.
The dust absorption opacities appropriate for thermal dust continuum emission from the diffuse Milky Way ISM and dense cores are summarized in Table \ref{tab:dustopacity} for different telescope instrument filters with $5$ $\mu$m $< \lambda < 7$ mm.

\startlongtable
\begin{deluxetable*}{lllccccccccc}
\tablecaption{Interpolated\tablenotemark{a} and extrapolated\tablenotemark{b}  $\kappa_{\nu,\rm{abs}}$ (cm$^2$/gram of dust) for Milky Way ISM and Dense Cores.   \label{tab:dustopacity}}
\tablewidth{0pt}
\tablehead{
\colhead{Telescope} &
\colhead{Instrument Filter} &
\colhead{$\lambda_{\rm{ref}}$\tablenotemark{c}} &
\multicolumn{2}{c}{Milky Way ISM} &
\colhead{} &
\multicolumn{4}{c}{Dense Cores} &
\colhead{} \\
\cline{4-5}
\cline{7-10}
\colhead{} &
\colhead{} &
\colhead{[$\mu$m]}  &
\colhead{HD23} &
\colhead{Y24} &
\colhead{} &
\colhead{OHSC} &
\colhead{OH5} &
\colhead{OH8} &
\colhead{KP5} & 
\colhead{} }
\startdata
 \textit{JWST} & MIRI F560W
 & 5.64 & 1285 & 448.5 & & 1191 & 1330 & 2031 & 553.8 & & \\ 
 \textit{Spitzer} & IRAC Chan 3
 & 5.74 & 1282 & 441.4 & & 1210 & 1348 & 2414 & 585.1 & & \\ 
 \textit{JWST} & MIRI F770W
 & 7.64 & 1247 & 469.5 & & 986.5 & 1089 & 1369 & 649.0 & & \\ 
 \textit{Spitzer} & IRAC Chan 4
 & 7.80 & 1231 & 479.2 & & 972.3 & 1072 & 1279 & 711.4 & & \\ 
	\textit{JWST} & MIRI F1000W
 & 9.95 & 2594 & 705.5 & & 2445 & 2605 & 3630 & 2023 & & \\ 
 \textit{IRAS} & 12$\mu$m 
 & 11.04 & 1873 & 604.6 & & 2310 & 2522 & 6152 & 1696 & & \\ 
 \textit{SOFIA} & FORECAST SWC F111
 & 11.08 & 1849 & 602.4 & & 2315 & 2529 & 6345 & 1688 & & \\ 
	 \textit{JWST} & MIRI F1130W
 & 11.31 & 1701 & 584.9 & & 2313 & 2545 & 7343 & 1644 & & \\ 
  \textit{SOFIA} & FORECAST LWC F113
 & 11.34 & 1679 & 579.0 & & 2308 & 2541 & 7439 & 1638 & & \\ 
	 \textit{WISE} & W3
 & 11.56 & 1535 & 534.4 & & 2268 & 2517 & 8172 & 1598 & & \\ 
	 \textit{JWST} & MIRI F1280W
 & 12.81 & 1068 & 298.2 & & 1663 & 1875 & 8555 & 1257 & & \\ 
	 \textit{JWST} & MIRI F1500W
 & 15.06 & 1065 & 315.7 & & 1194 & 1340 & 5388 & 1252 & & \\ 
	 \textit{JWST} & MIRI F1800W
 & 17.98 & 1356 & 448.2 & & 1383 & 1558 & 3823 & 1008 & & \\ 
 \textit{SOFIA} & FORECAST SWC F197
 & 19.65 & 1228 & 427.2 & & 1193 & 1388 & 3123 & 969.9 & & \\ 
	 \textit{JWST} & MIRI F2100W
 & 20.79 & 1110 & 404.4 & & 1022 & 1206 & 2584 & 913.1 & & \\ 
 \textit{WISE} & W4
 & 22.09 & 1001 & 391.5 & & 869.4 & 1036 & 2062 & 843.0 & & \\ 
 \textit{IRAS} & 25$\mu$m
 & 23.07 & 910.4 & 383.2 & & 782.1 & 936.6 & 1740 & 792.8 & & \\ 
	 \textit{Spitzer} & MIPS 24 $\mu$m
 & 23.59 & 862.8 & 378.5 & & 740.8 & 889.1 & 1595 & 770.6 & & \\ 
 \textit{SOFIA} & FORECAST LWC F242
 & 24.21 & 810.8 & 372.2 & & 695.4 & 836.9 & 1441 & 745.2 & & \\ 
	 \textit{JWST} & MIRI F2550W
 & 25.36 & 726.9 & 355.0 & & 622.1 & 751.8 & 1191 & 703.2 & & \\ 
 \textit{SOFIA} & FORECAST SWC F253
 & 25.38 & 725.6 & 354.7 & & 621.0 & 750.5 & 1186 & 702.5 & & \\ 
 \textit{SOFIA} & FORECAST LWC F315
 & 31.39 & 446.2 & 273.9 & & 414.0 & 500.7 & 686.4 & 557.0 & & \\ 
 \textit{SOFIA} & FORECAST LWC F336
 & 33.58 & 383.1 & 249.2 & & 382.0 & 460.8 & 803.4 & 527.0 & & \\ 
 \textit{SOFIA} & FORECAST LWC F348
 & 34.78 & 354.0 & 236.6 & & 371.0 & 447.1 & 889.1 & 518.2 & & \\ 
 \textit{SOFIA} & FORECAST LWC F371
 & 37.15 & 305.5 & 213.6 & & 359.9 & 433.2 & 1109 & 508.9 & & \\ 
	\textit{SOFIA} & HAWC+ A
 & 53.6 & 136.7 & 119.8 & & 231.5 & 285.4 & 1216 & 343.2 & & \\ 
	\textit{IRAS} & 60 $\mu$m 
 & 58.2 & 114.5 & 104.7 & & 197.7 & 244.8 & 998.5 & 294.8 & & \\ 
	\textit{SOFIA} & HAWC+ B
 & 63.5 & 95.13 & 90.55 & & 167.8 & 208.7 & 806.7 & 251.3 & & \\ 
	\textit{AKARI} & N60
 & 65.4 & 89.41 & 86.60 & & 158.6 & 197.4 & 744.7 & 236.7 & & \\ 
	\textit{Herschel} & PACS
 & 70.8 & 75.63 & 76.16 & & 136.1 & 169.9 & 600.3 & 201.0 & & \\ 
	\textit{Spitzer} & MIPS 70 $\mu$m
 & 70.9 & 75.41 & 75.98 & & 135.7 & 169.4 & 598.0 & 200.4 & & \\ 
	\textit{AKARI} & WIDE-S
 & 85.1 & 51.59 & 57.39 & & 95.55 & 120.0 & 347.8 & 131.4 & & \\ 
 \textit{SOFIA} & HAWC+ C
 & 89.7 & 46.31 & 53.15 & & 85.80 & 107.9 & 278.2 & 116.3 & & \\ 
	\textit{IRAS} & 100 $\mu$m 
 & 99.5 & 37.47 & 45.70 & & 69.26 & 87.36 & 175.4 & 90.37 & & \\ 
 \textit{Herscel} & PACS
 & 101 & 36.36 & 44.65 & & 67.45 & 85.10 & 167.0 & 87.10 & & \\ 
 \textit{AKARI} & WIDE-L
 & 146 & 17.46 & 22.96 & & 36.73 & 46.72 & 67.21 & 36.16 & & \\ 
 \textit{Spitzer} & MIPS 160 $\mu$m
 & 155 & 15.53 & 20.47 & & 33.25 & 42.38 & 58.63 & 31.60 & & \\ 
 \textit{SOFIA} & HAWC+ D
 & 157 & 15.15 & 19.95 & & 32.55 & 41.50 & 56.94 & 30.70 & & \\ 
 \textit{AKARI} & N160
 & 162 & 14.26 & 18.76 & & 30.74 & 39.22 & 52.85 & 28.60 & & \\ 
 \textit{Herschel} & PACS
 & 162 & 14.26 & 18.76 & & 30.74 & 39.22 & 52.85 & 28.60 & & \\ 
 \textit{SOFIA} & HAWC+ E
 & 217 & 8.165 & 10.22 & & 18.14 & 23.26 & 27.34 & 14.24 & & \\ 
 \textit{Herschel} & SPIRE
 & 249 & 6.318 & 7.592 & & 14.26 & 18.30 & 20.99 & 10.52 & & \\ 
 \textit{Herschel} & SPIRE
 & 348 & 3.438 & 3.933 & & 7.915 & 10.18 & 11.34 & 5.248 & & \\ 
 \textit{Planck}/FYST & HFI/Prime-Cam
 & 353 & 3.351 & 3.838 & & 7.707 & 9.914 & 11.03 & 5.121 & & \\ 
 APEX & SABOCA
 & 357 & 3.284 & 3.763 & & 7.540 & 9.700 & 10.78 & 5.024 & & \\ 
 JCMT & SCUBA2
 & 449 & 2.194 & 2.441 & & 4.831 & 6.215 & 6.821 & 3.364 & & \\ 
 \textit{Herschel} & SPIRE
 & 501 & 1.816 & 1.985 & & 3.904 & 5.023 & 5.480 & 2.769 & & \\ 
 \textit{Planck} & HFI
 & 545 & 1.573 & 1.706 & & 3.301 & 4.245 & 4.634 & 2.419 & & \\ 
 FYST & Prime-Cam
 & 731 & 0.9638 & 1.026 & & 1.855 & 2.383 & 2.604 & 1.514 & & \\ 
 \textit{Planck} & HFI
 & 841 & 0.7676 & 0.8277 & & 1.447 & 1.858 & 2.023 & 1.208 & & \\ 
 FYST & Prime-Cam
 & 857 & 0.7447 & 0.8046 & & 1.400 & 1.797 & 1.955 & 1.174 & & \\ 
 JCMT & SCUBA2
 & 861 & 0.7390 & 0.7989 & & 1.388 & 1.782 & 1.939 & 1.165 & & \\ 
 APEX & LABOCA
 & 876 & 0.7189 & 0.7782 & & 1.346 & 1.729 & 1.880 & 1.135 & & \\ 
 FYST & Prime-Cam
 & 1070 & 0.5237 & 0.5759 & & 0.9560 & 1.227 & 1.325 & 0.8525 & & \\ 
	LMT & TolTEC
 & 1090 & 0.5088 & 0.5602 & & 0.9281 & 1.191 & 1.285 & 0.8312 & & \\ 
 CSO & MUSIC Band 2
 & 1100 & 0.5015 & 0.5528 & & 0.9147 & 1.174 & 1.266 & 0.8210 & & \\ 
 CSO & Bolocam
 & 1113 & 0.4925 & 0.5433 & & 0.8977 & 1.152 & 1.242 & 0.8080 & & \\ 
 LMT & AZTEC
 & 1120 & 0.4878 & 0.5383 & & 0.8888 & 1.141 & 1.229 & 0.8011 & & \\ 
 CSO & MUSIC Band 1
 & 1140 & 0.4745 & 0.5245 & & 0.8640 & 1.109 & 1.194 & 0.7821 & & \\ 
	IRAM & NIKA2
 & 1200 & 0.4384 & 0.4870 & & 0.7961 & 1.022 & 1.097 & 0.7295 & & \\ 
 IRAM & MAMBO2
 & 1210 & 0.4328 & 0.4813 & & 0.7856 & 1.008 & 1.083 & 0.7212 & & \\ 
 FYST & Prime-Cam
 & 1362 & 0.3612 & 0.4082 & & \textbf{0.6407} & \textbf{0.8222} & \textbf{0.8793} & 0.6085 & & \\ 
	\textit{Planck}/LMT & HFI/TolTEC
 & 1370 & 0.3581 & 0.4048 & & \textbf{0.6344} & \textbf{0.8141} & \textbf{0.8704} & 0.6032 & & \\ 
	LMT & TolTEC
 & 2040 & 0.1987 & 0.2365 & & \textbf{0.3222} & \textbf{0.4133} & \textbf{0.4357} & \textbf{0.3197} & & \\ 
	IRAM & NIKA2
 & 2050 & 0.1973 & 0.2349 & & \textbf{0.3196} & \textbf{0.4099} & \textbf{0.4320} & \textbf{0.3173} & & \\ 
 CSO & MUSIC Band 0
 & 2110 & 0.1888 & 0.2259 & & \textbf{0.3043} & \textbf{0.3902} & \textbf{0.4109} & \textbf{0.3037} & & \\ 
	\textit{Planck} & HFI
 & 2130 & 0.1861 & 0.2230 & & \textbf{0.2994} & \textbf{0.3840} & \textbf{0.4042} & \textbf{0.2994} & & \\ 
	\textit{Planck} & HFI
 & 3000 & 0.1105 & 0.1442 & & \textbf{0.1672} & \textbf{0.2143} & \textbf{0.2229} & \textbf{0.1781} & & \\ 
	GBT & MUSTANG2
 & 3380 & 0.09213 & 0.1240 & & \textbf{0.1365} & \textbf{0.1749} & \textbf{0.1811} & \textbf{0.1487} & & \\ 
	\textit{Planck} & LFI
 & 4300 & 0.06386 & 0.09204 & & \textbf{0.09063} & \textbf{0.1161} & \textbf{0.1192} & \textbf{0.1032} & & \\ 
	\textit{Planck} & LFI
 & 6830 & 0.03158 & 0.05312 & & \textbf{0.04125} & \textbf{0.05281} & \textbf{0.05333} & \textbf{0.05118} & & \\ 
\enddata
\tablenotetext{a}{$\kappa_{\nu,\rm{abs}}$ values are logarithmically interpolated from the published values using the closest wavelengths that bracket $\lambda_{\rm{pivot}} = c/\nu_{\rm{pivot}}$. If the closest bracketing wavelengths are $\lambda_1 \leq \lambda_{\rm{pivot}} \leq \lambda_2$, then $\log_{10} \kappa_{\nu_{\rm{pivot}},\rm{abs}} = \log_{10} \kappa_{\nu_{1},\rm{abs}} + \frac{\log_{10} \kappa_{\nu_{2},\rm{abs}} - \log_{10}\kappa_{\nu_{1},\rm{abs}}}{\log_{10} \lambda_2 - \log_{10} \lambda_1}(\log_{10} \lambda_{\rm{pivot}} - \log_{10} \lambda_1 )$.  Note that $\kappa_{\nu,\rm{abs}} = \kappa_{\lambda,\rm{abs}}$ for all $\lambda \nu = c$. }
\tablenotetext{b}{\textbf{$\kappa_{\nu,\rm{abs}}$ values in bold} are extrapolated from the published values using a power-law determined from a linear regression of the logarithmic values between $700$ $\mu$m and the longest published wavelength ($1300$ $\mu$m for OH5, OH8, and OHSC and $2000$ $\mu$m for KP5). 
For OHSC $ \log_{10} \kappa_{\nu_{\rm{pivot}},\rm{abs}} = -1.701228 \log_{10} \lambda_{\rm{pivot}} + 5.138625  $.
For OH5 $ \log_{10} \kappa_{\nu_{\rm{pivot}},\rm{abs}} = -1.702661 \log_{10} \lambda_{\rm{pivot}} + 5.251426  $. For OH8 $ \log_{10} \kappa_{\nu_{\rm{pivot}},\rm{abs}} = -1.738152 \log_{10} \lambda_{\rm{pivot}} + 5.391809  $. For KP5 $ \log_{10} \kappa_{\nu_{\rm{pivot}},\rm{abs}} = -1.516068 \log_{10} \lambda_{\rm{pivot}} + 4.522308  $.}
\tablenotetext{c}{The reference wavelengths for the filters are $ \lambda_{\rm{ref}} = \lambda_{\rm{pivot}}$ for all instruments on \textit{JWST}, \textit{Spitzer}, \textit{WISE}, \textit{IRAS}, \textit{SOFIA}, \textit{AKARI}, \textit{Herschel}, \textit{Planck}, APEX, JCMT, and the LMT. For MUSTANG2 on the GBT, the reference wavelength also accounts for variation in the aperture efficiency of the GBT with wavelength and is calculated for a source spectral index of $\alpha_s = 0$ (see \url{https://gbtdocs.readthedocs.io/en/latest/references/receivers/mustang2/mustang2\_bandpass.html}).  For the future Prime-Cam on the FYST, the reference wavelengths are published in  \cite{2023ApJS..264....7C}.  For the MUSIC instrument, the reference wavelengths are given on \url{http://cso.caltech.edu/wiki/cso/instruments/music/music}. We caution for the future and in-development instruments that $\lambda_{ref}$ may not be the $\lambda_{\rm{pivot}}$ of the eventual filters.}
\end{deluxetable*}

The diffuse ISM grain models are parameterized in terms of the ratio of V-band extinction to the B-V color excess, 
\begin{equation}
R_V = \frac{A_V}{(m_{B} - m_{V}) - (m_{B} - m_{V})_{intr.}} = \frac{A_V}{E(B-V)}  \; ,
\end{equation}
where $(m_{B} - m_{V})$ is the color (difference in magnitudes in the filters B and V) with dust extinction and $(m_{B} - m_{V})_{intr.}$ the intrinsic observed $B - V$ color without dust extinction. The visual $B$ and $V$ filters correspond to 440 nm and 550 nm, respectively\footnote{The color excess is usually determined for line-of-sight toward background stars whose spectral type may be used to constrain their intrinsic B-V color. See \cite{1970A&A.....4..234F} for the intrinsic $B - V$ colors of stars of different spectral types.}.
A typical line-of-sight through the diffuse ISM has $R_V = 3.1$ \citep{2019ApJ...886..108F} with some modest variability (e.g., from 2 to 5.3; \citealt{2022ApJ...930...15D,2023A&A...676A.132S}) for lines-of-sight with $A_V < 3$ mag.
Lines-of-sight through denser regions in the ISM tend to have larger values of $R_V$  \citep{1989ApJ...345..245C}.
Both of the diffuse ISM models shown in Figure~\ref{fig:Opacity1} are calculated for $R_V = 3.1$.

Some dust opacity models are comprised of multiple dust grain constituents and populations.
In general, the total dust mass opacity will be the mass-weighted average of each grain population.
For example, \cite{2023ApJ...948...55H} and \cite{2024A&A...684A..34Y} report the ratio of dust optical depth to H atom column density,  $(\tau_{\nu}/N_{\rm{H}})_i$, which is equivalent to the dust cross-section per H atom (cm$^2$/H atom) for each grain population $i$.  The dust mass opacity (absorption, scattering, or extinction) for each grain population is then given by
\begin{equation}
\kappa_{\nu, i} = \frac{(\tau_{\nu}/N_{\rm{H}})_i}{m_{\rm{H}}} \frac{M_{\rm{H}}}{M_{\rm{dust},i}} \;\;\;\;  \rm{cm^2} \; (\rm{g} \; \rm{of} \; \rm{dust} \; \rm{of} \; \rm{type} \; i)^{-1} \;\;,
\end{equation}
where $(M_{\rm{H}}/M_{\rm{dust}_i})$ is the gas mass to dust mass ratio of grain population $i$. 
The combined opacity for all grain populations is then the weighted average of $\kappa_{\nu, i}$ by the mass fraction of each grain population,
\begin{equation}
    \kappa_{\nu} = \sum_{i} \left( \frac{ \frac{M_{\rm{dust},i}}{M_{\rm{H}}} }{\sum_{i} \frac{M_{\rm{dust},i}}{M_{\rm{H}}}} \right)  \kappa_{\nu, i} \;\;\;\;  \rm{cm^2} \; (\rm{g} \; \rm{of} \; \rm{dust})^{-1} \;\;.
\end{equation}
Combining these two equations gives the conversion
\begin{equation}
    \kappa_{\nu} = \sum_{i} \frac{(\tau_{\nu}/N_{\rm{H}})_i}{m_{\rm{H}} \sum_{i} \frac{M_{\rm{dust},i}}{M_{\rm{H}}} }  \;\;\;\;  \rm{cm^2} \; (\rm{g} \; \rm{of} \; \rm{dust})^{-1} \;\;.
\end{equation}

The \cite{2023ApJ...948...55H} AstroDust+PAH models shown in Figure~\ref{fig:Opacity1} are from two dust grain populations\footnote{The Astrodust+PAH models are available at Hensley, Brandon, 2022, ``Astrodust+PAH Model Output", \url{https://doi.org/10.7910/DVN/3B6E6S}, Harvard Dataverse, V2. }.
The first is comprised of single-composition  composite grains, called ``Astrodust", that simultaneously reproduce the observed extinction and emission constraints from dust in the diffuse ISM, including polarization.
Grains larger than $a_{grain} \gtrsim 20$ nm contain small domains (and voids) of the typical refractory elements \citep[\ie\ C, O, Si, Mg, Fe, etc.; see][for abundances]{2021ApJ...906...73H} with roughly similar compositions.
The second population consists of smaller grains that have variable composition and are made of constituents such as polycyclic-aromatic hydrocarbons (PAHs).
The dust to hydrogen gas mass ratio for each dust species is $(M_{\rm{dust}_{\rm{Ad}}}/M_{\rm{H}}) = 0.00642$ and $(M_{\rm{dust}_{\rm{PAH}}}/M_{\rm{H}}) = 0.000659$.

The \cite{2024A&A...684A..34Y} THEMIS 2.0 models are composed of three populations of grains: spherical carbonaceous grains ($a_{grain} \leq 20$ nm) consisting purely of aromatic-rich H-poor amorphous carbon, larger spheroidal (prolates with e = 2) carbonaceous grains where the core is made of aliphatic-rich H-rich carbon with a 20 nm thick mantle of aromatic-rich H-poor amorphous carbon, and spheroidal silicates (prolates with e = 2) consisting of amorphous silicate cores and 5 nm mantles of aromatic-rich H-poor amorphous carbon.\footnote{The THEMIS 2.0 models are available at \url{https://www.ias.u-psud.fr/themis/downloads_2.html}.}
The dust to hydrogen gas mass ratios for the different grain populations are: $(M_{\rm{dust}_{\rm{cm20}}}/M_{\rm{H}}) = 0.001319$, $(M_{\rm{dust}_{\rm{carb5nm}}}/M_{\rm{H}}) = 0.0008$ and $(M_{\rm{dust}_{\rm{aSil2}}}/M_{\rm{H}}) = 0.003266$.
One distinct characteristic of the \cite{2024A&A...684A..34Y} models is that the opacities in the silicate features are lower than the other plotted models which comes from optical constants measured in the laboratory by \cite{2022A&A...666A.192D}.
Additional differences between the diffuse ISM models are a result of the models using different sets of observational constraints.

Figure \ref{fig:Opacity1} also shows coagulated grain models with ice mantles for two of the most popular opacity models in the literature, OH5 and KP5, for dense cores \citep{1994A&A...291..943O,2024RNAAS...8...68P}.
OH5 and OH8 are shorthand for the opacities in column 5 and 8 respectively of Table 1 in \cite{1994A&A...291..943O} which report $\kappa_{\nu,abs}$ for grains that have coagulated (based on the calculations of \citealt{1993A&A...280..617O}), starting with a MRN distribution, for $10^5$ years at a gas density of $10^6$ cm$^{-3}$ that have accreted ``thin" and ``thick" H$_2$O ice mantles.\footnote{OH5 has thin ice mantles where thin means the ice mass is about 0.17 times the mass of the refractory grain mass matching the maximum ratios observed in the 3.1 $\mu$m ice band to $9.7$ $\mu$m silicate band. OH8 has coagulated at the same density and for the same amount of time as OH5, but has thick ice mantles where thick means all gas phase volatiles have accreted before coagulation begins.}
The density chosen for OH5 and OH8 opacities is high for a typical starless core (see Figure 7 of \citealt{2014prpl.conf...27A}).
In Table \ref{tab:dustopacity}, we have also tabulated \cite{1994A&A...291..943O} opacities for grains that have starting with a MRN distribution and coagulated for $10^5$ years at a gas density of at a density of $10^5$ cm$^{-3}$ and that have accreted ``thin" ice mantles. 
These ``OHSC" opacities may be more appropriate for a typical starless core.

While the OH5 opacities, in particular, are popular in the literature for dense cores and have properties that are consistent with submillimeter observations in environments when scattering is not important \citep{2011ApJ...728..143S}, their opacity calculations did not include scattering\footnote{Scattering opacity calculations from \cite{1994ApJ...421..615P} have been scaled and added to the OH5 models for radiative transfer calculations (see Section 2.1 of \citealt{2005ApJ...627..293Y} for details).}.
The KP5 opacity model (short for Klaus Pontoppidan grain model 5; \citealt{2024RNAAS...8...68P}), however, does include scattering calculations for coagulated populations of silicate and carbonaceous grains with mixtures of H$_2$O, CO$_2$, and CO ice. 
They are a good fit to mid-infrared \citep{2009ApJ...690..496C} and submillimeter opacity constraints \citep{2011ApJ...728..143S}.
One distinct character of KP5 opacities compared to OH5 is the flatness at mid-infrared wavelengths.

\begin{figure}[h]
\includegraphics[scale=0.45,trim= 0mm 50mm 0mm 30mm, clip]{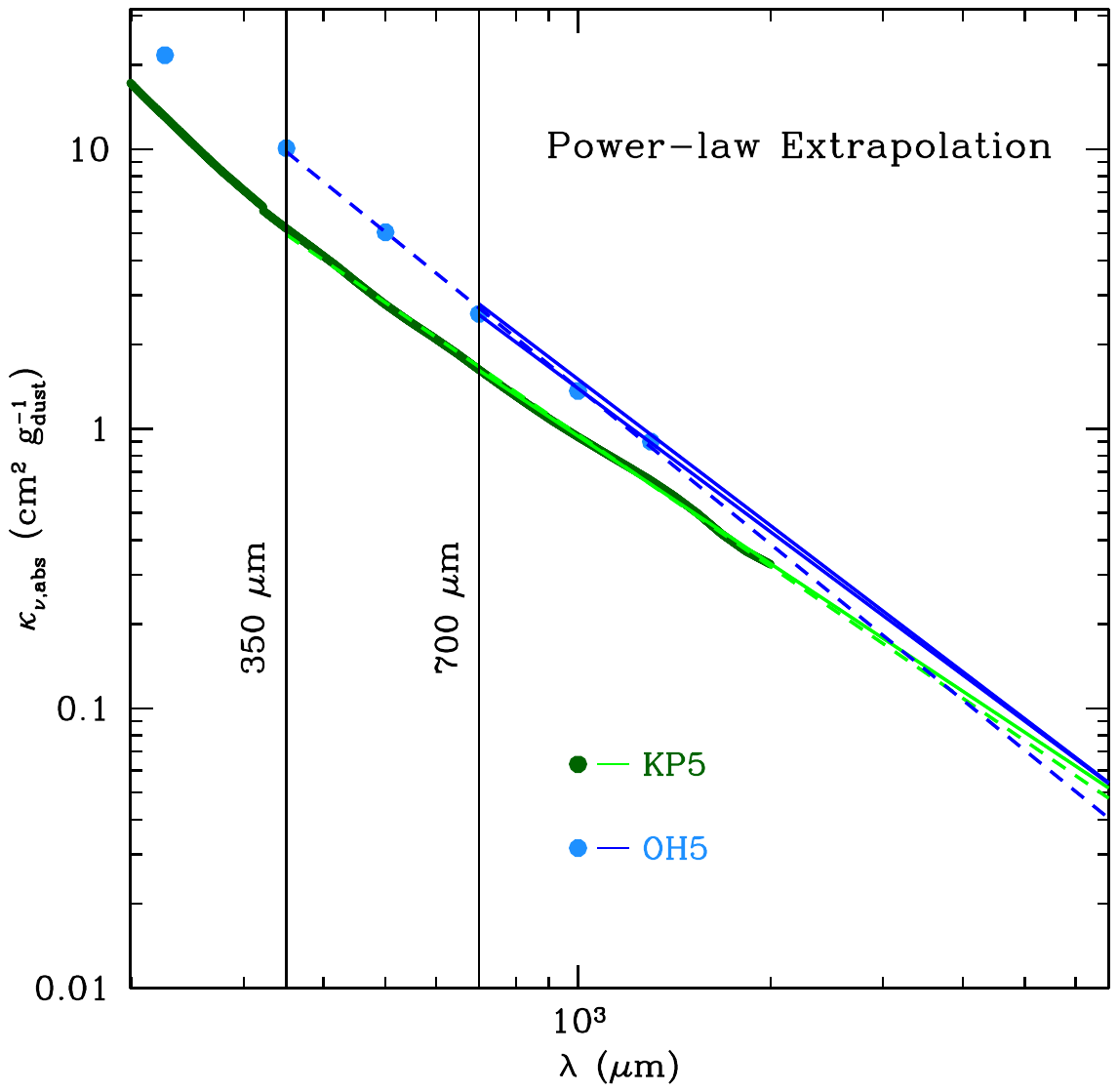}
\caption{Examples of power-law extrapolations of $\kappa_{\nu,\rm{abs}}$ for OH5 and KP5 dust opacities plotted to a maximum wavelength corresponding to $\lambda_{\rm{pivot}} = 6830$ $\mu$m for the \textit{Planck} LFI filter.  The published dust opacities are plotted as points with OH5 shown in light blue and KP5 shown in dark green (N.B. the wavelength spacing between KP5 opacities is so small that the points appear as a line).  Two power-law extrapolations are plotted for each model depending on whether the fits begins at $350$ $\mu$m (dashed lines) or at $700$ $\mu$m (solid lines).  The values reported in Table \ref{tab:dustopacity} use the extrapolations that start at $700$ $\mu$m (solid lines).  OH8 opacities are not plotted to minimize confusion. }
\label{fig:KapExtrap}
\end{figure}

Some of the published opacities such as \cite{1994A&A...291..943O} and \cite{2024RNAAS...8...68P} do not extend their calculation beyond 1300 $\mu$m and 2000 $\mu$m respectively. 
There are several bolometer cameras, however, that operate at longer wavelengths.
Opacities in Table \ref{tab:dustopacity} for $\lambda_{\rm{pivot}}$ longer than the longest published wavelength are extrapolated using a power-law fit.
The power-law extrapolation was determined from a linear regression of the published $\log_{10} \lambda$ and $\log_{10} \kappa_{\nu,\rm{abs}}$ values over some range of wavelengths.
Figure \ref{fig:KapExtrap} shows examples of the fits with starting wavelengths of $350$ $\mu$m and $700$ $\mu$m and extending to the maxium published wavelength.

For OH5 opacities fit between $350$ $\mu$m and 1300 $\mu$m, we find that $ \log_{10} \kappa_{\nu_{\rm{pivot}},\rm{abs}} = -1.853598 \log_{10} \lambda_{\rm{pivot}} + 5.706233  $.
If we instead fit between $700$ $\mu$m and 1300 $\mu$m, we find that $ \log_{10} \kappa_{\nu_{\rm{pivot}},\rm{abs}} = -1.702661 \log_{10} \lambda_{\rm{pivot}} + 5.251426  $. 
There is a substantial difference in the $\beta$ of $1.85$ compared to $1.70$ depending on whether we start the fit at $350$ $\mu$m or $700$ $\mu$m.
Looking at the inset in Figure \ref{fig:Opacity1}, it is apparent that $\beta$ is dropping from wavelengths of $700$ $\mu$m to 1300 $\mu$m.
Other dust opacities models tend to have smaller values of $\beta$ at wavelengths longer than 1300 $\mu$m.
Therefore, we calculate the extrapolated values for the fit to OH5, OH8, and OHSC between $700$ $\mu$m and 1.3 mm in Table \ref{tab:dustopacity}.
A single-power law extrapolation, however, does not capture the variation in $\beta$ as a function of wavelength and this extrapolation should be used with caution.

For KP5 the extrapolated opacities are given by $ \log_{10} \kappa_{\nu_{\rm{pivot}},\rm{abs}} = -1.567326 \log_{10} \lambda_{\rm{pivot}} + 4.680700  $ determined from a linear regression of the logarithmic values between $350$ $\mu$m $\leq \lambda_{\rm{pivot}} \leq 1300$ $\mu$m and $ \log_{10} \kappa_{\nu_{\rm{pivot}},\rm{abs}} = -1.516068 \log_{10} \lambda_{\rm{pivot}} + 4.522308  $ determined from a linear regression of the logarithmic values between $700$ $\mu$m $\leq \lambda_{\rm{pivot}} \leq 1300$ $\mu$m.
The difference between the fits starting at $350$ $\mu$m and $700$ $\mu$m is not as large as it is for the \cite{1994A&A...291..943O} models.
For consistency with the fits to OH5 and OH8 we calculate the extrapolated values for the fit to KP5 between $700$ $\mu$m and 2000 $\mu$m in Table \ref{tab:dustopacity}.

We do not tabulate opacities for protoplanetary disks in Table \ref{tab:dustopacity} because there are too many free parameters to make a useful tabulation.
$\kappa_{\nu,\rm{abs}}$ is a function of the grain composition, size distribution power-law ($q$), and the maximum grain size and protoplanetary disks can have a significant range in each of these variables.
For the DSHARP opacities presented in \cite{2018ApJ...869L..45B}, the opacities can be calculated from scripts that are provided by the authors\footnote{The code to construct DSHARP opacities may be found at \url{https://github.com/birnstiel/dsharp_opac}}.
Examples of the DSHARP opacities are shown in Figure \ref{fig:OpacityDSHARP}.

\section{Bolometric Temperature of the Spectral Energy Distribution}\label{AppendixTbol}

One useful quantity derived from the spectral energy distribution (SED) of dust emission that has been used for classifying protostars is the bolometric temperature of the source, $T_{\rm{bol}}$ \citep{1995ApJ...445..377C}.
Bolometric temperature is defined as the temperature of a blackbody that has the same mean frequency as the observed spectral energy distribution.
The observed mean frequency of the SED is
\begin{equation}
    \langle \nu \rangle = \frac{\int_{\nu = 0}^{\infty} \nu F_{\nu,tot} d\nu}{\int_{\nu = 0}^{\infty} F_{\nu,tot} d\nu} \;\; \rm{Hz} \;,
\end{equation}
where $F_{\nu,tot}$ mean the total flux from the source (\ie\ $F_{\nu,tot} = F_{\nu,src}$ over a solid angle of $\Omega_{src}$ that fully encompasses the source).
The mean frequency of a blackbody at temperature $T_{\rm{bol}}$ is
\begin{eqnarray}
    \langle \nu \rangle_{\rm{BB}} & = & \frac{\int_{\nu = 0}^{\infty} \nu B_{\nu}(T_{\rm{bol}}) d\nu}{\int_{\nu = 0}^{\infty} B_{\nu}(T_{\rm{bol}}) d\nu}  \\ \langle \nu \rangle_{\rm{BB}} & = & \frac{ \left( \frac{kT_{\rm{bol}}}{h} \right)^5 \Gamma(5) \zeta(5)}{ \left( \frac{kT_{\rm{bol}}}{h} \right)^4 \Gamma(4) \zeta(4)}  \\ \langle \nu \rangle_{\rm{BB}} & = & \frac{4 k T_{\rm{bol}} \zeta(5)}{h \zeta(4)} \;\; \rm{Hz} \;,
\end{eqnarray}
where $\Gamma$ and $\zeta$ and the Gamma Function and Reimann Zeta Function respectively.
The Reimann Zeta function has an exact value for $\zeta(4) = \pi^4/90$ while $\Gamma(5) = 24$ and $\Gamma(4) = 6$. 
No simple numerical formula is known for $\zeta(5) \approx 1.0369$.
Setting these average frequencies equal to each other ($\langle \nu \rangle = \langle \nu \rangle_{\rm{BB}}$), we find that
\begin{equation}
    T_{\rm{bol}} = \frac{\pi^4}{360 \zeta(5)} \frac{h \langle \nu \rangle}{k} \;\; \rm{K} \;.
\end{equation}

\section{Spectral Index of the Spectral Energy Distribution}\label{AppendixSpectralIndex}

The spectral index of a modified blackbody SED is found by taking the logarithmic derivative of Equation \ref{eq:sedtaupow},
\begin{eqnarray}
    \alpha_s & = & \frac{\partial \ln{F_{\nu}}}{\partial \ln{\nu}} \nonumber \\
     & = & \left( \frac{\nu}{F_{\nu}} \right) \frac{\partial F_{\nu}}{\partial \nu} \nonumber \\
    & = & \left( \frac{\nu}{F_{\nu}} \right) \left[ \frac{\partial B_{\nu}(T_d)}{\partial \nu}\left( 1 - e^{-\overline{\tau_{\nu}}} \right) +  B_{\nu}(T_d) \frac{\partial \left( 1 - e^{-\overline{\tau_{\nu}}} \right)}{\partial \overline{\tau_{\nu}}} \frac{\partial \overline{\tau_{\nu}}}{\partial \nu} \right] \Omega_{ap} \nonumber \\
    & = & \left( \frac{e^{h\nu/kT_d} - 1}{\nu^{2} (1 - e^{-\overline{\tau_{\nu}}}) } \right) \left[ 3\frac{\nu^{2} (1 - e^{-\overline{\tau_{\nu}}})}{e^{h\nu/kT_d} - 1} - \frac{h}{k T_d} \frac{\nu^{3}e^{h\nu/kT_d}(1 - e^{-\overline{\tau_{\nu}}})}{(e^{h\nu/kT_d} - 1)^2} + \beta \frac{\tau_0}{\nu_0^{\beta}}\frac{\nu^{2 + \beta} e^{-\overline{\tau_{\nu}}}}{e^{h\nu/kT_d} - 1} \right] \nonumber \\ 
     & = & 3  - \frac{\frac{h\nu}{kT_d}e^{h\nu/kT_d}}{e^{h\nu/kT_d} - 1} + \beta\frac{\overline{\tau_{\nu}}e^{-\overline{\tau_{\nu}}}}{1 - e^{-\overline{\tau_{\nu}}}}  \;\;\;\; ,
\end{eqnarray}
where we have used Equations~\ref{eq:sedtaupow}, \ref{eq:planck}, and \ref{eq:taunubar} for $F_\nu$, $B_\nu(T_d)$, and $\overline{\tau_{\nu}}$, respectively.  
Multiplying the second and third terms by 1 in the form of  $(e^{-h\nu/kT_d}/e^{-h\nu/kT_d})$ and $(e^{\overline{\tau_{\nu}}}/e^{\overline{\tau_{\nu}}})$, respectively, and rearranging the second and third terms gives the isothermal SED spectral index as
\begin{equation}
\alpha_s(\nu, \overline{\tau_{\nu}}, T_d)  =  3 + \beta\frac{\overline{\tau_{\nu}}}{e^{\overline{\tau_{\nu}}} - 1} - \frac{\frac{h\nu}{kT_d}}{1 - e^{-h\nu/kT_d}} \;\;\;\; .
\label{eq:sedspecindex}
\end{equation}
In the limit when the emission is optically thin ($\overline{\tau_{\nu}} \ll 1$), then the Maclaurin expansion of $e^{\overline{\tau_{\nu}}} \approx 1 + \overline{\tau_{\nu}}$ reduces the second term in Equation \ref{eq:sedspecindex} to $\beta$, eliminates the dependence on $\overline{\tau_{\nu}}$, and the spectral index is given by
\begin{equation}
  \lim_{\overline{\tau_{\nu}} \ll 1}  \;  \alpha_s(\nu, T_d) = 2 + \beta + \gamma_{\nu}(T_d) \;\;\;\; ,
\label{eq:specindplanck}
\end{equation}
where $\gamma_{\nu}(T_d)$ is a frequency and temperature dependent correction factor given by
\begin{equation}
   \gamma_{\nu}(T_d) =  1 - \frac{\frac{h\nu}{kT_d}}{1 - e^{-h\nu/kT_d}} \;\;\;\; .
\label{eq:gammatheoretical}
\end{equation}

\begin{figure}[h]
\includegraphics[scale=0.43,trim= 0mm 50mm 0mm 30mm, clip]{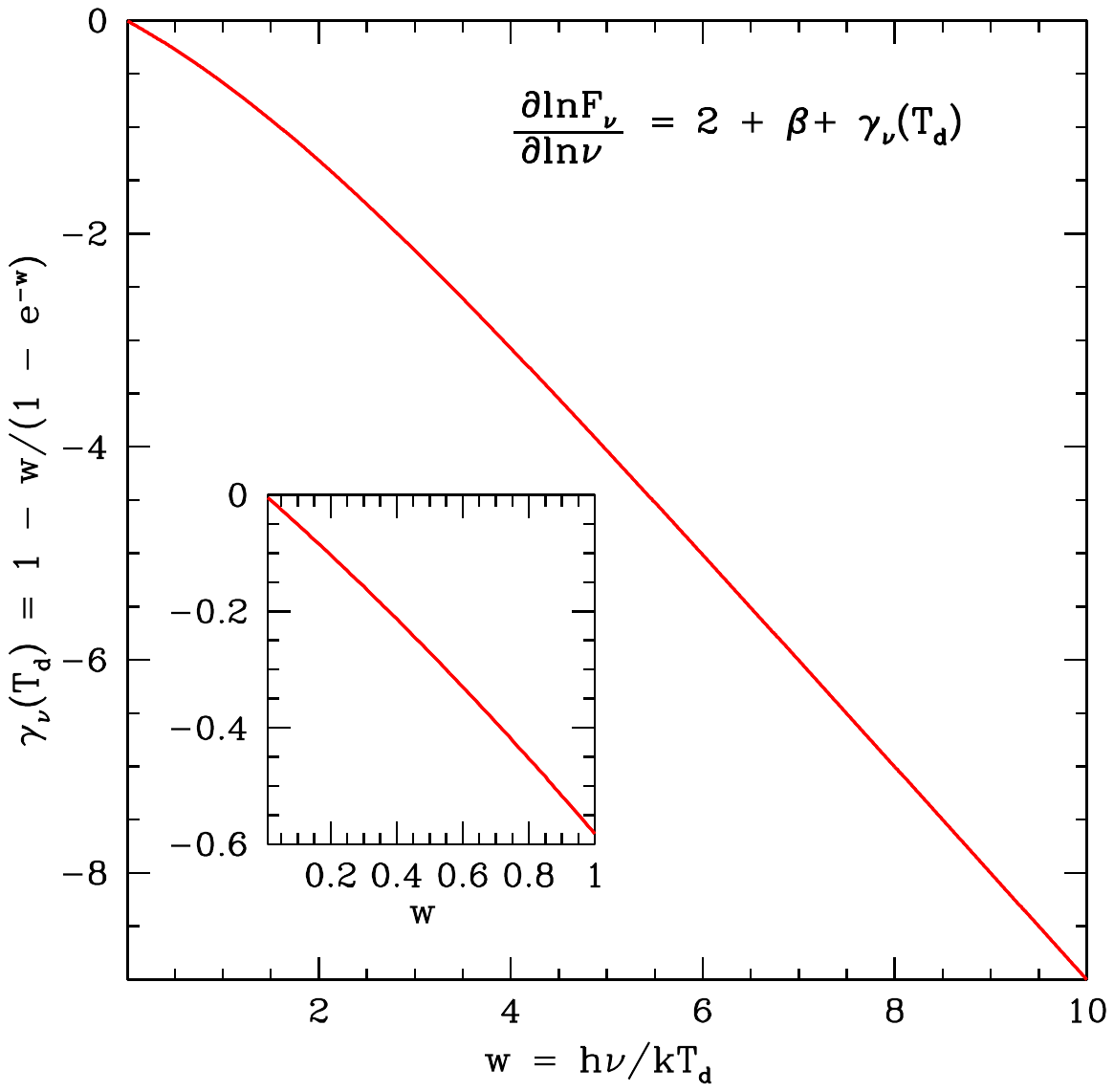}
\includegraphics[scale=0.43,trim= 0mm 50mm 0mm 30mm, clip]{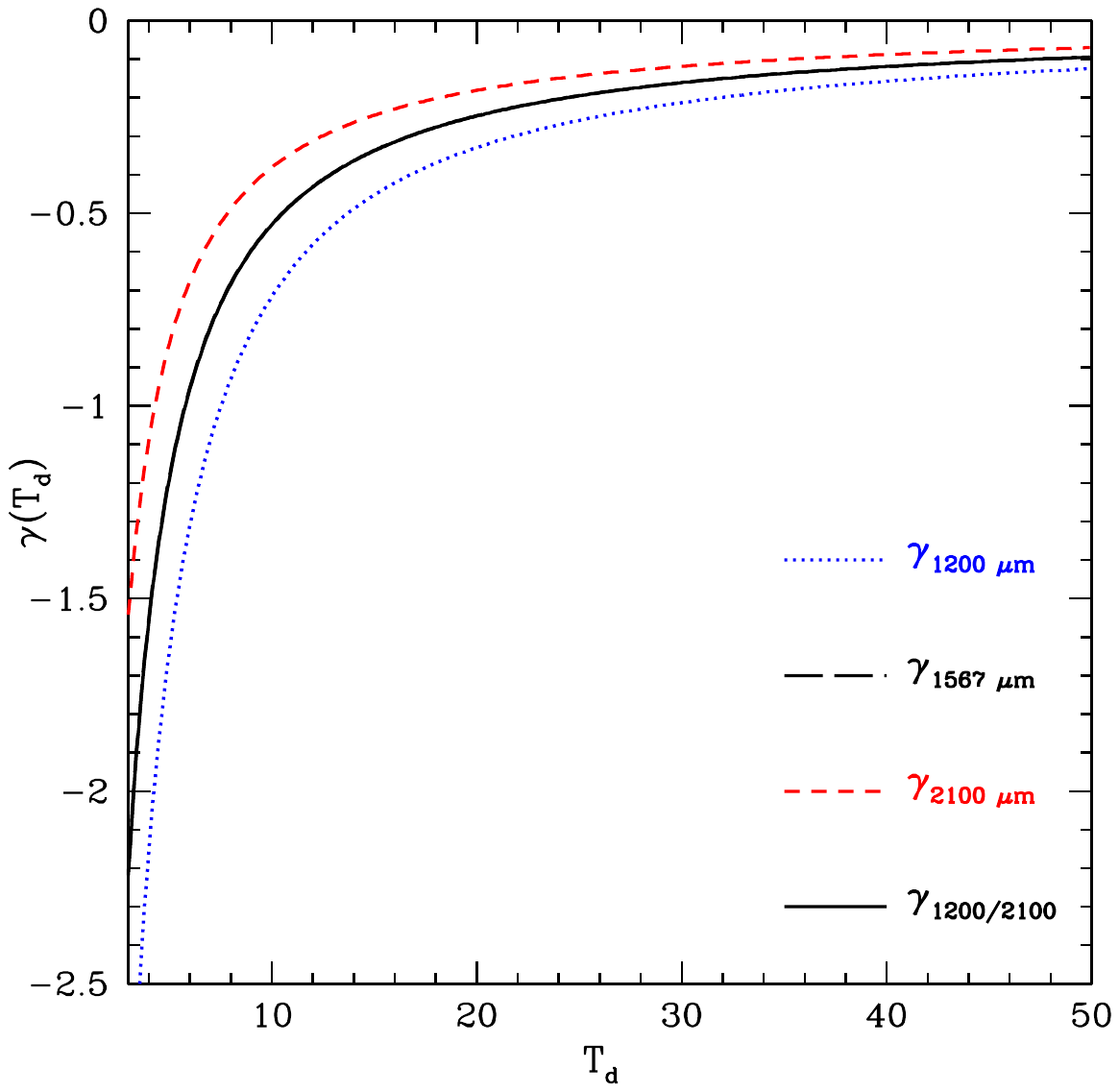}
\caption{\textbf{Left}: The Rayleigh-Jeans correction factor, $\gamma_{\nu}(T_d)$ (Equation \ref{eq:gammatheoretical}) plotted versus $w = h\nu/kT_d$ for an optically thin, isothermal dust continuum.  The inset  on the lower left shows a blow up of the same function over the range $w = h\nu/kT_d \leq 1$.  \textbf{Right}: The empirical Rayleigh-Jeans corrections factor for observations at $1200$ $\mu$m and $2100$ $\mu$m (solid black line, Equation \ref{eq:gammaempirical}) is plotted versus isothermal dust temperature.  The dotted blue line and the dashed red line are the Rayleigh-Jeans correction factors (Equation \ref{eq:gammatheoretical}) calculated at $1200$ $\mu$m and $2100$ $\mu$m respectively.  The long dashed black line is the Rayleigh-Jeans correction factor (Equation \ref{eq:gammatheoretical}) calculated at the logarithmic mean frequency, corresponding to $1567$ $\mu$m (= $c/\langle \nu \rangle_L)$.  This long dashed black curve nearly perfectly overlaps the solid black line with a maximum deviation of $0.009$ over the plotted $T_d$ range.}
\label{fig:GammaFig}
\end{figure}

Since $\alpha_s \propto \beta$ in the optically thin limit, the spectral index may be used to constrain $\beta$. 
Nevertheless, the dust temperature results in a correction factor, $\gamma_{\nu}(T_d)$, that must be included.
In the Rayleigh-Jeans limit, when $h\nu / kT_d \ll 1$, then $e^{-h\nu/kT_d} \approx 1 - (h\nu/kT_d)$, $\gamma_{\nu}(T_d) \rightarrow 0$ and $\alpha_s \rightarrow 2 + \beta$, indicating that an optically thin dust emission SED is steeper than a blackbody spectrum, which has $\alpha_s = 2$ in Rayleigh-Jean limit, for all $\beta > 0$.
Figure \ref{fig:GammaFig} shows $\gamma_{\nu}(T_d)$.
When $h\nu/kT_d = 0.2$, a value that is considered in the Rayleigh-Jeans limit, $\gamma_{\nu}(T_d) = -0.10$ and the dust temperature correction factor is small. 
By $h\nu/kT_d = 0.5$, $\gamma_{\nu}(T_d) = -0.27$, a more modest correction. 
When $h\nu/kT_d \gg 1$, $\gamma_{\nu}(T_d)$ is approximately a linear function with a value of $\gamma_{\nu}(T_d) \approx 1 - h\nu/kT_d$.
For example, at $500$ $\mu$m, if $T_d = 10$ K, then $h\nu/kT_d = 2.9$ and $\gamma_{500 \mu\rm{m}}(10 \;\rm{K}) = -2.0$.
In order to minimize the dust temperature correction factor, it is best to measure the spectral index at wavelengths as close to the Rayleigh-Jeans limit as possible.

An empirical spectral index is often measured between two frequencies, $\nu_1$ and $\nu_2$,
and is found by taking the logarithm\footnote{This expression will sometimes be defined in terms of $\log_{10}$ instead of the natural logarithm.  The expressions are equivalent.} of the ratio of Equation \ref{eq:SEDPL} at the two frequencies and solving for the spectral index,
\begin{equation}
    \alpha_{s,\nu_1/\nu_2} = \frac{\ln \left( \frac{F_{\nu_1}}{F_{\nu_2}} \right) }{\ln \left( \frac{\nu_1}{\nu_2} \right)} \;\;\;\; .
\label{eq:specind2freq}
\end{equation}
The order of $\nu_1/\nu_2$ or $\nu_2/\nu_1$ does not matter as long as it is the same order inside the logarithm in the numerator and in the denominator.
If an empirical spectral index was measured between two frequencies (Equation \ref{eq:specind2freq}), then this spectral index is related to the dust $\beta$ by
\begin{equation}
    \alpha_{s,\nu_1/\nu_2} = 2 + \beta + \gamma_{\nu_1/\nu_2}(T_d)   \;\;\;\; ,
\end{equation}
where $\gamma_{\nu_1/\nu_2}(T_d)$ is the temperature-dependent correction factor between the two frequencies
\begin{equation}
    \gamma_{\nu_1/\nu_2}(T_d) = 1 - \frac{\ln \left(\frac{e^{h\nu_1/kT_d} - 1}{e^{h\nu_2/kT_d} - 1}\right)}{\ln \left( \frac{\nu_1}{\nu_2} \right)} \;\;\;\; ,
\label{eq:gammaempirical}
\end{equation}
(see \citealt{1998MNRAS.301..585V}).

An example of this empirical Rayleigh-Jeans correction factor is shown in Figure \ref{fig:GammaFig} for observations at $1200$ $\mu$m and $2100$ $\mu$m\footnote{Bolometer filters are usuall named by their central wavelengths instead of frequencies.  We have converted to frequency $\nu = c/\lambda$ before plugging into Equations \ref{eq:gammatheoretical} and \ref{eq:gammaempirical}.} (\ie\ bands observable with the NIKA2 camera on the IRAM 30m telescope).
The empirical Rayleigh-Jeans correction factor is a strong function of the dust temperatures when $T_d < 20$ K, corresponding to $h\nu/kT_d = 0.6$ at 1200 $\mu$m.
The $\gamma_{\nu_1/\nu_2}(T_d)$ curve lies intermediate between the $\gamma_{\nu}(T_d)$ curves calculated at $1200$ $\mu$m and $2100$ $\mu$m.
This deviation occurs because the real SED has a different spectral index at the two frequencies of the observations (\ie\ the SED for $\lambda \in [1200, 2100]$ $\mu$m cannot be fit by a single power-law as is assumed in Equation \ref{eq:SEDPL}).
The empirical Rayleigh-Jeans correction factor (Equation \ref{eq:gammaempirical}) is a good approximation to the correction factor given in Equation \ref{eq:gammatheoretical} if it is calculated at the logarithmic mean of the two frequencies\footnote{It is possible to calculate the characteristic frequency, $\nu_c$ for Equation \ref{eq:gammatheoretical} at a given temperature that corresponds exactly to Equation \ref{eq:gammaempirical}.  Let $w = h\nu_c/kT_d$ and $u = 1 - \gamma_{\nu_1/\nu_2}(T_d)$.  Setting Equations \ref{eq:gammatheoretical} and \ref{eq:gammaempirical} equal to each other, we find that $w = u - ue^{-w}$.  This non-linear equation may be solved in terms of the principal branch of the Lambert $W_0$ function as $w = u + W_0(-u e^{-u})$, therefore $\nu_c = (kT_d/h)[u + W_0(-u e^{-u})]$.} given by
\begin{equation}
    \langle \nu \rangle_{L} = \frac{\nu_1 - \nu_2}{\ln{\frac{\nu_1}{\nu_2}} } \;\;\;\; \rm{Hz} \;\; .
\label{eq:logmean}
\end{equation} 
This formulation implies that $\alpha_{s,\nu_1/\nu_2} \approx \alpha_s( \langle \nu \rangle_{L})$\footnote{The logarithmic mean of two numbers is intermediate between the geometric mean and arithmetic mean of those numbers \citep{Burk1987}.}.

We must stress that equations in this section derived from Equation \ref{eq:sedspecindex} assume isothermality.
If the astrophysical source has temperature gradients, then there is no single dust temperature that applies to the formulae in this section at all frequencies.
It is possible to derive the temperature correction factor, $\gamma_{\nu}$, for an optically thin, constant opacity source with a temperature profile, $T_d(\vec{r})$ by logarithmically differentiating, with respect to frequency, the flux density Equation \ref{eq:Fmonosa} using the monochromatic specific intensity, $I_{\nu}(\theta,\phi)$, that is appropriate for the source geometry.
Since the integrals used to calculate $I_{\nu}$ and $F_{\nu}$ are integrated over geometric variables (\ie, line-of-sight coordinate z corresponding to a position on the sky given by $\theta$ and $\phi$), then the partial derivative with respect to frequency can be brought inside the integrals and operate directly on the frequency-dependent terms in the integrand,
\begin{equation}
    \alpha_s = \frac{\nu}{F_{\nu}} \int_{\phi = 0}^{2\pi} \int_{\theta = 0}^{\theta_{ap}} \frac{\partial I_{\nu}(
    \theta,\phi)}{\partial \nu} \theta d\theta d\phi \;\; .
\end{equation}
For example, in the case of an optically thin spherical source with radius R, with constant opacity, and with a purely radial dust temperature profile, then the monochromatic specific intensity profile is given by Equation \ref{eq:Inu_Abel} and the correction factor $\gamma_{\nu}[T_d(r)]$ in Equation \ref{eq:specindplanck} ($\alpha_s = 2 + \beta + \gamma_{\nu}$) now becomes
\begin{equation}
 \lim_{\kappa_{\nu}(\vec{r}) = \kappa_{\nu}}    \gamma_{\nu}[T_d(r)] = 1 - \frac{\int_{\theta = 0}^{\theta_{ap}} \int_{r = D\theta}^R  \frac{\frac{h\nu}{kT_d(r)}}{1 - e^{-h\nu/kT_d(r)}} \frac{\rho_d(r)}{e^{h\nu/kT_d(r)}  - 1} \frac{r\theta dr d\theta}{\sqrt{r^2 - D^2 \theta^2}}  }{\int_{\theta = 0}^{\theta_{ap}} \int_{r = D\theta}^R  \frac{\rho_d(r)}{e^{h\nu/kT_d(r)}  - 1} \frac{r \theta dr d\theta}{\sqrt{r^2 - D^2 \theta^2}}  } \;\; .
 \label{eq:gammasphereTr}
\end{equation}
Solutions to this equation require numerical techniques such as Simpson's Rule \citep{burden2011numerical}. 
Note that Equation \ref{eq:gammasphereTr} reduces to Equation \ref{eq:gammatheoretical} if the source is isothermal.
Also, if $h\nu/kT_d(r)$ is in the Rayleigh-Jeans limit at all radii ($h\nu/kT_d(r) \ll 1 \, \forall \, r \in [0,R]$), then the Macluarin expansion of the first fraction in the numerator integrand
will be $1$ (as we have seen above), the integrals in the numerator and denominator will then cancel, and $\gamma_{\nu}[T_d(r)] \rightarrow 0$ as expected.
For comparison of model spectral indicies to real observations, Equation \ref{eq:gammasphereTr} can be modified by placing the azimuthally-averaged telescope normalized beam pattern, $P_n(\theta)$, inside both the numerator and denominator integrals.

\section{Wien's Displacement Law for a Modified Blackbody  SED}\label{SEDPeakAppendix}

The frequency of the peak flux density of the SED, $\nu_{\rm{pk}}$ (Hz),  can be found by differentiating the observed flux density and setting this equal to zero
\begin{equation}
    \frac{d F_{\nu}}{d \nu} =  0 \;\; .
\end{equation}
For an isothermal source, the SED is singly peaked and therefore we can find the frequency that corresponds to a spectral index of $\alpha_s = 0$.
Starting with Equation \ref{eq:sedspecindex} for the spectral index of an isothermal, constant opacity source, we define two new variables, $w = h \nu/k T_d$ and $\epsilon = \overline{\tau_{\nu}}/(e^{\overline{\tau_{\nu}}} - 1)$, to recast the equation as 
\begin{equation}
    \alpha_s = 0 = 3 + \beta\epsilon - \frac{w}{1 - e^{-w}}  \;\; .
\label{eq:alphazero}
\end{equation}
With some algebraic manipulation, we can re-write Equation \ref{eq:alphazero} as a non-linear equation solving for $w$,
\begin{eqnarray}
     \frac{w}{1 - e^{-w}} & = & (3 + \beta\epsilon) \\
    w & = & (3 + \beta\epsilon)(1 - e^{-w})    \\
    w & = & (3 + \beta\epsilon) - (3 + \beta\epsilon)e^{-w}  \;\;.
\label{eq:Wienw}
\end{eqnarray}
The general solution to an equation of the form $w = c_1 + c_2e^{c_3w}$ is given in terms of the principal branch of the Lambert $W_0$ function (see \citealt{Corless1996OnTL} for a description of the Lambert W function) as
\begin{equation}
    w = c_1 - \frac{W_0(-c_2c_3e^{c_1c_3})}{c_3} \;\; .
\end{equation}
In Equation \ref{eq:Wienw} we have the constants $c_1 = 3+ \beta\epsilon$, $c_2 = -(3 + \beta\epsilon)$, and $c_3 = -1$ such that, after converting from $w$ back to frequency ($\nu_{\rm{pk}} = kT_dw/h$), the peak SED frequency for a constant opacity, isothermal source is given by
\begin{equation}
      \nu_{\rm{pk}} = \frac{k T_d}{h} \left[ (3 + \beta\epsilon) + W_0\left(- \frac{3 + \beta\epsilon}{e^{3 + \beta\epsilon}}\right)  \right] \; \rm{Hz} \;\;.
\label{eq:WienLawbeta}
\end{equation}

For a blackbody where $\tau_{\nu_{\rm{pk}}} \gg 1$, then $\epsilon \rightarrow 0$, $W_0(-3e^{-3}) \approx -0.17856$, and Wien's Displacement Law for a blackbody at temperature $T_d$, $\nu_{\rm{pk}}/T_d \approx 5.880 \times 10^{10}$ Hz, is recovered.
Equation \ref{eq:WienLawbeta} is the general form of Wien's Displacement Law for $\beta \geq 0$ and any optical depth.

When the peak frequency is in the optically thin limit ($\tau_{\nu_{\rm{pk}}} \ll 1$),
then $\epsilon \rightarrow 1$ and Equation \ref{eq:WienLawbeta} is well approximated by a linear fit to better than $3$\% for $\beta \in [0,4]$ with
\begin{equation}
     \lim_{\tau_{\nu} \ll 1} \; \nu_{\rm{pk}}  \left( \frac{10 \rm{K}} {T_d} \right) \approx 1.0348 + 0.2174(\beta - 2) \;\;\;\; \rm{THz} \;\; .
\label{eq:WienLawApprox}
\end{equation}
Equation \ref{eq:WienLawApprox} is only strictly valid in the optically thin limit.
For example, the $\beta = 3$ curve in Figure \ref{fig:SEDBetaFig} has a peak wavelength
($\lambda_{\rm{pk}} = c / \nu_{\rm{pk}}$) from Equation \ref{eq:WienLawbeta} of $\lambda_{\rm{pk}} = 126.780$ $\mu$m while the peak wavelength predicted by Equation \ref{eq:WienLawApprox} is $119.706$ $\mu$m.
These peak wavelengths are different by $6$\%, which is greater than the quoted accuracy of $3$\% for the linear approximation, because the optical depth at the SED peak is not strictly in the optically thin limit ($\tau_{\nu_{\rm{pk}}} = \tau_0 \left[\nu_{\rm{pk}}/\nu_0 \right]^{\beta} = 0.002[\nu_{\rm{pk}}/500 \,\rm{GHz}]^{3} = 0.21$).

\section{Equilibrium Grain Temperature from Diffuse ISRF}\label{AppendixISRF}

If the main source of incident flux density is the surrounding interstellar radiation field (ISRF), then it is more convenient to rewrite the incident flux density in terms of the energy density of the surrounding radiation field, $u_{\nu}$ (erg cm$^{-3}$ Hz$^{-1}$).
The energy density of a radiation field is related to the specific intensity by
\begin{equation}
    u_{\nu}(\vec{r}) = \frac{1}{c} \int_{\Omega} I_{\nu}(\vec{r},\hat{r}^{\prime}) d\Omega \;\;\;\; \rm{erg} \; \rm{cm}^{-3} \, \rm{Hz}^{-1} \;\;.
\label{eq:ISFEnergyDensity}
\end{equation}
In the case of the short wavelength portion of the ISRF (where dust grains are most efficient at absorbing radiation), the energy density is dominated by the light from the photospheres of a myriad of distant stars with small solid angles.
The energy density equation above becomes the sum of $\bar{I}_{\nu, *} \Omega_{*} / c$ for each star.
As the flux density from one of those stars is $f_{\nu, *} = \bar{I}_{\nu, *} \Omega_{*}$, we can relate the total incident flux density to the ISRF energy density by
\begin{equation}
    f_{\nu,\rm{ISRF}} = c u_{\nu,\rm{ISRF}}  \;\;\;\; \rm{erg} \; \rm{s}^{-1}\, \rm{cm}^{-2}\, \rm{Hz}^{-1} \;\; .
\end{equation}
Substituting for incident flux density and for the effective cross section for absorption, $\sigma_{\nu, abs} = m_{\rm{grain}} \kappa_{\nu,abs}$ in Equation~\ref{eq:dEdt_abs}, we find that
\begin{equation}
    \left(\frac{dE}{dt}\right)_{abs} = \int_{\nu = 0}^{\infty} c u_{\nu,\rm{ISRF}} m_{\rm{grain}} \kappa_{\nu,abs} d\nu \;\;\;\; \rm{erg} \; \rm{s}^{-1} \;\;.
\label{eq:dEdt_abs4}
\end{equation}
Following \cite{Draine2011}, it is convenient to define the energy-density weighted mean opacity, 
\begin{eqnarray}
    \langle \kappa_{\nu,abs} \rangle_{u} & = & \frac{\int_{\nu = 0}^{\infty} u_{\nu,\rm{ISRF}} \kappa_{\nu, abs} d\nu}{\int_{\nu = 0}^{\infty} u_{\nu,\rm{ISRF}} d\nu} \\
    \langle \kappa_{\nu,abs} \rangle_{u} & = & \frac{\int_{\nu = 0}^{\infty} u_{\nu,\rm{ISRF}} \kappa_{\nu, abs} d\nu}{ u_{\rm{ISRF}}} \;\;\;\;  \rm{cm^2} \; (\rm{g} \; \rm{of} \; \rm{dust})^{-1}  \;\; ,
\label{eq:kappa_mean2}
\end{eqnarray}
such that the energy absorption rate (Equation~\ref{eq:dEdt_abs2}) can be written as
\begin{equation}
    \left(\frac{dE}{dt}\right)_{abs} = c \, u_{\rm{ISRF}} \, m_{\rm{grain}} \, \langle \kappa_{\nu,abs} \rangle_{u} \;\;\;\; \rm{erg} \; \rm{s}^{-1} \;\; ,
\label{eq:dEdt_abs5}
\end{equation}
where $u_{\rm{ISRF}} = \int u_{\nu,\rm{ISRF}} d \nu$ is the total energy density of the surrounding radiation field (erg cm$^{-3}$).
Equating the energy absorption rate and the energy emission rate gives
\begin{equation}
       c \, u_{\rm{ISRF}} \, \langle \kappa_{\nu,abs} \rangle_{u} = 4  \langle \kappa_{\nu,abs} \rangle_{T_d} \, \sigma_{\rm{SB}} T_d^4 \;\;\;\; \rm{erg} \; \rm{s}^{-1} \;\; .
\label{eq:EnergyBalance3}
\end{equation}
We can solve for the equilibrium dust temperature which is given by
\begin{equation}
    T_d = \frac{h \nu_0}{k} \left( \frac{\langle \kappa_{\nu,abs} \rangle_{u} \, c^3 }{8 \pi \Gamma(4 + \beta) \zeta(4 + \beta) \kappa_{\nu_0} h \nu_0^{4} } \; u_{\rm{ISRF}}  \right)^{\frac{1}{(4+\beta)}} \;\;\;\; K \;\;.
\label{eq:TdISRF}
\end{equation}
The equilibrium dust temperature is $T_d \propto u_{\rm{ISRF}}^{-(4+\beta)}$.
Just as was found for the dependence of the equilibrium dust temperature on the central luminosity of a source, it take substantial changes in the strength of the ISRF to significantly affect equilibrium dust grain temperature.

We note that there are studies in the literature of how the equilibrium dust temperature varies with the amount of extinction parametrized by $A_{\rm{V}}$ (mag).
For example, \cite{2017A&A...604A..58H}, find an empirical formula of 
\begin{equation}
    T_d \approx [11 + 5.7\tanh (0.61 - \log_{10}(A_{\rm{V}}))] \chi_{\rm{UV}}^{\frac{1}{5.9}} \;\;\;\; K \;\; ,
\label{eq:TdAvEmp}
\end{equation}
where $\chi_{\rm{UV}}$ is a parameter characterizing the interstellar radiation field strength in the ultraviolet ($6 - 13$ eV) from \cite{1978ApJS...36..595D}.
A $\chi_{\rm{UV}} = 1$ corresponds to G$_0 = 1.7$ which is $1.7$ times the Habing measured flux of $1.6 \times 10^{-3}$ erg s$^{-1}$ cm$^{-2}$ \citep{1968BAN....19..421H}.
Note that for $\beta = 1.9$ in Equation \ref{eq:TdISRF}, these expressions have a similar functional dependence on their respective measures of the integrated strength of the ISRF.
See \cite{2017A&A...604A..58H} Table H.1 and Equation H.1 for empirical fits to different types of dust.

\bibliography{DustTutorial}{}
\bibliographystyle{aasjournalv7}



\end{document}